Nocturnality, seasonality and the SARS-CoV-2 Ecological Niche

Geraldine Finlayson[1,2,3]*, Stewart Finlayson[1,4], Clive Finlayson[1,2,3,5], Keith Bensusan[6,3], Rhian Guillem[6,3], Tyson L. Holmes[1,3], Francisco Giles Guzman[1], José S. Carrión[7], Cristóbal Belda[8], Lawrence Sawchuk[5]

1. The Gibraltar National Museum, 18-20 Bomb House Lane, Gibraltar

2. Department of Life Sciences, Liverpool John Moores University, United Kingdom

3. Institute of Life and Earth Sciences, The University of Gibraltar, Gibraltar

4. Department of Life Sciences, Anglia Ruskin University, Cambridge, United Kingdom

5. Department of Social Sciences, University of Toronto Scarborough, Canada

6. Gibraltar Botanic Gardens, 'The Alameda', Gibraltar

7. Departamento de Biologia, Universidad de Murcia, Spain.

8. Instituto de Salud Carlos III, Ministerio de Ciencia e Innovación, Spain.

* Corresponding author

E-mail: geraldine.finlayson@gibmuseum.gi (GF)

## Abstract

Understanding the behaviour of hosts of SARS-CoV-2 is crucial to our understanding of the virus. A comparison of environmental features related to the incidence of SARS-CoV-2 with those of its potential hosts is critical. We examine the distribution of coronaviruses among bats. We analyse the distribution of SARS-CoV-2 in a nine-week period following lockdown in Italy, Spain, and Australia. We correlate its incidence with environmental variables particularly ultraviolet radiation, temperature, and humidity. We establish a clear negative relationship between COVID-19 and ultraviolet radiation, modulated by temperature and humidity. We relate our results with data showing that the bat species most vulnerable to coronavirus infection are those which live in environmental conditions that are similar to those that appear to be most favourable to the spread of COVID-19. The SARS-CoV-2 ecological niche has been the product of long-term coevolution of coronaviruses with their host species. Understanding the key parameters of that niche in host species allows us to predict circumstances where its spread will be most favourable. Such conditions can be summarised under the headings of nocturnality and seasonality. High ultraviolet radiation, in particular, is proposed as a key limiting variable. We therefore expect the risk of spread of COVID-19 to be highest in winter conditions, and in low light environments. Human activities resembling those of highly social cave-dwelling bats (e.g. large nocturnal gatherings or high density indoor activities) will only serve to compound the problem of COVID-19.

## Introduction

A range of coronaviruses have been identified in over 100 bat species in Asia, Europe, Africa, Australia and America (Ge *et al*., 2015). The similarity of SARS-CoV-2 (cause of the current COVID-19 pandemic, Li *et al*., 2020) to bat SARS-CoV-like coronaviruses makes it likely that bats may have been reservoir hosts for the SARS-CoV-2 progenitor (Andersen *et al*. 2020, Boni *et al*. 2020, Zhou *et al*. 2020). Coronaviruses are an ancient viral lineage with an estimated mean time of the most recent common ancestor (tMRCA) of approximately 293 million years ago (range 190-489 mya), which roughly coincides with the inferred tMRCA of birds and mammals (Wertheim *et al*. 2013). It is therefore likely that co-evolutionary relationships exist between coronaviruses and their natural hosts (Vijaykrishna *et al*. 2007, Wang *et al*. 2011, Zhang *et al*. 2013). Coronaviruses and bats would be expected to have ecological features in common and the coronavirus environment should mirror that of its natural hosts. A number of recent papers have highlighted possible links between the incidence of SARS-CoV and climatic variables at various spatial scales (Tan *et al*. 2005, Yuan *et al*. 2006) and specifically, SARS-CoV-2 (Araujo and Naimi 2020, Moriyama *et al*. 2020). Here we examine the relationship between climatic variables and the incidence of COVID-19 (and hence SARS-CoV-2) using data from Italy, Spain and Australia. We link the observed associations with the ecology of the natural coronavirus hosts, specifically bats, and argue that these reflect deep time co-evolutionary ecological relationships.

## Materials and Methods

We compiled a database of bat species known to have been infected by coronaviruses (from Ge *et al*. 2015). We also generated a database of bat species to which we added ecological and behavioural features: 545 species for which data were available (Wilson and Mittermeier 2019). This allowed us to examine if bats known to have been infected by coronaviruses had particular ecological and behavioural characteristics.

We selected Italy and Spain for an analysis of COVID-19 incidence, being two countries which had been significantly affected by COVID-19 at an early stage and which had a number of regions within each country: (a) 20 for Italy; and (b) 19 for Spain. We analysed daily reported new cases for a period of seven weeks following lockdown (data sources: http://www.protezionecivile.gov.it/ and https://elpais.com/sociedad/2020/03/30/actualidad/1585589827_546714.html). The logic behind the exercise was to examine the effect of a national lockdown (acting as a standardized control between regions) on COVID-19 incidence, the assumption being that any differences between regions would have to implicate variables other than lockdown itself. As seasonal climatic and environmental variables were being examined, it was decided to contrast the results with those of a southern hemisphere country. We chose Australia as it offered a range of regions (8 states and territories) that would permit analysis. In this case we examined daily new cases listed on the website of the John Hopkins University Center for Systems Science and Engineering (CSSE, 2020). Several studies have examined cases using four time delays in relation to weather conditions: zero, three, seven and fourteen days (Chen *et al*. 2020, Liu *et al*. 2020). We adopted the same approach to time lags for this part of our study. We understand that there may have been errors in reporting of daily cases, which may have added noise to our analyses. Mortality data may have been more accurate than incidence data, although also subject to reporting errors, but would not have reflected the reach of the virus in each region as the majority of cases do not end in death.

For each region within the three countries, we compiled a database of daily climatological variables obtained from the OpenWeather website (OpenWeather, 2020), coinciding with the seven-week period from the commencement of lockdown in each case. Climatological data included daily Ultraviolet Index (UVI) data, temperature (mean, maximum and minimum), relative humidity, wind speed and direction, rainfall, and cloud cover for each country/territory. These data were derived and summarised from hourly data. For each territory, the station closest to the capital of the territory was chosen.

Multivariate statistics, using Microsoft Excel and SPSS, were used to analyse the data. Where zero cases were reported for a particular day, we added a constant (0.01) to each value prior to log transformation, in order to deal with the log of zero in our stepwise multiple regression models. With regards to temperature, we used Kelvin scale to avoid zeros when log transforming. Details of models used and variables entered are provided, as appropriate, in Appendix 1.

## Null Hypothesis

It could be inferred from the long period of coronavirus-bat coevolution, spanning millions of years (Mao *et al*. 2010, 2013, You *et al*. 2010, Latinne *et al*. 2020, Wertheim *et al*. 2013), that bats and coronaviruses would share common features of climatic tolerance shaped by a co-shared environment. Among the features of the bat-coronavirus environment, given the generalised nocturnal behaviour of bats, absence of solar radiation would appear to be the most widespread and prevalent variable.

Of the ultraviolet (UV) radiation reaching the Earth's surface it I, s UV-B (280 to 315 nm) which causes damage and mutations to living organisms (Flenley 2007). UV-B is therefore of particular interest to us and its absence would be the constant feature of all organisms living in darkness (Figure 1).

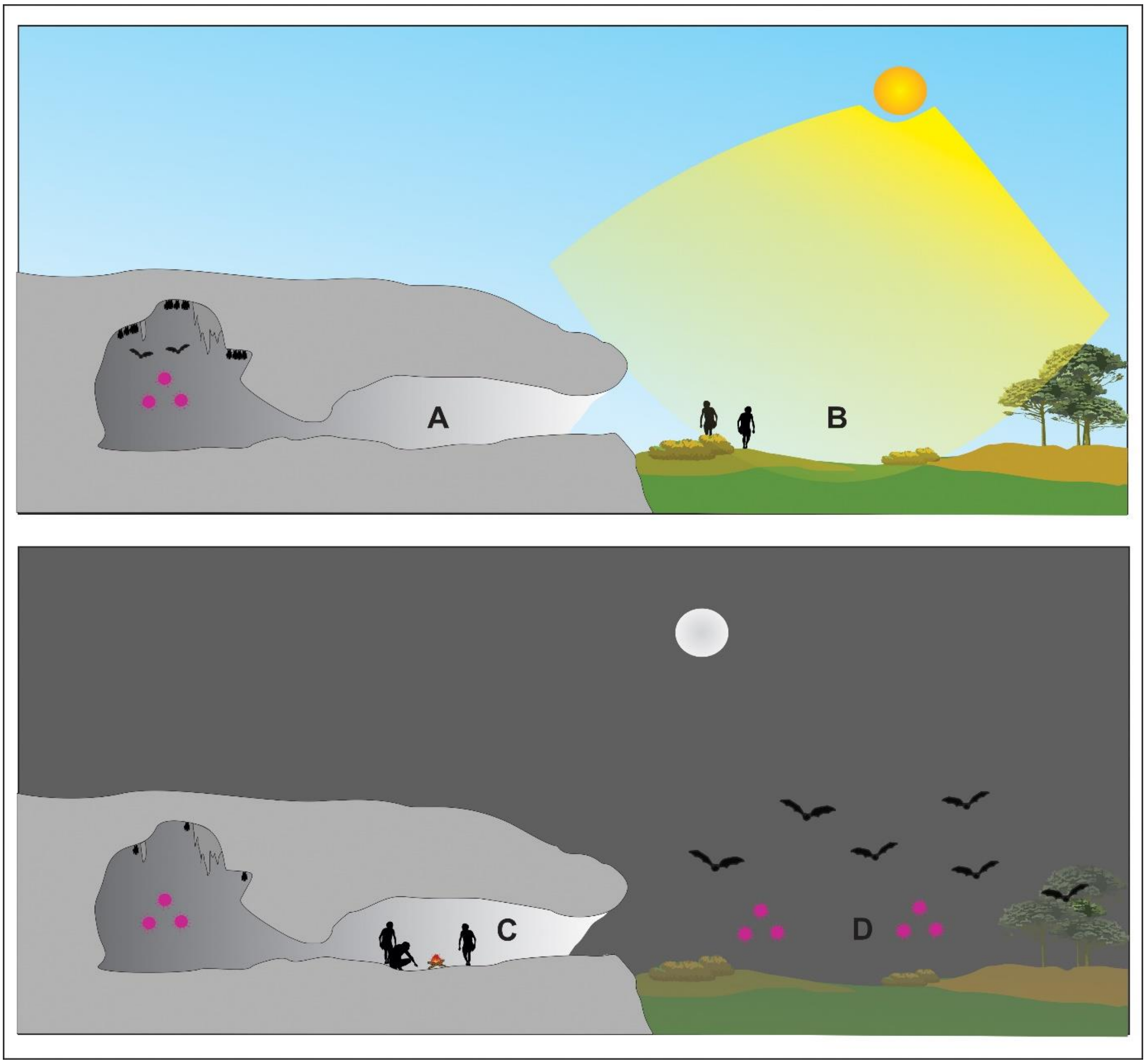

*Figure 1: Absence of UV-B would be a constant feature of all organisms living in the darkness of caves, and emerging at night. Presence of UV-B is a key factor defining the ecological niche of bats and corona viruses.* (A – presence of bats and viruses in caves during the day, B – absence of bats and viruses in the exterior during the day, C – reduced presence of bats and viruses in caves during the night, D – presence of bats and viruses in the exterior during the night.)

In addition to darkness (Rowse *et al*. 2016), a wide range of temperatures (but not exceeding ~35ºC), and high relative humidity (between 60 and 100%; Perry 2013) also appear as regular features of bat cave environments. In seeking patterns linking SARS-CoV-2 to environmental variables, we would expect the closest relationships to be between the distribution of the virus and these particular variables, especially UV-B. Our null hypothesis is therefore that there is no relationship between (a) ultraviolet radiation (UVR); (b) a broad temperature range, excluding very high temperatures, and (c) high relative humidity, and the incidence of SARS-CoV-2. Failure to uphold the null hypothesis would, instead, implicate some or all of these climatic and environmental variables with the incidence of SARS-CoV-2.

# Results

## Bats, coronaviruses and the shared environment

Ge *et al*. (2015) list the bat species recorded with coronavirus infection. We have used this list to examine the broad ecological and behavioural characteristics of bats which have been identified with coronavirus infection.

A comparison by social status of number of species detected with coronavirus infection, with those that were not infected, shows that the gregarious species were more likely to be infected than the less social species (one-way $\chi^2_4 = 44.241$; $p = 0.000$). Similarly, bat species inhabiting caves were more likely to be infected than those not inhabiting caves (one-way $\chi^2_1 = 24.387$; $p = 0.000$). Bats harbouring coronaviruses are therefore predominantly highly gregarious cave dwellers that gather to roost in large numbers and at high densities (Tables 1 & 2).

*Table 1: The number and proportion of cave-dwelling bats by family and their relationship with the incidence of coronavirus infection. Bat ecological features from Wilson & Mittermaier (2019) and coronavirus incidence from Ge et al. (2015). [1] sub-order Yingpterochiroptera, [2] Yangpterochiroptera; [a] superfamily Pteropodoidea, [b] Rhinolophoidea, [c] Emballonuroidea, [d] Noctilionoidea, [e] Vespertilionoidea.*

| FAMILY | CAVE SPECIES | NON-CAVE SPECIES | TOTAL SPECIES | PRO-PORTION (%) OF SPECIES THAT ARE CAVE DWELLERS | SPECIES DETECTED WITH INFECTION | PROPORTION (%) INFECTED | PRO-PORTION (%) OF INFECTED THAT ARE CAVE DWELLERS | PRO-PORTION (%) OF INFECTED THAT ARE CAVE OR FISSURE DWELLERS |
|---|---|---|---|---|---|---|---|---|
| *Pteropodidae*[1,a] | 4 | 19 | 23 | 17.4 | 8 | 34.8 | 50 | 62.5 |
| *Rhinopomatidae*[1,b] | 4 | 0 | 4 | 100 | 1 | 25 | 100 | 100 |
| *Megadermetidae*[1,b] | 5 | 1 | 6 | 83.3 | 1 | 16.7 | 100 | 100 |
| *Rhinonycteridae*[1,b] | 7 | 0 | 7 | 100 | 1 | 14.3 | 100 | 100 |
| *Hipposideridae*[1,b] | *52* | *4* | *56* | *92.9* | *6* | *10.7* | *100* | *100* |
| *Rhinolophidae*[1,b] | *50* | *3* | *53* | *94.3* | *13* | *24.5* | *100* | *100* |
| *Emballonuridae*[2,c] | *30* | *9* | *39* | *76.9* | *1* | *2.6* | *100* | *100* |
| *Mystacinidae*[2,d] | *0* | *1* | *1* | *0* | *1* | *100* | *0* | *0* |
| *Mormoopidae*[2,d] | *10* | *0* | *10* | *100* | *2* | *20* | *100* | *100* |
| *Phyllostomatidae*[2,d] | *43* | *33* | *76* | *56.7* | *9* | *11.8* | *100* | *100* |
| *Molossidae*[2,e] | *25* | *46* | *71* | *35.2* | *8* | *11.3* | *50* | *100* |
| *Miniopteridae*[2,e] | *18* | *0* | *18* | *100* | *8* | *44.4* | *100* | *100* |
| *Vespertilionidae*[2,e] | *69* | *112* | *181* | *61.6* | *28* | *15.5* | *75* | *100* |

We detected no difference in rate of infection among bats of the two major taxonomic subdivisions, suborders Yingpterochiroptera and Yangpterochiroptera (Table 1; two-way Pearson $\chi^2_1$ = 1.775; p= 0.183). There were no observable differences between the four major superfamilies Pteropodoidea, Rhinolophoidea, Noctilionoidea and Vespertilionoidea either (Table 1; two-way Pearson $\chi^2_3$ = 5.903; p= 0.116).

We detected no differences in the number of species infected when compared by social status (two-way Pearson $\chi^2_4$ = 2.777; p= 0.596) or caves (two-way Pearson $\chi^2_1$ = 1.804; p= 0.179) between the two suborders. In addition, the bat families most implicated in coronavirus-bat coevolution (Rhinolophidae and Miniopteridae/Vespertilionidae; Latinne *et al*. 2020) belong in different sub-orders (Table 1). The propensity to infection is therefore phylogenetically independent.

Caves offer a sheltered environment capable of accepting large numbers of bats. The large gatherings and dense clustering presumably provide the context for viral transmission within the caves (Kuzmin *et al*. 2011, Willoughby *et al*. 2017) and the use of a cave by various species, in some cases in very close proximity, raises the possibility of interspecific viral exchange (Messenger *et al*. 2003, Kuzmin *et al*. 2011). It is notable that among the fruit bats (Pteropodidae), a family noted for typically roosting on trees, the four known cave dwelling species have all been reported with coronavirus infections, three of these species being well-known for gathering in very large numbers (up to a million individuals; Wilson and Mittermeier 2019) in roosts (Table 2).

*Table 2: The number and proportion of bats by family and social status and their relationship with the incidence of coronavirus infection. Bat behavioural features from Wilson & Mittermaier (2019) and coronavirus incidence from Ge et al. (2015). Social status categories are not mutually exclusive: A solitary; B forms gatherings of up to 100 individuals; C of 101-1000 individuals; D of 1001-10,000 individuals; E of over 10,001 individuals.*

| | SOCIAL STATUS CATEGORY | | | | | NUMBER INFECTED IN SOCIAL STATUS CATEGORY | | | | | PROPORTION INFECTED IN SOCIAL STATUS CATEGORY | | | | |
|---|---|---|---|---|---|---|---|---|---|---|---|---|---|---|---|
| FAMILY | A | B | C | D | E | AI | BI | CI | DI | EI | AI% | BI% | CI% | DI% | EI% |
| Pteropodidae | 12 | 15 | 7 | 6 | 5 | 3 | 4 | 2 | 2 | 4 | 25.0 | 26.7 | 28.6 | 33.3 | 80.0 |
| Rhinopomatidae | 0 | 3 | 3 | 2 | 1 | 0 | 1 | 1 | 1 | 0 | | 33.3 | 33.3 | 50.0 | 0.0 |
| Megadermetidae | 1 | 3 | 4 | 2 | 0 | 0 | 1 | 1 | 0 | 0 | 0.0 | 33.3 | 25.0 | 0.0 | |
| Rhinonycteridae | 0 | 4 | 3 | 4 | 3 | 0 | 1 | 1 | 1 | 0 | | 25.0 | 33.3 | 25.0 | 0.0 |
| Hipposideridae | 8 | 46 | 32 | 11 | 3 | 0 | 4 | 4 | 2 | 0 | 0.0 | 8.7 | 12.5 | 18.2 | 0.0 |
| Rhinolophidae | 16 | 34 | 29 | 15 | 0 | 4 | 9 | 10 | 7 | 0 | 25.0 | 26.5 | 34.5 | 46.7 | |
| Emballonuridae | 5 | 31 | 17 | 5 | 1 | 0 | 1 | 0 | 0 | 0 | 0.0 | 3.2 | 0.0 | 0.0 | 0.0 |
| Mystacinidae | 0 | 0 | 0 | 1 | 0 | 0 | 0 | 0 | 1 | 0 | | | | 100.0 | |
| Mormoopidae | 0 | 3 | 3 | 5 | 2 | 0 | 2 | 2 | 0 | 0 | | 66.7 | 66.7 | 0.0 | 0.0 |
| Phyllostomatidae | 4 | 61 | 18 | 6 | 5 | 1 | 7 | 8 | 0 | 0 | 25.0 | 11.5 | 44.4 | 0.0 | 0.0 |
| Molossidae | 5 | 54 | 19 | 5 | 4 | 0 | 5 | 4 | 1 | 2 | 0.0 | 9.3 | 21.1 | 20.0 | 50.0 |
| Miniopteridae | 0 | 2 | 6 | 13 | 6 | 0 | 0 | 3 | 7 | 4 | | 0.0 | 50.0 | 53.8 | 66.7 |
| Vespertilionidae | 63 | 56 | 52 | 18 | 10 | 4 | 19 | 18 | 9 | 3 | 6.3 | 33.9 | 34.6 | 50.0 | 30.0 |

## SARS-CoV-2 and UV: Regional Comparisons: Italy and Spain

We examined two countries which had been significantly affected by COVID-19 in the initial stages of the pandemic (Ceylan 2020, Gatto 2020). In Italy, national lockdown commenced on 9th March, 2020 and, in Spain on 14th March, 2020. National lockdowns offered a "natural" experiment that permitted testing the spread of SARS-CoV-2 within regions in each country by providing a control. Our null hypothesis was that there would be no differences in the spread of the virus, or of its rate of control, between regions within each country as the conditions of national lockdown applied across the regions and could therefore be regarded as a constant.

The mean number of new daily cases (MNNDC) reported for Italian regions during the seven-week period following lockdown varied considerably (Figure 2). These regions separated into three clusters that had non-overlapping incidence: (a) a north to north-west cluster (b) a north-east and east-central cluster; and (c) a southern cluster with low incidence (Figure 2a). These results suggest that factors other than lockdown must have been at work and that these varied regionally.

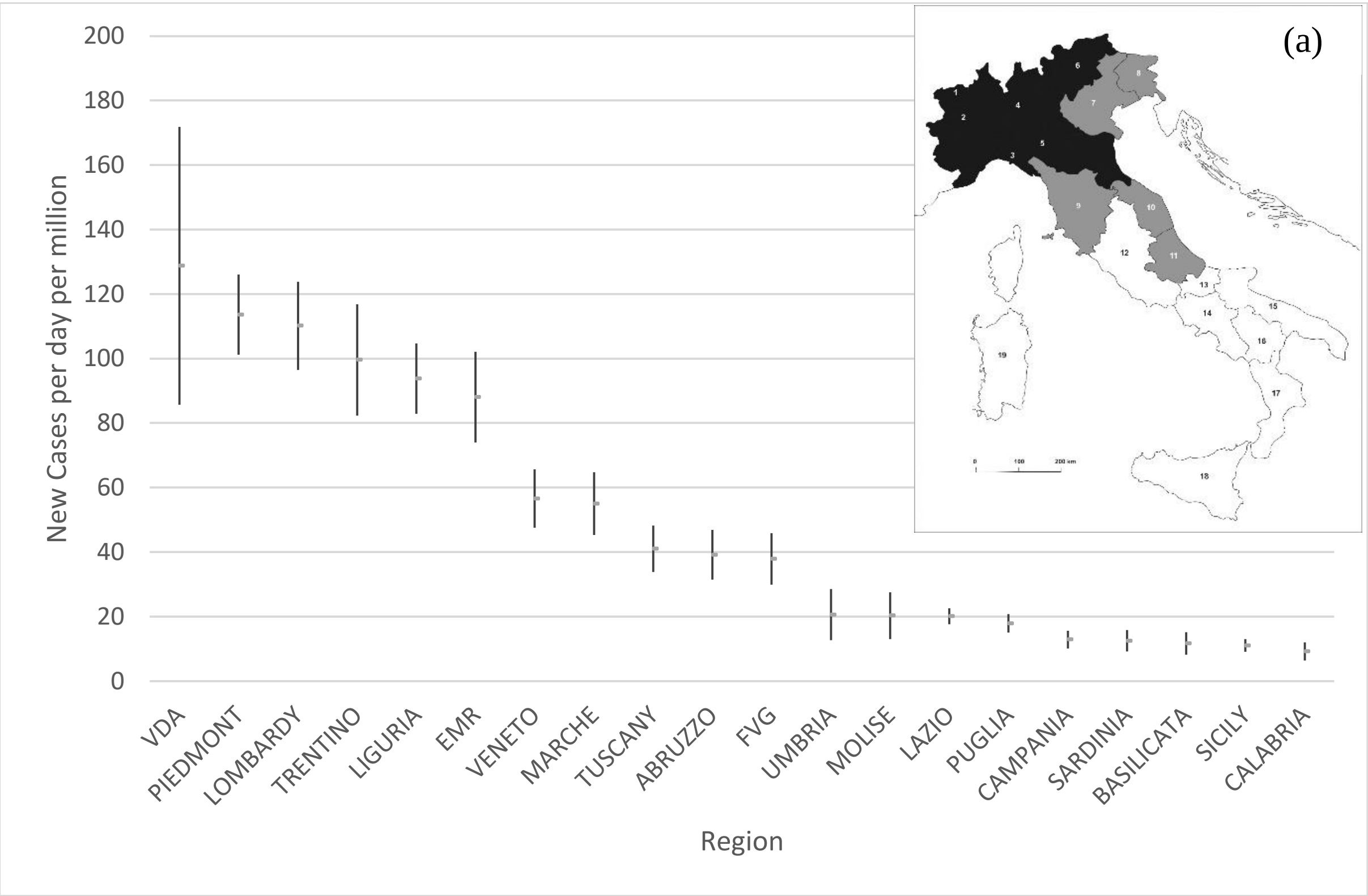

*Figure 2(a): Italy - distribution of COVID-19 incidence for the seven-week period after lockdown by region. Map inset: black regions with highest incidence; grey intermediate incidence and white low incidence.* 1-Aosta Valley (VDA); 2-Piedmont; 3-Liguria; 4-Lombardy; 5-Emilia-Romagna (EMR); 6-Trentino; 7-Veneto; 8-Friuli-Venezia-Giulia (FVG); 9-Tuscany; 10-Marche; 11-Abruzzo; 12-Lazio; 13-Molise; 14-Campania; 15-Puglia; 16-Basilicata; 17-Calabria; 18-Sicily; 19-Sardinia.

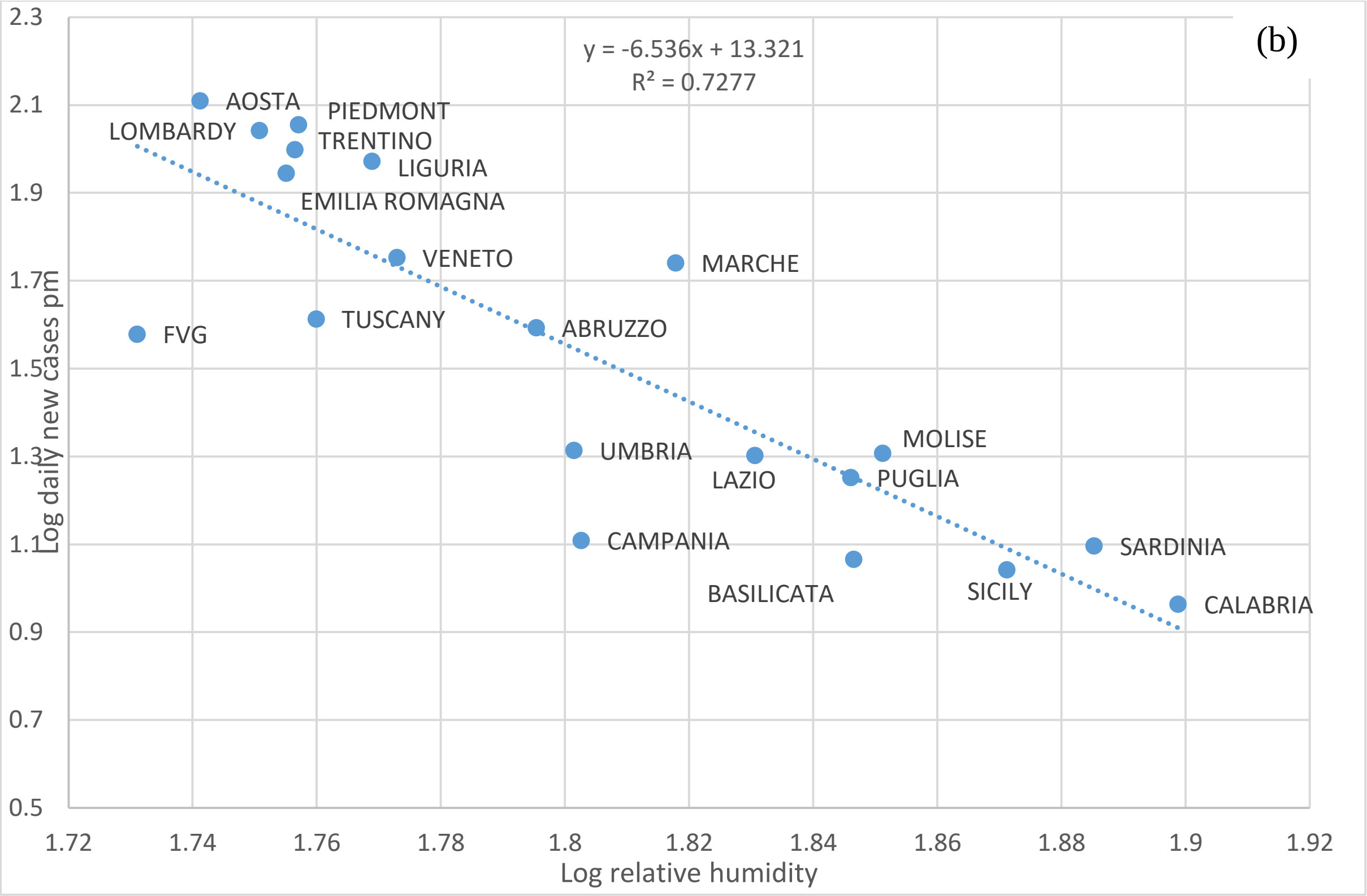

*Figure 2(b): Relationship between new daily cases and relative humidity for Italian regions for the seven-week period post-Lockdown. The relationship was statistically significant for zero, three-day, seven-day and fourteen-day time lags. Figure illustrates fourteen-day time lag.*

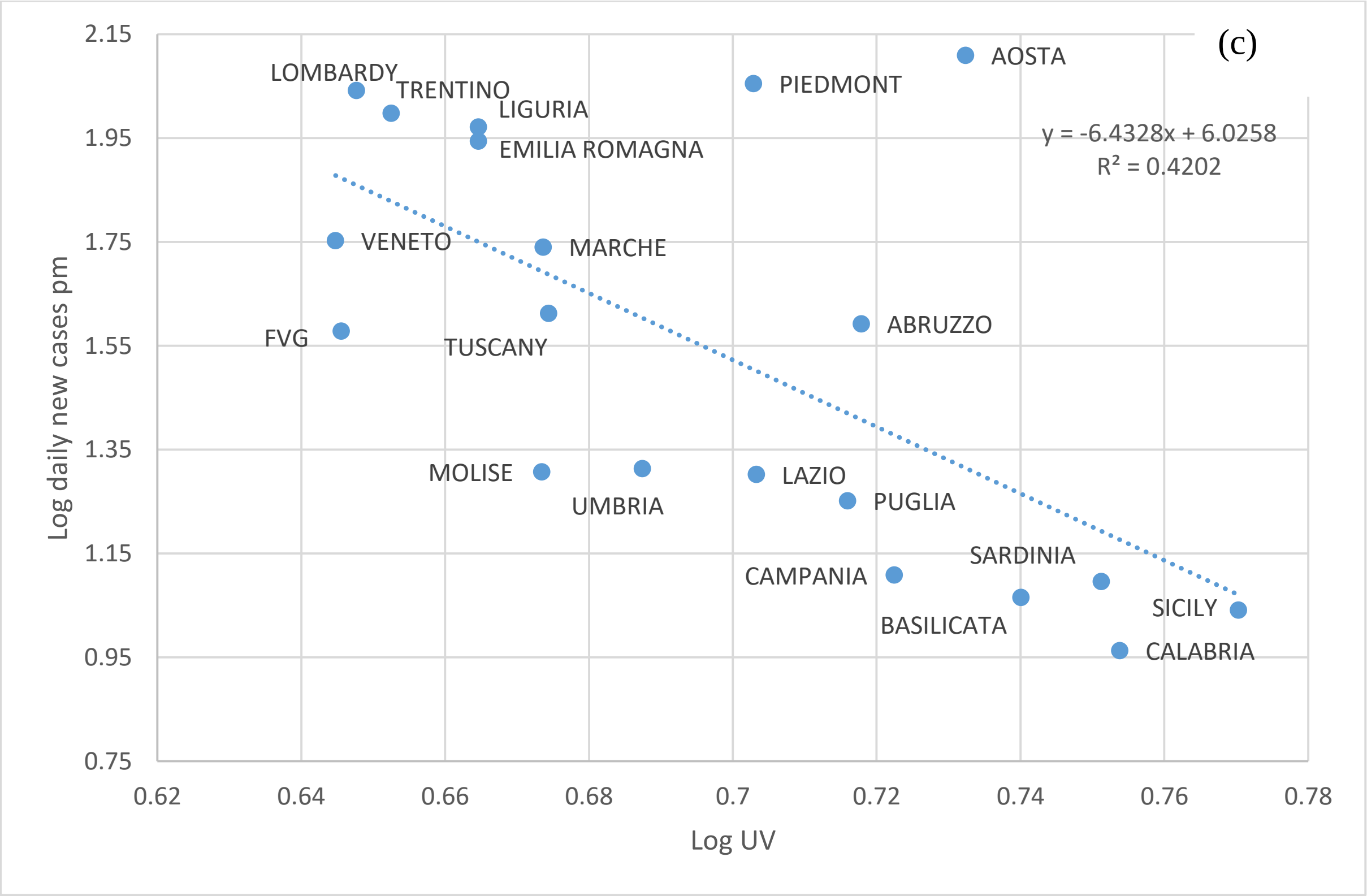

*Figure 2(c): Relationship between new daily cases and UVI for Italian regions for the seven-week period post-Lockdown. The relationship was statistically significant for zero, three-day, seven-day and fourteen-day time lags. Figure illustrates fourteen-day time lag.*

MNNDC per region was strongly correlated with latitude (Pearson correlation = 0.909; P = 0.000) and with longitude (Pearson correlation = -0.708; P = 0.000) but not with altitude (Pearson correlation = 0.094; P = 0.694). The trend was for high to low MNNDC on a north-west to south-east axis. The strongest environmental correlates of latitude were relative humidity (RH; Pearson correlation = -0.933; P = 0.000) and UV (Pearson correlation = 0.814, P = 0.000). No environmental variables were correlated with longitude. Using stepwise multiple regression, the model that best explained the regional distribution of daily new cases had relative humidity as sole explanatory variable (Figure 2b). The relationship held valid for zero, three-day, seven-day and fourteen-day time lags (Appendix 1). RH and UV were strongly correlated (Pearson correlation = 0.706, P = 0.000). Removing RH from the model, UV emerged as the single explanatory variable (Figure 2c). In Italy, the highest regional MNNDC were therefore associated with low RH and low UV.

In Spain MNNDC for Spanish regions during the seven-week period following lockdown also varied considerably (Figure 3a). The regions separated into three clusters: (a) a central inland cluster with high incidence; (b) a northern and eastern cluster with intermediate incidence comprising; and (c) a southern cluster with low incidence (Figure 3a). As in Italy, factors other than lockdown must have been at work at regional scales.

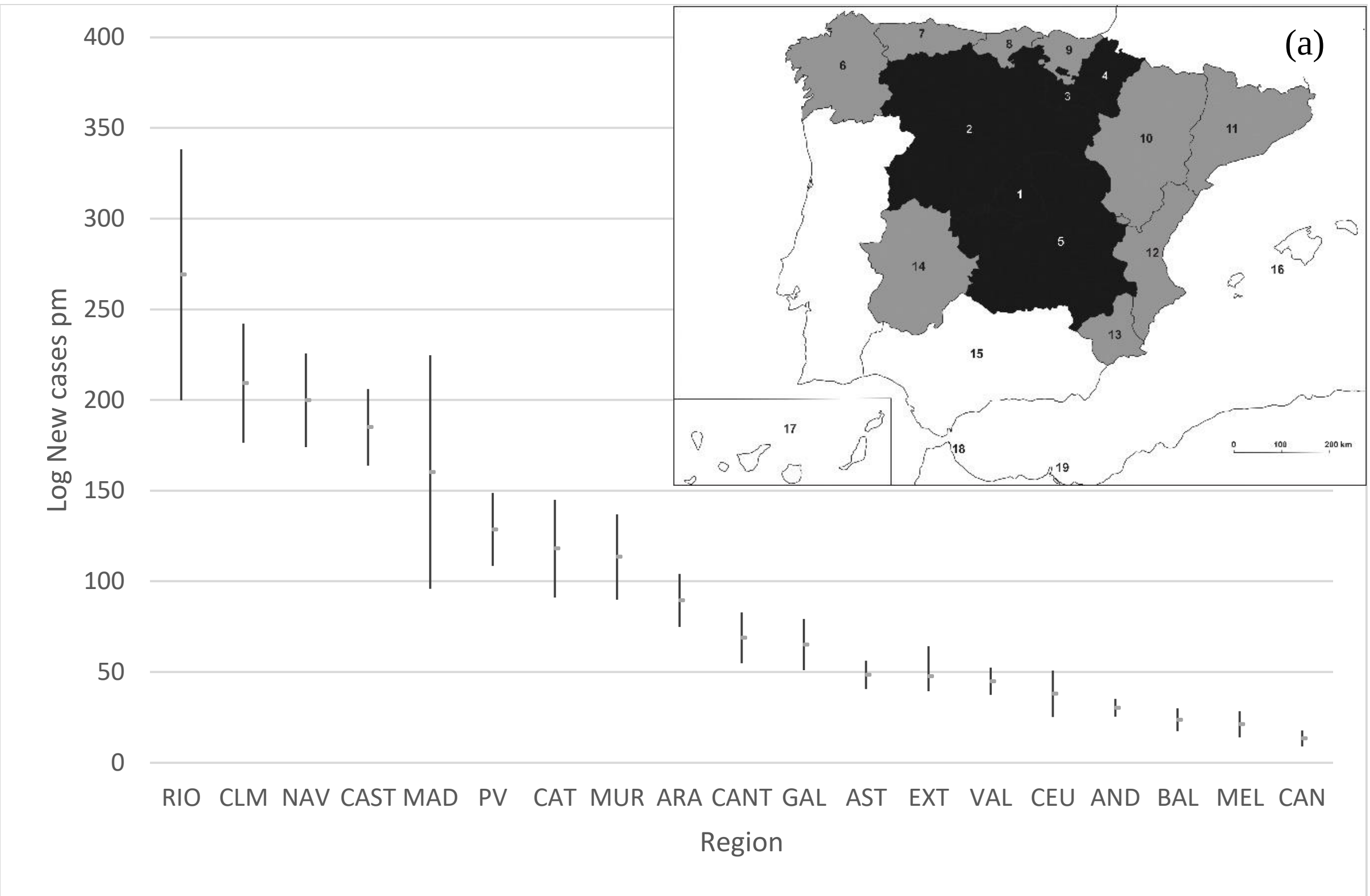

*Figure 3(a): Spain - distribution of COVID-19 incidence for the seven-week period after lockdown by region. Map inset: black regions with highest incidence; grey intermediate incidence and white low incidence.* 1-Madrid; 2-Castilla y Leon; 3-La Rioja; 4-Navarra; 5-Catilla La Mancha; 6-Galicia; 7-Asturias; 8-Cantabria; 9-Pais Basco; 10-Aragon; 11-Catalunya; 12-Valencia; 13-Murcia; 14-Extremadura; 15-Andalucia; 16-Baleares; 17-Canaries; 18-Ceuta; 19-Melilla.

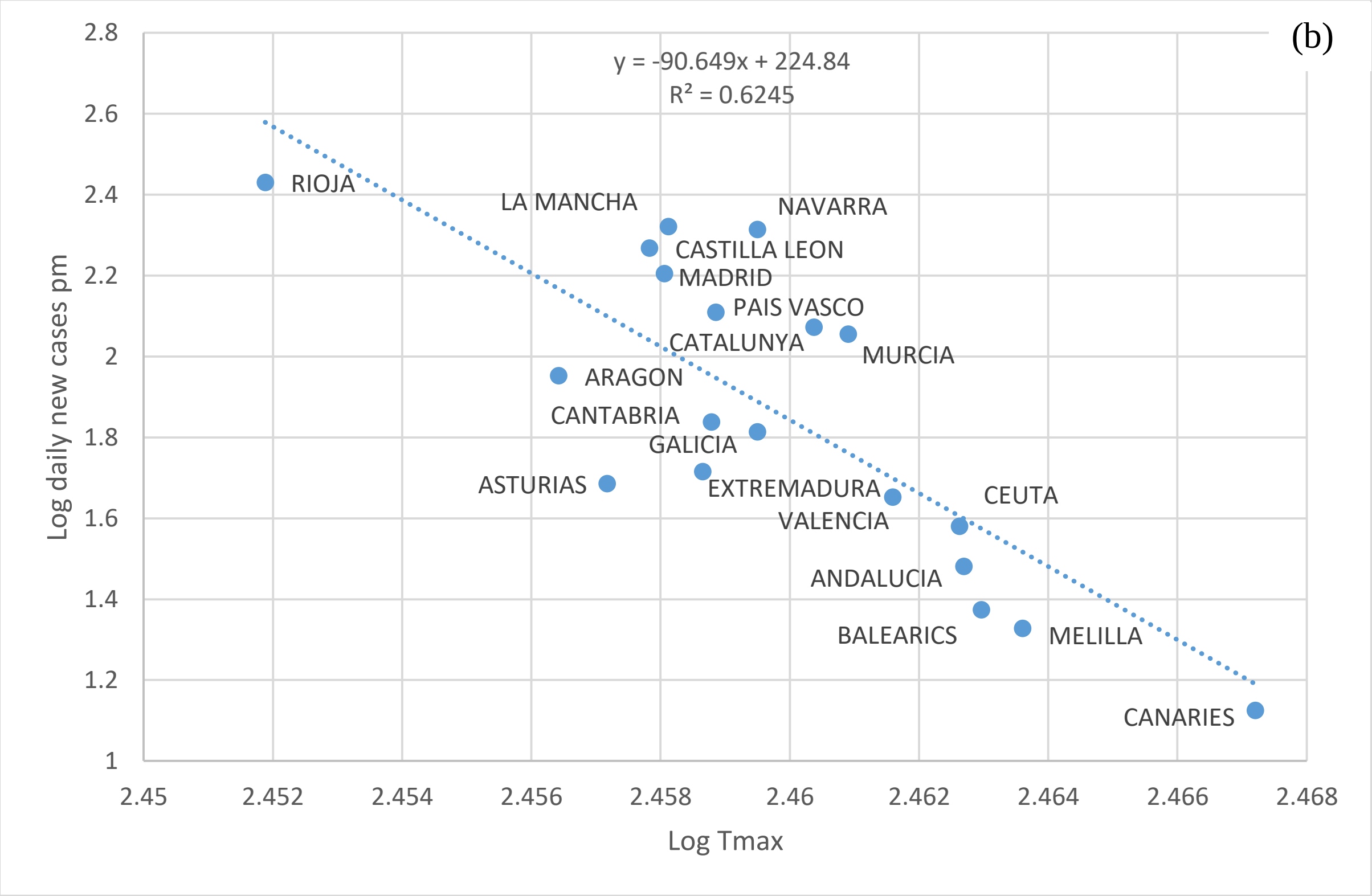

*Figure 3(b): Relationship between new daily cases and mean maximum daily temperature for Spanish regions for the seven-week period post-Lockdown. The relationship was statistically significant for zero, three-day, seven-day and fourteen-day time lags. Figure illustrates fourteen-day time lag.*

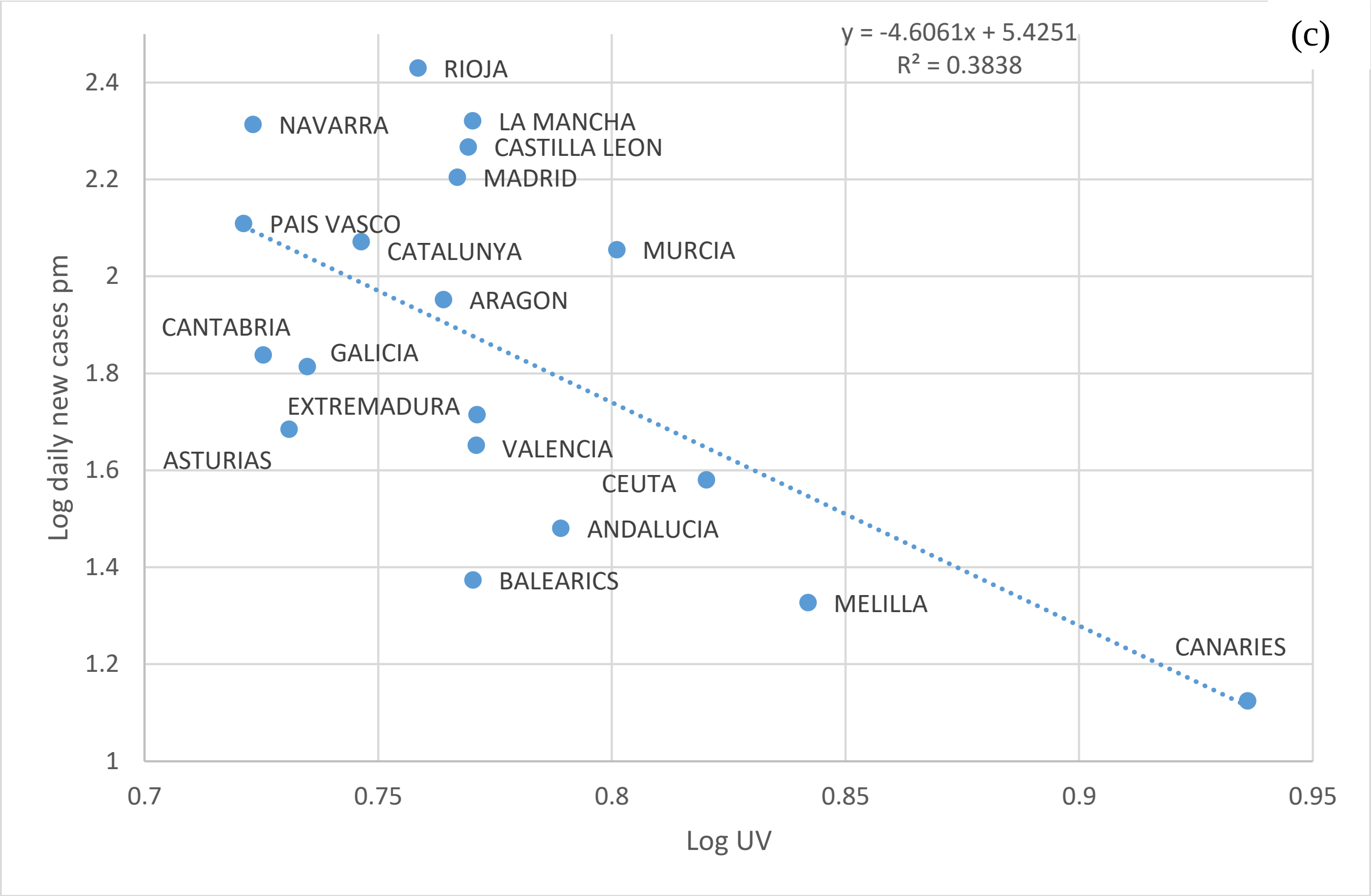

*Figure 3(c): Relationship between new daily cases and UV for Spanish regions for the seven-week period post-Lockdown. The relationship was statistically significant for zero, three-day, seven-day and fourteen-day time lags. Figure illustrates fourteen-day time lag.*

MNNDC was strongly correlated with altitude (Pearson correlation = 0.755; P = 0.000), latitude (Pearson correlation = 0.647; P = 0.003) and, less so, longitude (Pearson correlation = 0.464; P = 0.045). The strongest environmental correlate of altitude was temperature, particularly mean maximum daily temperature (MMDT; Pearson correlation = -0.758; P = 0.000). The strongest correlate of latitude was UV (Pearson correlation = -0.954; P = 0.000), followed by temperature (the strongest correlation being with mean daily temperature, MDT; Pearson correlation = -0.765; P = 0.000). The strongest correlate of longitude was RH (Pearson correlation = 0.716, P = 0.001) followed by UV (Pearson correlation = -0.698; P = 0.001). Using stepwise multiple regression, the model that best explained the regional distribution of daily new cases had MMDT as sole explanatory variable (Figure 3b). The relationship held valid for zero, three-day, seven-day and fourteen-day time lags (Appendix 1). MMDT and UV were strongly correlated (Pearson correlation = 0.685, P = 0.001). Removing temperature variables from the model, UV emerged as the single explanatory variable (Figure 3c). In Spain, the highest regional daily rates of new cases were therefore associated with low MMDT and low UV. Our results for Italy and Spain can be expected from a virus associated with nocturnal and cave-dwelling host species.

**SARS-CoV-2 and UV: Regional Comparisons: Australia**

The observed correlations between UV and temperature, in particular, follow the gradient from lockdown forwards as they occurred during the northern hemisphere spring, which is time of rising UV and increasing temperature. Even though each region's post-lockdown curve is different, it could be suggested that our observations simply reflect the natural progression from lockdown. For this reason, we look at a southern hemisphere country, which would have faced autumnal conditions of decreasing UV and temperatures during lockdown. We selected Australia as it offered a diversity of territories which all came under lockdown on 23rd March, 2020. MNNDC reported for Australian territories during the seven-week period following lockdown varied considerably (Figure 4a). These territories separated into three clusters: (a) Tasmania in the extreme south with high incidence; (b) a northern pair of Northern Territory and Queensland with low incidence; and (c) the remaining territories with intermediate incidence (Figure 4a).

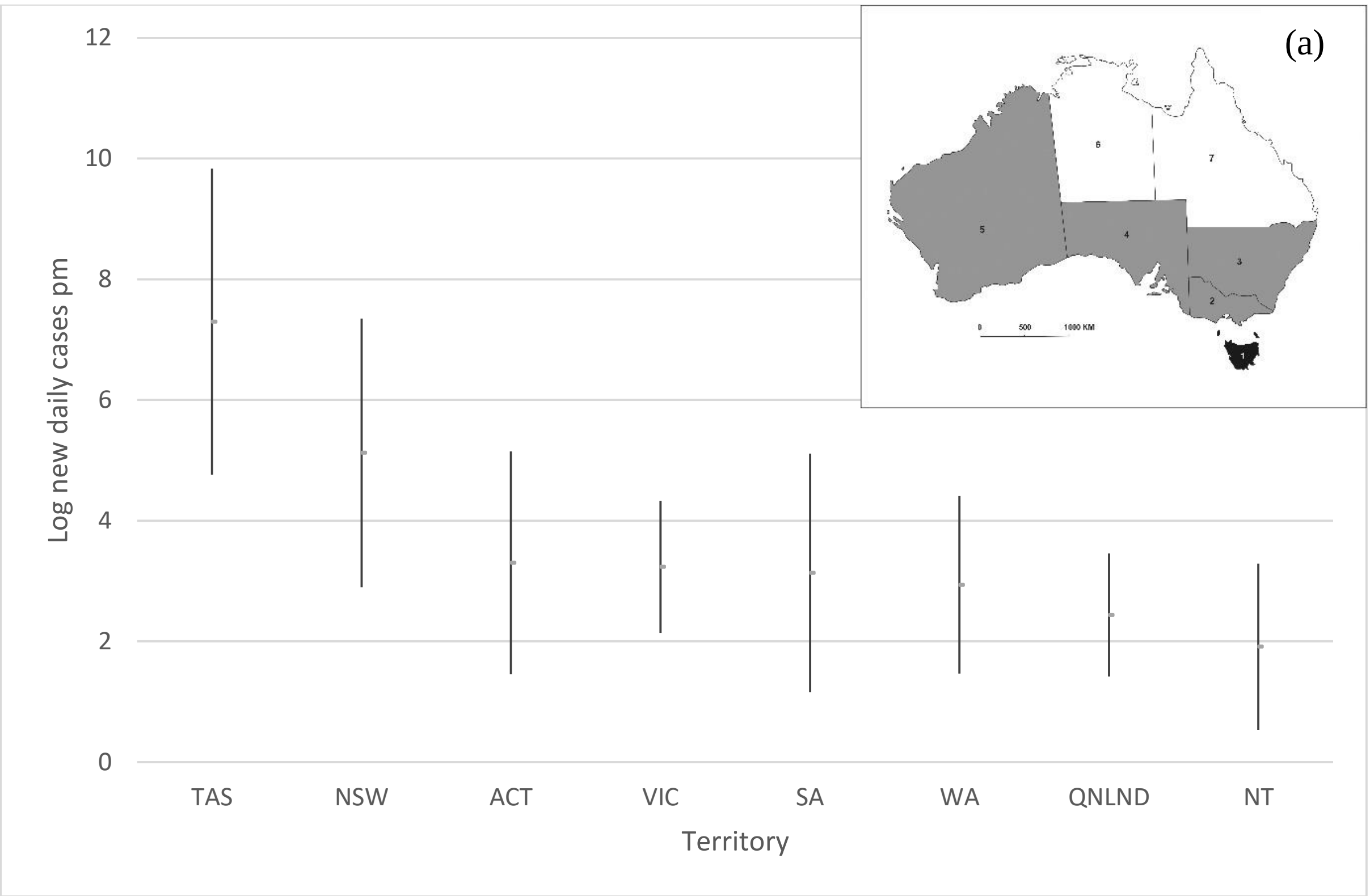

*Figure 4(a): Australia - distribution of COVID-19 incidence for the seven-week period after lockdown by region. Map inset: black regions with highest incidence; grey intermediate incidence and white low incidence.* 1-Tasmania; 2-Victoria; 3-New South Wales; 4-South Australia; 5-Western Australia; 6-Northern Territory; 7-Queensland.

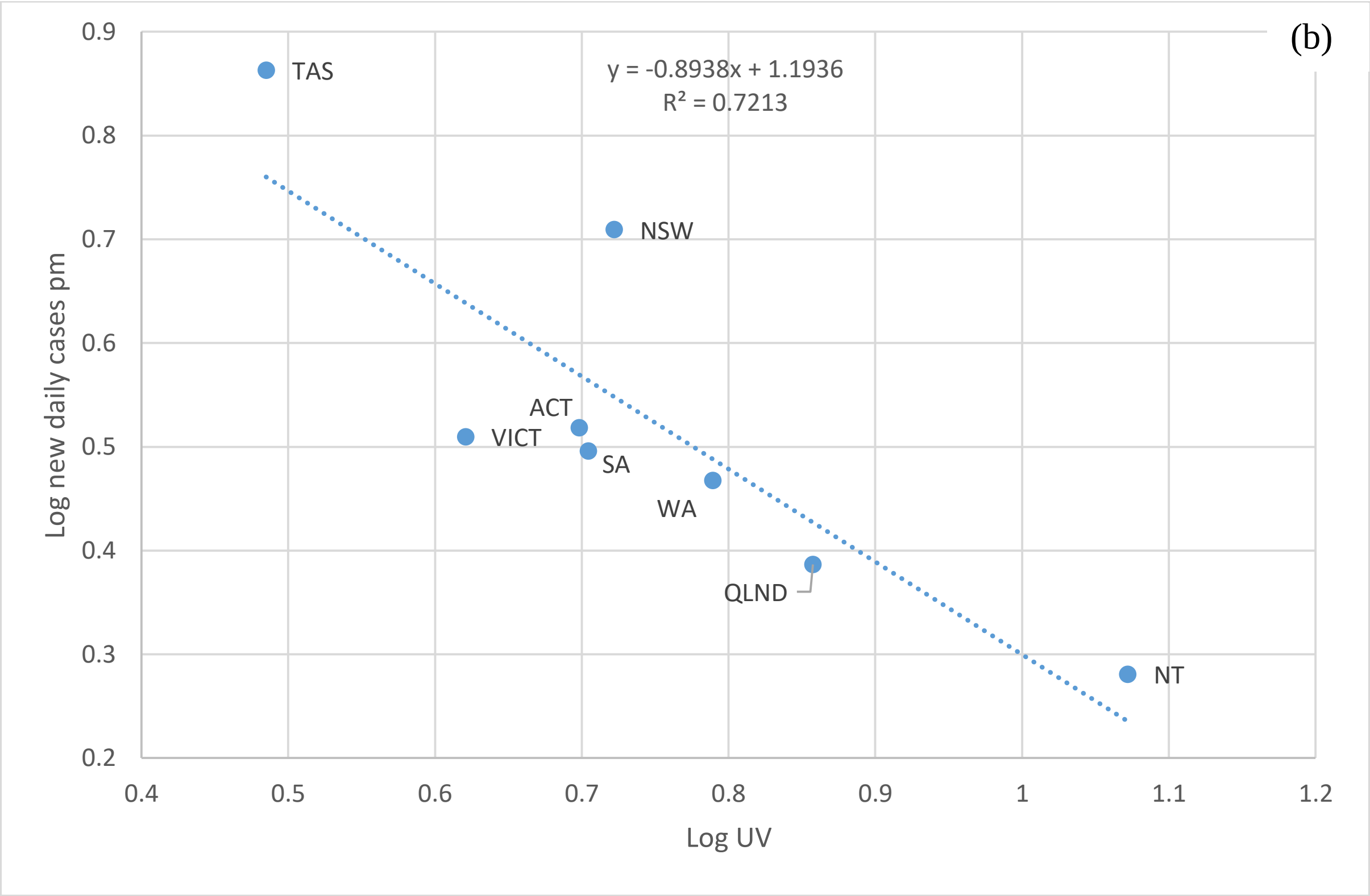

*Figure 4(b): Relationship between new daily cases and UV for Australian regions for the seven-week period post-Lockdown. The relationship was statistically significant for zero, three-day, seven-day and fourteen-day time lags. Figure illustrates three-day time lag.*

MNNDC per region was strongly correlated with latitude (Pearson correlation = 0.732; P = 0.039) but not with longitude (Pearson correlation = 0.382; P = 0.351) or altitude (Pearson correlation = 0.221; P = 0.6). There was therefore a trend from large numbers of daily new cases to low numbers on a south to north gradient. The strongest environmental correlates of latitude were UV (Pearson correlation = -0.934; P = 0.001) and temperature (highest with mean minimum daily temperature, Pearson correlation = 0.927, P = 0.001). Using stepwise multiple regression, the model that best explained the regional distribution of daily new cases had UV as sole explanatory variable (Figure 4b). The relationship held valid for zero, three-day, seven-day and fourteen-day time lags (Appendix 1). In Australia, the highest regional daily rates of new cases were therefore associated with low UV, conditions which we would expect the virus to favour.

## Discussion

**Bats, coronaviruses and the bat-coronavirus ecological niche**

Alpha- (α-) and beta (β-) coronaviruses regularly infect bats and other mammals, including humans, the latter having been identified from fewer hosts and showing less genetic diversity than the former (Ge *at al*. 2015). SARS-CoV-2 is a betacoronavirus (Wassenaar and Zou 2020) of the subgenus *Sarbecoronavirus* (Boni *et al.* 2020). Rhinolophidae and Hipposeridae– appear to have had an important role in the evolution of β-coronaviruses and Vespertilionidae and Miniopteridae with α-coronaviruses (Latinne *et al*. 2020).

Bats are reservoirs of coronaviruses (Guan *et al*. 2003, Li *et al*. 2005, Corman *et al*. 2014, Ge *et al*. 2015, Schneeberger and Voigt 2016, Anthony *et al*. 2017, Forni *et al*. 2017, Tao *et al*. 2017, Lau *et al*. 2018, Cui *et al*. 2019) and host the highest coronavirus diversity among mammals (Drexler *et al*. 2014, Wong *et al*. 2019). It seems that bats may well have been the progenitors of SARS-CoV-2 with phylogenetic analysis implicating horseshoe bats (*Rhinolophus* spp.) in eastern Asia as the most likely candidates (Latinne *et al*. 2020, Boni *et al.* 2020).

Cave-dwelling bats are often faithful to their roost sites and will occupy a cave for life and even for hundreds of generations (Altringham 2011). Temperature, humidity and cavity size are considered the most important factors determining the choice of caves as roosts by bats and microclimate is regulated further by the behaviour of the bats themselves, and these vary according to species, geography and season (Twente 1955, Dwyer 1971, Raesly and Gates 1987, Perry 2013). In tropical bats, physical protection from predators, relative constancy of temperatures (typically lower than in the exterior and moving to cooler parts of the roost during heat stress), lower levels of illumination and high humidity (though not the prime determining factor) determine the choice of roosts (Usman 1988). Although warm roosts provide significant benefits to bats (Dechmann *et al*. 2004), extremely high temperatures are stressful (Downs *et al*. 2015) and are avoided except in extraordinary circumstances (Bondarenco *et al*. 2014). In cave-dwelling bats, ambient temperatures in the >38$^{o}$C would appear to be on the limit of tolerance (Czenze *et al*. 2019). Given the use of caves with ambient temperatures ranging between 2 and 10$^{o}$C for hibernation in temperate environments (Perry 2013), it would appear that some bats (including Rhinolophidae, Vespertilionidae and Miniopteridae; Dwyer 1971) have a broad ambient temperature tolerance (aided by torpor) from ~0$^{o}$C to ~35$^{o}$C. In temperate North American vespertilionid bats, caves range from 60 to 100% relative humidity (Perry 2013).

Darkness is a major factor in choice of roost sites as it affords protection from predators, and the darkest parts of caves may be used by some species provided they have temperature constancy (Usman 1988). Among the few fruit bats (Pteropodidae) known to use caves, the echolocating *Rousettus* penetrate into the darker regions of caves; Vespertilionidae and Miniopteridae use dark areas of caves and Hipposideridae and Rhinolophidae are typical of the deepest areas of caves (Altringham 2011).

Our results are therefore in keeping with the observed characteristics of the bat-coronavirus ecological niche, specifically showing a consistent aversion to daylight (UV-B), modulated, in some cases, by a preference for cool and moderately humid climatic conditions.

**The SARS-CoV-2 ecological niche - ultraviolet radiation, temperature and relative humidity**

Coronaviruses have been found to display marked winter seasonality comparable to the pattern seen with influenza viruses (Gaunt *et al*. 2010). Recently, Schuit and colleagues (2020) confirmed epidemiological findings that sunlight levels were inversely correlated with influenza transmission, a finding that was suggested could assist in improved understanding of the spread of the virus under varied environmental conditions. Solar radiation, principally through the action of UVR, is known to have a direct effect on pathogen fitness including viral infections (Abhimanyu and Coussens 2017). It is a major factor threatening the life and activity of many microorganisms suspended in the atmosphere (Madronich *et al*. 2018). Airborne infectious animal viruses, including a coronavirus and an adenovirus, have been shown to have UV susceptibility, being higher in viral aerosols than in viral liquid suspensions (Walker and Ko 2007). SARS-CoV was inactivated by UV light at 254nm under laboratory conditions (Darnell *et al*. 2004). UV light irradiation for 60 min on the SARS-CoV in culture medium resulted in the destruction of viral infectivity (Duan *et al*. 2003).

There is an emerging literature on the impact of climatic factors such as temperature (e.g. Huang *et al.* 2020, Prata *et al.* 2020) and UV on SARS-CoV-2 at country (Sehra *et al.* 2020, Takagi *et al.* 2020) and global levels (Gunthe *et al.* 2020). Our data support and expand these preliminary results, and provide an evolutionary backdrop to the nature of the SARS-CoV-2 ecological niche and its origins.

An early worldwide study linked the outbreak of the COVID-19 to temperature, wind speed and relative humidity in combination as predictors of the pandemic situation. SARS-CoV-2 transmission reached a peak when the air temperature was 8.07 ℃, or when the wind speed was 16.1 mile/hr, or when the visibility was 2.99 statute miles to nearest tenth, or when the relative humidity was 64.6% (Chen *et al*. 2020). Liu *et al*. (2020) found that low temperatures, a mild diurnal temperature range and low humidity favoured the transmission of SARS-CoV-2. Another study in China showed that the incidence of the COVID-19 outbreak decreased as temperature increased, peaking at 10$^{o}$C (Shi *et al*. 2020). In a worldwide analysis, Sjadi *et al*. (2020) found that the eight cities with substantial community spread of COVID-19 as of 10$^{th}$ March, 2020, were located on a narrow band, roughly on the 30° N to 50° N corridor. These cities had consistently similar weather patterns, consisting of mean temperatures of between 5 and 11°C, combined with low specific humidity (3-6 g/kg) and low absolute humidity (4-7 g/m3). COVID-19 deaths in Wuhan, China, were positively associated with diurnal temperature range and negatively with absolute humidity (Ma *et al*. 2020). A global comprehensive ecogeographical analysis demonstrated that although cases of COVID-19 were reported all over the world, most outbreaks displayed a pattern of clustering in relatively cool and dry areas (Araujo and Naimi 2020). Our results support these conclusions.

Our results demonstrate a clear association between UVR and the incidence of SARS-CoV-2, with temperature and relative humidity being significant but not generalised variables. Our results are therefore consistent with the direct effects of UVR on viral inactivation. In this regard, the long evolutionary association between coronaviruses and nocturnal mammals may be a reflection of the ecological niche of the coronaviruses, natural selection having effectively delimited their niche outside the scope of ultraviolet radiation (Wertheim *et al*. 2013).

**Environmental versus behavioural effects**

If coronaviruses and bats have coevolved over millions of years, what has changed? We suggest that what has really changed has been human population growth and the vastly increased opportunities for cross-species viral jumps that result from increased human exploitation of the natural environment (e.g. Olivero *et al.* 2017 for Ebola). The problem is not with the bats, nor with the human-bat consumption at small-scale sustainable levels; the problem is with large-scale exploitation and human social behaviour at high density (Skórka *et al*. 2020).

In this paper we have shown clear relationships between SARS-CoV-2 and UVR, temperature and humidity but it is also clear that lockdown, an imposed form of social distancing, has been the key overriding factor in flattening the disease growth curve. Thus the significance of UVR and other climatic variables must be seen in a co-evolutionary context and providing the backdrop (Araujo and Naimi 2020). The lesson could be learnt, however, from the behaviour of bats. Thus human activities resembling those of highly social cave-dwelling bats (e.g. large nocturnal gatherings or high density indoor activities) will only serve to exacerbate the problem of COVID-19.

## Declarations

All authors contributed to the manuscript and results. Geraldine Finlayson and Clive Finlayson participated in the conceptual design, data analysis, and drafting of the manuscript. Stewart Finlayson contributed to the data on bats and coronaviruses, and data analysis. Keith Bensusan contributed to data analysis. Rhian Guillem contributed with the literature review and references. Tyson L Holmes contributed to data on bats. Francisco Giles Guzmán contributed in the preparation of Figures. José S Carrión contributed on evolutionary ecology aspects. Lawrence Sawchuk contributed on epidemiological aspects. All authors participated in revisions of drafts of the manuscript including review of data analyses. All authors declare that there are no conflicts of interest.

# Appendix 1

---------

*a) Results of stepwise multiple regression of cases for Italy by climatic variables with zero, three-day, seven-day and fourteen-day time lags.*

**ITALY ZERO DAY LAG**

### DESCRIPTIVE STATISTICS

| | MEAN | STD. DEVIATION | N |
|---|---|---|---|
| CASES | 1.6172 | .40142 | 20 |
| UV | .6969 | .03957 | 20 |
| T | 2.4540 | .00225 | 20 |
| TMIN | 2.4484 | .00428 | 20 |
| TMAX | 2.4583 | .00157 | 20 |
| HUM | 1.8020 | .05126 | 20 |
| WINDS | .4250 | .10027 | 20 |
| WINDD | 2.1220 | .08658 | 20 |
| RAIN | -1.5349 | .36233 | 20 |
| CLOUDS | .8558 | .33865 | 20 |

### CORRELATIONS

| | | CASES | UV | T | TMIN | TMAX | HUM | WINDS | WINDD | RAIN | CLOUDS |
|---|---|---|---|---|---|---|---|---|---|---|---|
| PEARSON CORRELATION | CASES | 1.000 | -.643 | -.261 | -.445 | -.105 | -.857 | -.329 | -.295 | -.343 | -.398 |
| | UV | -.643 | 1.000 | .187 | .175 | .076 | .706 | .408 | .496 | .427 | .616 |
| | T | -.261 | .187 | 1.000 | .864 | .760 | .308 | .180 | .125 | -.380 | -.229 |
| | TMIN | -.445 | .175 | .864 | 1.000 | .427 | .492 | .115 | .327 | -.146 | -.288 |
| | TMAX | -.105 | .076 | .760 | .427 | 1.000 | .153 | .109 | -.088 | -.468 | .012 |
| | HUM | -.857 | .706 | .308 | .492 | .153 | 1.000 | .357 | .517 | .336 | .389 |
| | WINDS | -.329 | .408 | .180 | .115 | .109 | .357 | 1.000 | .147 | -.061 | .182 |
| | WINDD | -.295 | .496 | .125 | .327 | -.088 | .517 | .147 | 1.000 | .483 | .238 |
| | RAIN | -.343 | .427 | -.380 | -.146 | -.468 | .336 | -.061 | .483 | 1.000 | .190 |
| | CLOUDS | -.398 | .616 | -.229 | -.288 | .012 | .389 | .182 | .238 | .190 | 1.000 |
| SIG. (1-TAILED) | CASES | . | .001 | .133 | .025 | .329 | .000 | .078 | .104 | .069 | .041 |
| | UV | .001 | . | .215 | .230 | .375 | .000 | .037 | .013 | .030 | .002 |
| | T | .133 | .215 | . | .000 | .000 | .093 | .223 | .300 | .049 | .166 |
| | TMIN | .025 | .230 | .000 | . | .030 | .014 | .315 | .080 | .270 | .109 |
| | TMAX | .329 | .375 | .000 | .030 | . | .260 | .324 | .357 | .019 | .479 |
| | HUM | .000 | .000 | .093 | .014 | .260 | . | .061 | .010 | .074 | .045 |
| | WINDS | .078 | .037 | .223 | .315 | .324 | .061 | . | .268 | .400 | .221 |
| | WINDD | .104 | .013 | .300 | .080 | .357 | .010 | .268 | . | .016 | .156 |
| | RAIN | .069 | .030 | .049 | .270 | .019 | .074 | .400 | .016 | . | .211 |
| | CLOUDS | .041 | .002 | .166 | .109 | .479 | .045 | .221 | .156 | .211 | . |
| N | CASES | 20 | 20 | 20 | 20 | 20 | 20 | 20 | 20 | 20 | 20 |
| | UV | 20 | 20 | 20 | 20 | 20 | 20 | 20 | 20 | 20 | 20 |
| | T | 20 | 20 | 20 | 20 | 20 | 20 | 20 | 20 | 20 | 20 |
| | TMIN | 20 | 20 | 20 | 20 | 20 | 20 | 20 | 20 | 20 | 20 |
| | TMAX | 20 | 20 | 20 | 20 | 20 | 20 | 20 | 20 | 20 | 20 |
| | HUM | 20 | 20 | 20 | 20 | 20 | 20 | 20 | 20 | 20 | 20 |
| | WINDS | 20 | 20 | 20 | 20 | 20 | 20 | 20 | 20 | 20 | 20 |
| | WINDD | 20 | 20 | 20 | 20 | 20 | 20 | 20 | 20 | 20 | 20 |
| | RAIN | 20 | 20 | 20 | 20 | 20 | 20 | 20 | 20 | 20 | 20 |
| | CLOUDS | 20 | 20 | 20 | 20 | 20 | 20 | 20 | 20 | 20 | 20 |

## VARIABLES ENTERED/REMOVED[A]

| MODEL | VARIABLES ENTERED | VARIABLES REMOVED | METHOD |
|---|---|---|---|
| 1 | HUM | . | Stepwise (Criteria: Probability-of-F-to-enter <= .050, Probability-of-F-to-remove >= .100). |

a. Dependent Variable: CASES

## Model Summary[b]

| Model | R | R Square | Adjusted R Square | Std. Error of the Estimate | Change Statistics | | | | |
|---|---|---|---|---|---|---|---|---|---|
| | | | | | R Square Change | F Change | df1 | df2 | Sig. F Change |
| 1 | .857[a] | .734 | .720 | .21254 | .734 | 49.774 | 1 | 18 | .000 |

a. Predictors: (Constant), HUM

b. Dependent Variable: CASES

## ANOVA[a]

| Model | | Sum of Squares | df | Mean Square | F | Sig. |
|---|---|---|---|---|---|---|
| 1 | Regression | 2.248 | 1 | 2.248 | 49.774 | .000[b] |
| | Residual | .813 | 18 | .045 | | |
| | Total | 3.062 | 19 | | | |

a. Dependent Variable: CASES

b. Predictors: (Constant), HUM

## Coefficients[a]

| | | Unstandardized Coefficients | | Standardized Coefficients | | | Correlations | | | Collinearity Statistics | |
|---|---|---|---|---|---|---|---|---|---|---|---|
| Model | | B | Std. Error | Beta | t | Sig. | Zero-order | Partial | Part | Tolerance | VIF |
| 1 | (Constant) | 13.711 | 1.715 | | 7.995 | .000 | | | | | |
| | HUM | -6.711 | .951 | -.857 | -7.055 | .000 | -.857 | -.857 | -.857 | 1.000 | 1.000 |

a. Dependent Variable: CASES

## Excluded Variables[a]

| | | | | | | Collinearity Statistics | | |
|---|---|---|---|---|---|---|---|---|
| Model | | Beta In | t | Sig. | Partial Correlation | Tolerance | VIF | Minimum Tolerance |
| 1 | UV | -.076[b] | -.431 | .672 | -.104 | .501 | 1.996 | .501 |
| | T | .003[b] | .022 | .983 | .005 | .905 | 1.105 | .905 |
| | TMIN | -.031[b] | -.218 | .830 | -.053 | .758 | 1.319 | .758 |
| | TMAX | .026[b] | .209 | .837 | .051 | .977 | 1.024 | .977 |
| | WINDS | -.026[b] | -.196 | .847 | -.048 | .872 | 1.147 | .872 |
| | WINDD | .202[b] | 1.472 | .159 | .336 | .733 | 1.365 | .733 |
| | RAIN | -.062[b] | -.471 | .644 | -.114 | .887 | 1.127 | .887 |
| | CLOUDS | -.075[b] | -.560 | .583 | -.134 | .848 | 1.179 | .848 |

a. Dependent Variable: CASES

b. Predictors in the Model: (Constant), HUM

**Collinearity Diagnostics[a]**

| Model | Dimension | Eigenvalue | Condition Index | Variance Proportions (Constant) | Variance Proportions HUM |
|---|---|---|---|---|---|
| 1 | 1 | 2.000 | 1.000 | .00 | .00 |
| | 2 | .000 | 72.153 | 1.00 | 1.00 |

a. Dependent Variable: CASES

**Residuals Statistics[a]**

| | Minimum | Maximum | Mean | Std. Deviation | N |
|---|---|---|---|---|---|
| Predicted Value | .9677 | 2.0934 | 1.6172 | .34401 | 20 |
| Std. Predicted Value | -1.888 | 1.384 | .000 | 1.000 | 20 |
| Standard Error of Predicted Value | .048 | .104 | .066 | .015 | 20 |
| Adjusted Predicted Value | .9385 | 2.1666 | 1.6159 | .34956 | 20 |
| Residual | -.44058 | .38176 | .00000 | .20687 | 20 |
| Std. Residual | -2.073 | 1.796 | .000 | .973 | 20 |
| Stud. Residual | -2.127 | 1.848 | .003 | 1.023 | 20 |
| Deleted Residual | -.48505 | .40398 | .00130 | .22871 | 20 |
| Stud. Deleted Residual | -2.389 | 1.995 | -.017 | 1.092 | 20 |
| Mahal. Distance | .000 | 3.566 | .950 | .927 | 20 |
| Cook's Distance | .000 | .393 | .053 | .087 | 20 |
| Centered Leverage Value | .000 | .188 | .050 | .049 | 20 |

a. Dependent Variable: CASES

## Italy Zero Day Lag without Humidity

### Descriptive Statistics

| | Mean | Std. Deviation | N |
|---|---|---|---|
| CASES | 1.6172 | .40142 | 20 |
| UV | .6969 | .03957 | 20 |
| T | 2.4540 | .00225 | 20 |
| TMIN | 2.4484 | .00428 | 20 |
| TMAX | 2.4583 | .00157 | 20 |
| WINDS | .4250 | .10027 | 20 |
| WINDD | 2.1220 | .08658 | 20 |
| RAIN | -1.5349 | .36233 | 20 |
| CLOUDS | .8558 | .33865 | 20 |

### Correlations

| | | CASES | UV | T | TMIN | TMAX | WINDS | WINDD | RAIN | CLOUDS |
|---|---|---|---|---|---|---|---|---|---|---|
| Pearson Correlation | CASES | 1.000 | -.643 | -.261 | -.445 | -.105 | -.329 | -.295 | -.343 | -.398 |
| | UV | -.643 | 1.000 | .187 | .175 | .076 | .408 | .496 | .427 | .616 |
| | T | -.261 | .187 | 1.000 | .864 | .760 | .180 | .125 | -.380 | -.229 |
| | TMIN | -.445 | .175 | .864 | 1.000 | .427 | .115 | .327 | -.146 | -.288 |
| | TMAX | -.105 | .076 | .760 | .427 | 1.000 | .109 | -.088 | -.468 | .012 |
| | WINDS | -.329 | .408 | .180 | .115 | .109 | 1.000 | .147 | -.061 | .182 |
| | WINDD | -.295 | .496 | .125 | .327 | -.088 | .147 | 1.000 | .483 | .238 |
| | RAIN | -.343 | .427 | -.380 | -.146 | -.468 | -.061 | .483 | 1.000 | .190 |
| | CLOUDS | -.398 | .616 | -.229 | -.288 | .012 | .182 | .238 | .190 | 1.000 |
| Sig. (1-tailed) | CASES | . | .001 | .133 | .025 | .329 | .078 | .104 | .069 | .041 |
| | UV | .001 | . | .215 | .230 | .375 | .037 | .013 | .030 | .002 |
| | T | .133 | .215 | . | .000 | .000 | .223 | .300 | .049 | .166 |
| | TMIN | .025 | .230 | .000 | . | .030 | .315 | .080 | .270 | .109 |
| | TMAX | .329 | .375 | .000 | .030 | . | .324 | .357 | .019 | .479 |
| | WINDS | .078 | .037 | .223 | .315 | .324 | . | .268 | .400 | .221 |
| | WINDD | .104 | .013 | .300 | .080 | .357 | .268 | . | .016 | .156 |
| | RAIN | .069 | .030 | .049 | .270 | .019 | .400 | .016 | . | .211 |
| | CLOUDS | .041 | .002 | .166 | .109 | .479 | .221 | .156 | .211 | . |
| N | CASES | 20 | 20 | 20 | 20 | 20 | 20 | 20 | 20 | 20 |
| | UV | 20 | 20 | 20 | 20 | 20 | 20 | 20 | 20 | 20 |
| | T | 20 | 20 | 20 | 20 | 20 | 20 | 20 | 20 | 20 |
| | TMIN | 20 | 20 | 20 | 20 | 20 | 20 | 20 | 20 | 20 |
| | TMAX | 20 | 20 | 20 | 20 | 20 | 20 | 20 | 20 | 20 |
| | WINDS | 20 | 20 | 20 | 20 | 20 | 20 | 20 | 20 | 20 |
| | WINDD | 20 | 20 | 20 | 20 | 20 | 20 | 20 | 20 | 20 |
| | RAIN | 20 | 20 | 20 | 20 | 20 | 20 | 20 | 20 | 20 |
| | CLOUDS | 20 | 20 | 20 | 20 | 20 | 20 | 20 | 20 | 20 |

**Variables Entered/Removed[a]**

| Model | Variables Entered | Variables Removed | Method |
|---|---|---|---|
| 1 | UV | . | Stepwise (Criteria: Probability-of-F-to-enter <= .050, Probability-of-F-to-remove >= .100). |

a. Dependent Variable: CASES

## Model Summary[b]

| Model | R | R Square | Adjusted R Square | Std. Error of the Estimate | Change Statistics | | | | |
|---|---|---|---|---|---|---|---|---|---|
| | | | | | R Square Change | F Change | df1 | df2 | Sig. F Change |
| 1 | .643[a] | .414 | .381 | .31577 | .414 | 12.705 | 1 | 18 | .002 |

a. Predictors: (Constant), UV

b. Dependent Variable: CASES

## ANOVA[a]

| Model | | Sum of Squares | df | Mean Square | F | Sig. |
|---|---|---|---|---|---|---|
| 1 | Regression | 1.267 | 1 | 1.267 | 12.705 | .002[b] |
| | Residual | 1.795 | 18 | .100 | | |
| | Total | 3.062 | 19 | | | |

a. Dependent Variable: CASES

b. Predictors: (Constant), UV

**Coefficients[a]**

| Model | | Unstandardized Coefficients | | Standardized Coefficients | t | Sig. | Correlations | | | Collinearity Statistics | |
|---|---|---|---|---|---|---|---|---|---|---|---|
| | | B | Std. Error | Beta | | | Zero-order | Partial | Part | Tolerance | VIF |
| 1 | (Constant) | 6.165 | 1.278 | | 4.825 | .000 | | | | | |
| | UV | -6.525 | 1.831 | -.643 | -3.564 | .002 | -.643 | -.643 | -.643 | 1.000 | 1.000 |

a. Dependent Variable: CASES

**Excluded Variables[a]**

| Model | | Beta In | t | Sig. | Partial Correlation | Collinearity Statistics | | |
|---|---|---|---|---|---|---|---|---|
| | | | | | | Tolerance | VIF | Minimum Tolerance |
| 1 | T | -.146[b] | -.787 | .442 | -.187 | .965 | 1.036 | .965 |
| | TMIN | -.343[b] | -2.026 | .059 | -.441 | .969 | 1.032 | .969 |
| | TMAX | -.057[b] | -.306 | .763 | -.074 | .994 | 1.006 | .994 |
| | WINDS | -.080[b] | -.397 | .697 | -.096 | .834 | 1.199 | .834 |
| | WINDD | .032[b] | .149 | .883 | .036 | .754 | 1.326 | .754 |
| | RAIN | -.084[b] | -.409 | .688 | -.099 | .818 | 1.223 | .818 |
| | CLOUDS | -.002[b] | -.008 | .994 | -.002 | .620 | 1.613 | .620 |

a. Dependent Variable: CASES

b. Predictors in the Model: (Constant), UV

## Collinearity Diagnostics[a]

| Model | Dimension | Eigenvalue | Condition Index | Variance Proportions | |
|---|---|---|---|---|---|
| | | | | (Constant) | UV |
| 1 | 1 | 1.998 | 1.000 | .00 | .00 |
| | 2 | .002 | 36.165 | 1.00 | 1.00 |

a. Dependent Variable: CASES

## Residuals Statistics[a]

| | Minimum | Maximum | Mean | Std. Deviation | N |
|---|---|---|---|---|---|
| Predicted Value | 1.1384 | 1.9578 | 1.6172 | .25821 | 20 |
| Std. Predicted Value | -1.854 | 1.319 | .000 | 1.000 | 20 |
| Standard Error of Predicted Value | .071 | .152 | .098 | .022 | 20 |
| Adjusted Predicted Value | 1.1535 | 1.9963 | 1.6187 | .25757 | 20 |
| Residual | -.49559 | .86468 | .00000 | .30735 | 20 |
| Std. Residual | -1.569 | 2.738 | .000 | .973 | 20 |
| Stud. Residual | -1.626 | 2.874 | -.002 | 1.018 | 20 |
| Deleted Residual | -.53208 | .95247 | -.00147 | .33625 | 20 |
| Stud. Deleted Residual | -1.711 | 3.797 | .044 | 1.173 | 20 |
| Mahal. Distance | .023 | 3.439 | .950 | .890 | 20 |
| Cook's Distance | .002 | .419 | .047 | .091 | 20 |
| Centered Leverage Value | .001 | .181 | .050 | .047 | 20 |

a. Dependent Variable: CASES

**Italy 3 Day Lag**

## Descriptive Statistics

| | Mean | Std. Deviation | N |
|---|---|---|---|
| CASES | 1.6201 | .40253 | 20 |
| UV | .6969 | .03957 | 20 |
| T | 2.4540 | .00225 | 20 |
| TMIN | 2.4484 | .00428 | 20 |
| TMAX | 2.4583 | .00157 | 20 |
| HUMIDITY | 1.8020 | .05126 | 20 |
| WINDS | .4250 | .10027 | 20 |
| WINDD | 2.1220 | .08658 | 20 |
| RAIN | -1.5349 | .36233 | 20 |
| CLOUDS | .8558 | .33865 | 20 |

## Correlations

| | | CASES | UV | T | TMIN | TMAX | HUMIDITY | WINDS | WINDD | RAIN | CLOUDS |
|---|---|---|---|---|---|---|---|---|---|---|---|
| Pearson Correlation | CASES | 1.000 | -.641 | -.260 | -.446 | -.100 | -.858 | -.329 | -.295 | -.341 | -.395 |
| | UV | -.641 | 1.000 | .187 | .175 | .076 | .706 | .408 | .496 | .427 | .616 |
| | T | -.260 | .187 | 1.000 | .864 | .760 | .308 | .180 | .125 | -.380 | -.229 |
| | TMIN | -.446 | .175 | .864 | 1.000 | .427 | .492 | .115 | .327 | -.146 | -.288 |
| | TMAX | -.100 | .076 | .760 | .427 | 1.000 | .153 | .109 | -.088 | -.468 | .012 |
| | HUMIDITY | -.858 | .706 | .308 | .492 | .153 | 1.000 | .357 | .517 | .336 | .389 |
| | WINDS | -.329 | .408 | .180 | .115 | .109 | .357 | 1.000 | .147 | -.061 | .182 |
| | WINDD | -.295 | .496 | .125 | .327 | -.088 | .517 | .147 | 1.000 | .483 | .238 |
| | RAIN | -.341 | .427 | -.380 | -.146 | -.468 | .336 | -.061 | .483 | 1.000 | .190 |
| | CLOUDS | -.395 | .616 | -.229 | -.288 | .012 | .389 | .182 | .238 | .190 | 1.000 |
| Sig. (1-tailed) | CASES | . | .001 | .134 | .024 | .337 | .000 | .078 | .103 | .070 | .043 |
| | UV | .001 | . | .215 | .230 | .375 | .000 | .037 | .013 | .030 | .002 |
| | T | .134 | .215 | . | .000 | .000 | .093 | .223 | .300 | .049 | .166 |
| | TMIN | .024 | .230 | .000 | . | .030 | .014 | .315 | .080 | .270 | .109 |
| | TMAX | .337 | .375 | .000 | .030 | . | .260 | .324 | .357 | .019 | .479 |
| | HUMIDITY | .000 | .000 | .093 | .014 | .260 | . | .061 | .010 | .074 | .045 |
| | WINDS | .078 | .037 | .223 | .315 | .324 | .061 | . | .268 | .400 | .221 |
| | WINDD | .103 | .013 | .300 | .080 | .357 | .010 | .268 | . | .016 | .156 |
| | RAIN | .070 | .030 | .049 | .270 | .019 | .074 | .400 | .016 | . | .211 |
| | CLOUDS | .043 | .002 | .166 | .109 | .479 | .045 | .221 | .156 | .211 | . |
| N | CASES | 20 | 20 | 20 | 20 | 20 | 20 | 20 | 20 | 20 | 20 |
| | UV | 20 | 20 | 20 | 20 | 20 | 20 | 20 | 20 | 20 | 20 |
| | T | 20 | 20 | 20 | 20 | 20 | 20 | 20 | 20 | 20 | 20 |
| | TMIN | 20 | 20 | 20 | 20 | 20 | 20 | 20 | 20 | 20 | 20 |
| | TMAX | 20 | 20 | 20 | 20 | 20 | 20 | 20 | 20 | 20 | 20 |
| | HUMIDITY | 20 | 20 | 20 | 20 | 20 | 20 | 20 | 20 | 20 | 20 |
| | WINDS | 20 | 20 | 20 | 20 | 20 | 20 | 20 | 20 | 20 | 20 |
| | WINDD | 20 | 20 | 20 | 20 | 20 | 20 | 20 | 20 | 20 | 20 |
| | RAIN | 20 | 20 | 20 | 20 | 20 | 20 | 20 | 20 | 20 | 20 |
| | CLOUDS | 20 | 20 | 20 | 20 | 20 | 20 | 20 | 20 | 20 | 20 |

**Variables Entered/Removed[a]**

| Model | Variables Entered | Variables Removed | Method |
| --- | --- | --- | --- |
| 1 | HUMIDITY | . | Stepwise (Criteria: Probability-of-F-to-enter <= .050, Probability-of-F-to-remove >= .100). |

a. Dependent Variable: CASES

## Model Summary

| Model | R | R Square | Adjusted R Square | Std. Error of the Estimate | Change Statistics | | | | |
| --- | --- | --- | --- | --- | --- | --- | --- | --- | --- |
| | | | | | R Square Change | F Change | df1 | df2 | Sig. F Change |
| 1 | .858[a] | .737 | .722 | .21216 | .737 | 50.393 | 1 | 18 | .000 |

a. Predictors: (Constant), HUMIDITY

## ANOVA[a]

| Model | | Sum of Squares | df | Mean Square | F | Sig. |
| --- | --- | --- | --- | --- | --- | --- |
| 1 | Regression | 2.268 | 1 | 2.268 | 50.393 | .000[b] |
| | Residual | .810 | 18 | .045 | | |
| | Total | 3.079 | 19 | | | |

a. Dependent Variable: CASES

b. Predictors: (Constant), HUMIDITY

## Coefficients[a]

| Model | | Unstandardized Coefficients B | Unstandardized Coefficients Std. Error | Standardized Coefficients Beta | t | Sig. | Correlations Zero-order | Correlations Partial | Correlations Part | Collinearity Statistics Tolerance | Collinearity Statistics VIF |
|---|---|---|---|---|---|---|---|---|---|---|---|
| 1 | (Constant) | 13.767 | 1.712 | | 8.043 | .000 | | | | | |
| | HUMIDITY | -6.741 | .950 | -.858 | -7.099 | .000 | -.858 | -.858 | -.858 | 1.000 | 1.000 |

a. Dependent Variable: CASES

## Excluded Variables[a]

| Model | | Beta In | t | Sig. | Partial Correlation | Collinearity Statistics Tolerance | Collinearity Statistics VIF | Collinearity Statistics Minimum Tolerance |
|---|---|---|---|---|---|---|---|---|
| 1 | UV | -.069[b] | -.396 | .697 | -.096 | .501 | 1.996 | .501 |
| | T | .005[b] | .038 | .970 | .009 | .905 | 1.105 | .905 |
| | TMIN | -.032[b] | -.224 | .826 | -.054 | .758 | 1.319 | .758 |
| | TMAX | .032[b] | .254 | .802 | .062 | .977 | 1.024 | .977 |
| | WINDS | -.025[b] | -.191 | .851 | -.046 | .872 | 1.147 | .872 |
| | WINDD | .203[b] | 1.487 | .155 | .339 | .733 | 1.365 | .733 |
| | RAIN | -.060[b] | -.455 | .655 | -.110 | .887 | 1.127 | .887 |
| | CLOUDS | -.071[b] | -.530 | .603 | -.127 | .848 | 1.179 | .848 |

a. Dependent Variable: CASES

b. Predictors in the Model: (Constant), HUMIDITY

## Collinearity Diagnostics[a]

| Model | Dimension | Eigenvalue | Condition Index | Variance Proportions (Constant) | HUMIDITY |
|---|---|---|---|---|---|
| 1 | 1 | 2.000 | 1.000 | .00 | .00 |
| | 2 | .000 | 72.153 | 1.00 | 1.00 |

a. Dependent Variable: CASES

## Italy 3 Day Lag without Humidity

### Descriptive Statistics

| | Mean | Std. Deviation | N |
|---|---|---|---|
| CASES | 1.6201 | .40253 | 20 |
| UV | .6969 | .03957 | 20 |
| T | 2.4540 | .00225 | 20 |
| TMIN | 2.4484 | .00428 | 20 |
| TMAX | 2.4583 | .00157 | 20 |
| WINDS | .4250 | .10027 | 20 |
| WINDD | 2.1220 | .08658 | 20 |
| RAIN | -1.5349 | .36233 | 20 |
| CLOUDS | .8558 | .33865 | 20 |

### Correlations

| | | CASES | UV | T | TMIN | TMAX | WINDS | WINDD | RAIN | CLOUDS |
|---|---|---|---|---|---|---|---|---|---|---|
| Pearson Correlation | CASES | 1.000 | -.641 | -.260 | -.446 | -.100 | -.329 | -.295 | -.341 | -.395 |
| | UV | -.641 | 1.000 | .187 | .175 | .076 | .408 | .496 | .427 | .616 |
| | T | -.260 | .187 | 1.000 | .864 | .760 | .180 | .125 | -.380 | -.229 |
| | TMIN | -.446 | .175 | .864 | 1.000 | .427 | .115 | .327 | -.146 | -.288 |
| | TMAX | -.100 | .076 | .760 | .427 | 1.000 | .109 | -.088 | -.468 | .012 |
| | WINDS | -.329 | .408 | .180 | .115 | .109 | 1.000 | .147 | -.061 | .182 |
| | WINDD | -.295 | .496 | .125 | .327 | -.088 | .147 | 1.000 | .483 | .238 |
| | RAIN | -.341 | .427 | -.380 | -.146 | -.468 | -.061 | .483 | 1.000 | .190 |
| | CLOUDS | -.395 | .616 | -.229 | -.288 | .012 | .182 | .238 | .190 | 1.000 |
| Sig. (1-tailed) | CASES | . | .001 | .134 | .024 | .337 | .078 | .103 | .070 | .043 |
| | UV | .001 | . | .215 | .230 | .375 | .037 | .013 | .030 | .002 |
| | T | .134 | .215 | . | .000 | .000 | .223 | .300 | .049 | .166 |
| | TMIN | .024 | .230 | .000 | . | .030 | .315 | .080 | .270 | .109 |
| | TMAX | .337 | .375 | .000 | .030 | . | .324 | .357 | .019 | .479 |
| | WINDS | .078 | .037 | .223 | .315 | .324 | . | .268 | .400 | .221 |
| | WINDD | .103 | .013 | .300 | .080 | .357 | .268 | . | .016 | .156 |
| | RAIN | .070 | .030 | .049 | .270 | .019 | .400 | .016 | . | .211 |
| | CLOUDS | .043 | .002 | .166 | .109 | .479 | .221 | .156 | .211 | . |
| N | CASES | 20 | 20 | 20 | 20 | 20 | 20 | 20 | 20 | 20 |
| | UV | 20 | 20 | 20 | 20 | 20 | 20 | 20 | 20 | 20 |
| | T | 20 | 20 | 20 | 20 | 20 | 20 | 20 | 20 | 20 |
| | TMIN | 20 | 20 | 20 | 20 | 20 | 20 | 20 | 20 | 20 |
| | TMAX | 20 | 20 | 20 | 20 | 20 | 20 | 20 | 20 | 20 |
| | WINDS | 20 | 20 | 20 | 20 | 20 | 20 | 20 | 20 | 20 |
| | WINDD | 20 | 20 | 20 | 20 | 20 | 20 | 20 | 20 | 20 |
| | RAIN | 20 | 20 | 20 | 20 | 20 | 20 | 20 | 20 | 20 |
| | CLOUDS | 20 | 20 | 20 | 20 | 20 | 20 | 20 | 20 | 20 |

### Variables Entered/Removed[a]

| Model | Variables Entered | Variables Removed | Method |
|---|---|---|---|
| 1 | UV | . | Stepwise (Criteria: Probability-of-F-to-enter <= .050, Probability-of-F-to-remove >= .100). |

a. Dependent Variable: CASES

### Model Summary

| Model | R | R Square | Adjusted R Square | Std. Error of the Estimate | Change Statistics | | | | |
|---|---|---|---|---|---|---|---|---|---|
| | | | | | R Square Change | F Change | df1 | df2 | Sig. F Change |
| 1 | .641[a] | .411 | .378 | .31741 | .411 | 12.557 | 1 | 18 | .002 |

a. Predictors: (Constant), UV

### ANOVA[a]

| Model | | Sum of Squares | df | Mean Square | F | Sig. |
|---|---|---|---|---|---|---|
| 1 | Regression | 1.265 | 1 | 1.265 | 12.557 | .002[b] |
| | Residual | 1.813 | 18 | .101 | | |
| | Total | 3.079 | 19 | | | |

a. Dependent Variable: CASES

b. Predictors: (Constant), UV

**Coefficients[a]**

| Model | | Unstandardized Coefficients | | Standardized Coefficients | t | Sig. | Correlations | | | Collinearity Statistics | |
|---|---|---|---|---|---|---|---|---|---|---|---|
| | | B | Std. Error | Beta | | | Zero-order | Partial | Part | Tolerance | VIF |
| 1 | (Constant) | 6.165 | 1.284 | | 4.800 | .000 | | | | | |
| | UV | -6.521 | 1.840 | -.641 | -3.544 | .002 | -.641 | -.641 | -.641 | 1.000 | 1.000 |

a. Dependent Variable: CASES

**Excluded Variables[a]**

| Model | | Beta In | t | Sig. | Partial Correlation | Collinearity Statistics | | |
|---|---|---|---|---|---|---|---|---|
| | | | | | | Tolerance | VIF | Minimum Tolerance |
| 1 | T | -.145[b] | -.779 | .447 | -.186 | .965 | 1.036 | .965 |
| | TMIN | -.345[b] | -2.031 | .058 | -.442 | .969 | 1.032 | .969 |
| | TMAX | -.052[b] | -.278 | .784 | -.067 | .994 | 1.006 | .994 |
| | WINDS | -.081[b] | -.400 | .694 | -.097 | .834 | 1.199 | .834 |
| | WINDD | .030[b] | .142 | .889 | .034 | .754 | 1.326 | .754 |
| | RAIN | -.083[b] | -.404 | .691 | -.098 | .818 | 1.223 | .818 |
| | CLOUDS | .001[b] | .004 | .997 | .001 | .620 | 1.613 | .620 |

a. Dependent Variable: CASES

b. Predictors in the Model: (Constant), UV

**Collinearity Diagnostics[a]**

| Model | Dimension | Eigenvalue | Condition Index | Variance Proportions (Constant) | UV |
|---|---|---|---|---|---|
| 1 | 1 | 1.998 | 1.000 | .00 | .00 |
| | 2 | .002 | 36.165 | 1.00 | 1.00 |

a. Dependent Variable: CASES

## Italy 7 Day Lag

**Descriptive Statistics**

| | Mean | Std. Deviation | N |
|---|---|---|---|
| CASES | 1.6113 | .39924 | 20 |
| UV | .6969 | .03957 | 20 |
| T | 2.4540 | .00225 | 20 |
| TMIN | 2.4484 | .00428 | 20 |
| TMAX | 2.4583 | .00157 | 20 |
| HUMIDITY | 1.8020 | .05126 | 20 |
| WINDS | .4250 | .10027 | 20 |
| WINDD | 2.1220 | .08658 | 20 |
| RAIN | -1.5349 | .36233 | 20 |
| CLOUD | .8558 | .33865 | 20 |

**Correlations**

| | | CASES | UV | T | TMIN | TMAX | HUMIDITY | WINDS | WINDD | RAIN | CLOUD |
|---|---|---|---|---|---|---|---|---|---|---|---|
| Pearson Correlation | CASES | 1.000 | -.631 | -.265 | -.450 | -.103 | -.858 | -.338 | -.280 | -.321 | -.386 |
| | UV | -.631 | 1.000 | .187 | .175 | .076 | .706 | .408 | .496 | .427 | .616 |
| | T | -.265 | .187 | 1.000 | .864 | .760 | .308 | .180 | .125 | -.380 | -.229 |
| | TMIN | -.450 | .175 | .864 | 1.000 | .427 | .492 | .115 | .327 | -.146 | -.288 |
| | TMAX | -.103 | .076 | .760 | .427 | 1.000 | .153 | .109 | -.088 | -.468 | .012 |
| | HUMIDITY | -.858 | .706 | .308 | .492 | .153 | 1.000 | .357 | .517 | .336 | .389 |
| | WINDS | -.338 | .408 | .180 | .115 | .109 | .357 | 1.000 | .147 | -.061 | .182 |
| | WINDD | -.280 | .496 | .125 | .327 | -.088 | .517 | .147 | 1.000 | .483 | .238 |
| | RAIN | -.321 | .427 | -.380 | -.146 | -.468 | .336 | -.061 | .483 | 1.000 | .190 |
| | CLOUD | -.386 | .616 | -.229 | -.288 | .012 | .389 | .182 | .238 | .190 | 1.000 |
| Sig. (1-tailed) | CASES | . | .001 | .129 | .023 | .333 | .000 | .073 | .116 | .084 | .046 |
| | UV | .001 | . | .215 | .230 | .375 | .000 | .037 | .013 | .030 | .002 |
| | T | .129 | .215 | . | .000 | .000 | .093 | .223 | .300 | .049 | .166 |
| | TMIN | .023 | .230 | .000 | . | .030 | .014 | .315 | .080 | .270 | .109 |
| | TMAX | .333 | .375 | .000 | .030 | . | .260 | .324 | .357 | .019 | .479 |
| | HUMIDITY | .000 | .000 | .093 | .014 | .260 | . | .061 | .010 | .074 | .045 |
| | WINDS | .073 | .037 | .223 | .315 | .324 | .061 | . | .268 | .400 | .221 |
| | WINDD | .116 | .013 | .300 | .080 | .357 | .010 | .268 | . | .016 | .156 |
| | RAIN | .084 | .030 | .049 | .270 | .019 | .074 | .400 | .016 | . | .211 |
| | CLOUD | .046 | .002 | .166 | .109 | .479 | .045 | .221 | .156 | .211 | . |
| N | CASES | 20 | 20 | 20 | 20 | 20 | 20 | 20 | 20 | 20 | 20 |
| | UV | 20 | 20 | 20 | 20 | 20 | 20 | 20 | 20 | 20 | 20 |
| | T | 20 | 20 | 20 | 20 | 20 | 20 | 20 | 20 | 20 | 20 |
| | TMIN | 20 | 20 | 20 | 20 | 20 | 20 | 20 | 20 | 20 | 20 |
| | TMAX | 20 | 20 | 20 | 20 | 20 | 20 | 20 | 20 | 20 | 20 |

| | | | | | | | | | | |
|---|---|---|---|---|---|---|---|---|---|---|
| HUMIDITY | 20 | 20 | 20 | 20 | 20 | 20 | 20 | 20 | 20 | 20 |
| WINDS | 20 | 20 | 20 | 20 | 20 | 20 | 20 | 20 | 20 | 20 |
| WINDD | 20 | 20 | 20 | 20 | 20 | 20 | 20 | 20 | 20 | 20 |
| RAIN | 20 | 20 | 20 | 20 | 20 | 20 | 20 | 20 | 20 | 20 |
| CLOUD | 20 | 20 | 20 | 20 | 20 | 20 | 20 | 20 | 20 | 20 |

**Variables Entered/Removed[a]**

| Model | Variables Entered | Variables Removed | Method |
|---|---|---|---|
| 1 | HUMIDITY | . | Stepwise (Criteria: Probability-of-F-to-enter <= .050, Probability-of-F-to-remove >= .100). |

a. Dependent Variable: CASES

**Model Summary**

| Model | R | R Square | Adjusted R Square | Std. Error of the Estimate | Change Statistics | | | | |
|---|---|---|---|---|---|---|---|---|---|
| | | | | | R Square Change | F Change | df1 | df2 | Sig. F Change |
| 1 | .858[a] | .736 | .721 | .21084 | .736 | 50.127 | 1 | 18 | .000 |

a. Predictors: (Constant), HUMIDITY

**ANOVA[a]**

| Model | | Sum of Squares | df | Mean Square | F | Sig. |
|---|---|---|---|---|---|---|
| 1 | Regression | 2.228 | 1 | 2.228 | 50.127 | .000[b] |
| | Residual | .800 | 18 | .044 | | |
| | Total | 3.029 | 19 | | | |

a. Dependent Variable: CASES

b. Predictors: (Constant), HUMIDITY

**Coefficients[a]**

| Model | | Unstandardized Coefficients | | Standardized Coefficients | t | Sig. | Correlations | | | Collinearity Statistics | |
|---|---|---|---|---|---|---|---|---|---|---|---|
| | | B | Std. Error | Beta | | | Zero-order | Partial | Part | Tolerance | VIF |
| 1 | (Constant) | 13.651 | 1.701 | | 8.024 | .000 | | | | | |
| | HUMIDITY | -6.681 | .944 | -.858 | -7.080 | .000 | -.858 | -.858 | -.858 | 1.000 | 1.000 |

a. Dependent Variable: CASES

**Excluded Variables[a]**

| Model | | Beta In | t | Sig. | Partial Correlation | Collinearity Statistics | | |
|---|---|---|---|---|---|---|---|---|
| | | | | | | Tolerance | VIF | Minimum Tolerance |
| 1 | UV | -.049[b] | -.281 | .782 | -.068 | .501 | 1.996 | .501 |
| | T | -.001[b] | -.009 | .993 | -.002 | .905 | 1.105 | .905 |
| | TMIN | -.037[b] | -.258 | .799 | -.062 | .758 | 1.319 | .758 |
| | TMAX | .029[b] | .230 | .821 | .056 | .977 | 1.024 | .977 |
| | WINDS | -.036[b] | -.269 | .791 | -.065 | .872 | 1.147 | .872 |
| | WINDD | .223[b] | 1.646 | .118 | .371 | .733 | 1.365 | .733 |
| | RAIN | -.037[b] | -.277 | .785 | -.067 | .887 | 1.127 | .887 |
| | CLOUD | -.062[b] | -.457 | .653 | -.110 | .848 | 1.179 | .848 |

a. Dependent Variable: CASES

b. Predictors in the Model: (Constant), HUMIDITY

**Collinearity Diagnostics[a]**

| Model | Dimension | Eigenvalue | Condition Index | Variance Proportions (Constant) | HUMIDITY |
|---|---|---|---|---|---|
| 1 | 1 | 2.000 | 1.000 | .00 | .00 |
| | 2 | .000 | 72.153 | 1.00 | 1.00 |

a. Dependent Variable: CASES

## Italy 7 Day Lag without Humidity

**Descriptive Statistics**

| | Mean | Std. Deviation | N |
|---|---|---|---|
| CASES | 1.6113 | .39924 | 20 |
| UV | .6969 | .03957 | 20 |
| T | 2.4540 | .00225 | 20 |
| TMIN | 2.4484 | .00428 | 20 |
| TMAX | 2.4583 | .00157 | 20 |
| WINDS | .4250 | .10027 | 20 |
| WINDD | 2.1220 | .08658 | 20 |
| RAIN | -1.5349 | .36233 | 20 |
| CLOUD | .8558 | .33865 | 20 |

**Correlations**

| | | CASES | UV | T | TMIN | TMAX | WINDS | WINDD | RAIN | CLOUD |
|---|---|---|---|---|---|---|---|---|---|---|
| Pearson Correlation | CASES | 1.000 | -.631 | -.265 | -.450 | -.103 | -.338 | -.280 | -.321 | -.386 |
| | UV | -.631 | 1.000 | .187 | .175 | .076 | .408 | .496 | .427 | .616 |
| | T | -.265 | .187 | 1.000 | .864 | .760 | .180 | .125 | -.380 | -.229 |
| | TMIN | -.450 | .175 | .864 | 1.000 | .427 | .115 | .327 | -.146 | -.288 |
| | TMAX | -.103 | .076 | .760 | .427 | 1.000 | .109 | -.088 | -.468 | .012 |
| | WINDS | -.338 | .408 | .180 | .115 | .109 | 1.000 | .147 | -.061 | .182 |
| | WINDD | -.280 | .496 | .125 | .327 | -.088 | .147 | 1.000 | .483 | .238 |
| | RAIN | -.321 | .427 | -.380 | -.146 | -.468 | -.061 | .483 | 1.000 | .190 |
| | CLOUD | -.386 | .616 | -.229 | -.288 | .012 | .182 | .238 | .190 | 1.000 |
| Sig. (1-tailed) | CASES | . | .001 | .129 | .023 | .333 | .073 | .116 | .084 | .046 |
| | UV | .001 | . | .215 | .230 | .375 | .037 | .013 | .030 | .002 |
| | T | .129 | .215 | . | .000 | .000 | .223 | .300 | .049 | .166 |
| | TMIN | .023 | .230 | .000 | . | .030 | .315 | .080 | .270 | .109 |
| | TMAX | .333 | .375 | .000 | .030 | . | .324 | .357 | .019 | .479 |
| | WINDS | .073 | .037 | .223 | .315 | .324 | . | .268 | .400 | .221 |
| | WINDD | .116 | .013 | .300 | .080 | .357 | .268 | . | .016 | .156 |
| | RAIN | .084 | .030 | .049 | .270 | .019 | .400 | .016 | . | .211 |
| | CLOUD | .046 | .002 | .166 | .109 | .479 | .221 | .156 | .211 | . |
| N | CASES | 20 | 20 | 20 | 20 | 20 | 20 | 20 | 20 | 20 |
| | UV | 20 | 20 | 20 | 20 | 20 | 20 | 20 | 20 | 20 |
| | T | 20 | 20 | 20 | 20 | 20 | 20 | 20 | 20 | 20 |
| | TMIN | 20 | 20 | 20 | 20 | 20 | 20 | 20 | 20 | 20 |
| | TMAX | 20 | 20 | 20 | 20 | 20 | 20 | 20 | 20 | 20 |
| | WINDS | 20 | 20 | 20 | 20 | 20 | 20 | 20 | 20 | 20 |
| | WINDD | 20 | 20 | 20 | 20 | 20 | 20 | 20 | 20 | 20 |
| | RAIN | 20 | 20 | 20 | 20 | 20 | 20 | 20 | 20 | 20 |
| | CLOUD | 20 | 20 | 20 | 20 | 20 | 20 | 20 | 20 | 20 |

**Variables Entered/Removed[a]**

| Model | Variables Entered | Variables Removed | Method |
|---|---|---|---|
| 1 | UV | . | Stepwise (Criteria: Probability-of-F-to-enter <= .050, Probability-of-F-to-remove >= .100). |

a. Dependent Variable: CASES

**Model Summary**

| Model | R | R Square | Adjusted R Square | Std. Error of the Estimate | Change Statistics | | | | |
|---|---|---|---|---|---|---|---|---|---|
| | | | | | R Square Change | F Change | df1 | df2 | Sig. F Change |
| 1 | .631[a] | .398 | .364 | .31832 | .398 | 11.888 | 1 | 18 | .003 |

a. Predictors: (Constant), UV

## ANOVA[a]

| Model | | Sum of Squares | df | Mean Square | F | Sig. |
|---|---|---|---|---|---|---|
| 1 | Regression | 1.205 | 1 | 1.205 | 11.888 | .003[b] |
| | Residual | 1.824 | 18 | .101 | | |
| | Total | 3.029 | 19 | | | |

a. Dependent Variable: CASES

b. Predictors: (Constant), UV

## Coefficients[a]

| | | Unstandardized Coefficients | | Standardized Coefficients | | | Correlations | | | Collinearity Statistics | |
|---|---|---|---|---|---|---|---|---|---|---|---|
| Model | | B | Std. Error | Beta | t | Sig. | Zero-order | Partial | Part | Tolerance | VIF |
| 1 | (Constant) | 6.046 | 1.288 | | 4.694 | .000 | | | | | |
| | UV | -6.363 | 1.845 | -.631 | -3.448 | .003 | -.631 | -.631 | -.631 | 1.000 | 1.000 |

a. Dependent Variable: CASES

## Excluded Variables[a]

| | | | | | | Collinearity Statistics | | |
|---|---|---|---|---|---|---|---|---|
| Model | | Beta In | t | Sig. | Partial Correlation | Tolerance | VIF | Minimum Tolerance |
| 1 | T | -.153[b] | -.812 | .428 | -.193 | .965 | 1.036 | .965 |
| | TMIN | -.350[b] | -2.043 | .057 | -.444 | .969 | 1.032 | .969 |
| | TMAX | -.055[b] | -.294 | .772 | -.071 | .994 | 1.006 | .994 |
| | WINDS | -.097[b] | -.473 | .642 | -.114 | .834 | 1.199 | .834 |
| | WINDD | .043[b] | .197 | .846 | .048 | .754 | 1.326 | .754 |
| | RAIN | -.063[b] | -.303 | .766 | -.073 | .818 | 1.223 | .818 |
| | CLOUD | .004[b] | .016 | .987 | .004 | .620 | 1.613 | .620 |

a. Dependent Variable: CASES

b. Predictors in the Model: (Constant), UV

**Collinearity Diagnostics[a]**

| | | | | Variance Proportions | |
|---|---|---|---|---|---|
| Model | Dimension | Eigenvalue | Condition Index | (Constant) | UV |
| 1 | 1 | 1.998 | 1.000 | .00 | .00 |
| | 2 | .002 | 36.165 | 1.00 | 1.00 |

a. Dependent Variable: CASES

## Italy 14 Day Lag

**Descriptive Statistics**

| | Mean | Std. Deviation | N |
|---|---|---|---|
| CASES | 1.5426 | .39272 | 20 |
| UV | .6969 | .03957 | 20 |
| T | 2.4540 | .00225 | 20 |
| TMIN | 2.4484 | .00428 | 20 |
| TMAX | 2.4583 | .00157 | 20 |
| HUMIDITY | 1.8020 | .05126 | 20 |
| WINDS | .4250 | .10027 | 20 |
| WINDD | 2.1220 | .08658 | 20 |
| RAIN | -1.5349 | .36233 | 20 |
| CLOUD | .8558 | .33865 | 20 |

**Correlations**

| | | CASES | UV | T | TMIN | TMAX | HUMIDITY | WINDS | WINDD | RAIN | CLOUD |
|---|---|---|---|---|---|---|---|---|---|---|---|
| Pearson Correlation | CASES | 1.000 | -.648 | -.261 | -.438 | -.085 | -.853 | -.386 | -.268 | -.285 | -.397 |
| | UV | -.648 | 1.000 | .187 | .175 | .076 | .706 | .408 | .496 | .427 | .616 |
| | T | -.261 | .187 | 1.000 | .864 | .760 | .308 | .180 | .125 | -.380 | -.229 |
| | TMIN | -.438 | .175 | .864 | 1.000 | .427 | .492 | .115 | .327 | -.146 | -.288 |
| | TMAX | -.085 | .076 | .760 | .427 | 1.000 | .153 | .109 | -.088 | -.468 | .012 |
| | HUMIDITY | -.853 | .706 | .308 | .492 | .153 | 1.000 | .357 | .517 | .336 | .389 |
| | WINDS | -.386 | .408 | .180 | .115 | .109 | .357 | 1.000 | .147 | -.061 | .182 |
| | WINDD | -.268 | .496 | .125 | .327 | -.088 | .517 | .147 | 1.000 | .483 | .238 |
| | RAIN | -.285 | .427 | -.380 | -.146 | -.468 | .336 | -.061 | .483 | 1.000 | .190 |
| | CLOUD | -.397 | .616 | -.229 | -.288 | .012 | .389 | .182 | .238 | .190 | 1.000 |
| Sig. (1-tailed) | CASES | . | .001 | .134 | .027 | .360 | .000 | .047 | .127 | .112 | .041 |
| | UV | .001 | . | .215 | .230 | .375 | .000 | .037 | .013 | .030 | .002 |
| | T | .134 | .215 | . | .000 | .000 | .093 | .223 | .300 | .049 | .166 |
| | TMIN | .027 | .230 | .000 | . | .030 | .014 | .315 | .080 | .270 | .109 |
| | TMAX | .360 | .375 | .000 | .030 | . | .260 | .324 | .357 | .019 | .479 |
| | HUMIDITY | .000 | .000 | .093 | .014 | .260 | . | .061 | .010 | .074 | .045 |
| | WINDS | .047 | .037 | .223 | .315 | .324 | .061 | . | .268 | .400 | .221 |
| | WINDD | .127 | .013 | .300 | .080 | .357 | .010 | .268 | . | .016 | .156 |
| | RAIN | .112 | .030 | .049 | .270 | .019 | .074 | .400 | .016 | . | .211 |
| | CLOUD | .041 | .002 | .166 | .109 | .479 | .045 | .221 | .156 | .211 | . |
| N | CASES | 20 | 20 | 20 | 20 | 20 | 20 | 20 | 20 | 20 | 20 |
| | UV | 20 | 20 | 20 | 20 | 20 | 20 | 20 | 20 | 20 | 20 |
| | T | 20 | 20 | 20 | 20 | 20 | 20 | 20 | 20 | 20 | 20 |
| | TMIN | 20 | 20 | 20 | 20 | 20 | 20 | 20 | 20 | 20 | 20 |
| | TMAX | 20 | 20 | 20 | 20 | 20 | 20 | 20 | 20 | 20 | 20 |
| | HUMIDITY | 20 | 20 | 20 | 20 | 20 | 20 | 20 | 20 | 20 | 20 |
| | WINDS | 20 | 20 | 20 | 20 | 20 | 20 | 20 | 20 | 20 | 20 |
| | WINDD | 20 | 20 | 20 | 20 | 20 | 20 | 20 | 20 | 20 | 20 |
| | RAIN | 20 | 20 | 20 | 20 | 20 | 20 | 20 | 20 | 20 | 20 |
| | CLOUD | 20 | 20 | 20 | 20 | 20 | 20 | 20 | 20 | 20 | 20 |

**Variables Entered/Removed[a]**

| Model | Variables Entered | Variables Removed | Method |
|---|---|---|---|
| 1 | HUMIDITY | . | Stepwise (Criteria: Probability-of-F-to-enter <= .050, Probability-of-F-to-remove >= .100). |

a. Dependent Variable: CASES

**Model Summary**

| Model | R | R Square | Adjusted R Square | Std. Error of the Estimate | Change Statistics | | | | |
|---|---|---|---|---|---|---|---|---|---|
| | | | | | R Square Change | F Change | df1 | df2 | Sig. F Change |
| 1 | .853[a] | .728 | .713 | .21053 | .728 | 48.112 | 1 | 18 | .000 |

a. Predictors: (Constant), HUMIDITY

**ANOVA[a]**

| Model | | Sum of Squares | df | Mean Square | F | Sig. |
|---|---|---|---|---|---|---|
| 1 | Regression | 2.132 | 1 | 2.132 | 48.112 | .000[b] |
| | Residual | .798 | 18 | .044 | | |
| | Total | 2.930 | 19 | | | |

a. Dependent Variable: CASES

b. Predictors: (Constant), HUMIDITY

**Coefficients[a]**

| Model | | Unstandardized Coefficients | | Standardized Coefficients | t | Sig. | Correlations | | | Collinearity Statistics | |
|---|---|---|---|---|---|---|---|---|---|---|---|
| | | B | Std. Error | Beta | | | Zero-order | Partial | Part | Tolerance | VIF |
| 1 | (Constant) | 13.321 | 1.699 | | 7.842 | .000 | | | | | |
| | HUMIDITY | -6.536 | .942 | -.853 | -6.936 | .000 | -.853 | -.853 | -.853 | 1.000 | 1.000 |

a. Dependent Variable: CASES

**Excluded Variables[a]**

| Model | | Beta In | t | Sig. | Partial Correlation | Collinearity Statistics | | |
|---|---|---|---|---|---|---|---|---|
| | | | | | | Tolerance | VIF | Minimum Tolerance |
| 1 | UV | -.091[b] | -.513 | .615 | -.123 | .501 | 1.996 | .501 |
| | T | .003[b] | .019 | .985 | .005 | .905 | 1.105 | .905 |
| | TMIN | -.024[b] | -.164 | .872 | -.040 | .758 | 1.319 | .758 |
| | TMAX | .046[b] | .361 | .722 | .087 | .977 | 1.024 | .977 |
| | WINDS | -.092[b] | -.691 | .499 | -.165 | .872 | 1.147 | .872 |
| | WINDD | .237[b] | 1.739 | .100 | .389 | .733 | 1.365 | .733 |
| | RAIN | .002[b] | .015 | .988 | .004 | .887 | 1.127 | .887 |
| | CLOUD | -.077[b] | -.563 | .580 | -.135 | .848 | 1.179 | .848 |

a. Dependent Variable: CASES

b. Predictors in the Model: (Constant), HUMIDITY

**Collinearity Diagnostics[a]**

| Model | Dimension | Eigenvalue | Condition Index | Variance Proportions | |
|---|---|---|---|---|---|
| | | | | (Constant) | HUMIDITY |
| 1 | 1 | 2.000 | 1.000 | .00 | .00 |
| | 2 | .000 | 72.153 | 1.00 | 1.00 |

a. Dependent Variable: CASES

**Italy 14 Day Lag without Humidity**

## Descriptive Statistics

| | Mean | Std. Deviation | N |
|---|---|---|---|
| CASES | 1.5426 | .39272 | 20 |
| UV | .6969 | .03957 | 20 |
| T | 2.4540 | .00225 | 20 |
| TMIN | 2.4484 | .00428 | 20 |
| TMAX | 2.4583 | .00157 | 20 |
| WINDS | .4250 | .10027 | 20 |
| WINDD | 2.1220 | .08658 | 20 |
| RAIN | -1.5349 | .36233 | 20 |
| CLOUD | .8558 | .33865 | 20 |

## Correlations

| | | CASES | UV | T | TMIN | TMAX | WINDS | WINDD | RAIN | CLOUD |
|---|---|---|---|---|---|---|---|---|---|---|
| Pearson Correlation | CASES | 1.000 | -.648 | -.261 | -.438 | -.085 | -.386 | -.268 | -.285 | -.397 |
| | UV | -.648 | 1.000 | .187 | .175 | .076 | .408 | .496 | .427 | .616 |
| | T | -.261 | .187 | 1.000 | .864 | .760 | .180 | .125 | -.380 | -.229 |
| | TMIN | -.438 | .175 | .864 | 1.000 | .427 | .115 | .327 | -.146 | -.288 |
| | TMAX | -.085 | .076 | .760 | .427 | 1.000 | .109 | -.088 | -.468 | .012 |
| | WINDS | -.386 | .408 | .180 | .115 | .109 | 1.000 | .147 | -.061 | .182 |
| | WINDD | -.268 | .496 | .125 | .327 | -.088 | .147 | 1.000 | .483 | .238 |
| | RAIN | -.285 | .427 | -.380 | -.146 | -.468 | -.061 | .483 | 1.000 | .190 |
| | CLOUD | -.397 | .616 | -.229 | -.288 | .012 | .182 | .238 | .190 | 1.000 |
| Sig. (1-tailed) | CASES | . | .001 | .134 | .027 | .360 | .047 | .127 | .112 | .041 |
| | UV | .001 | . | .215 | .230 | .375 | .037 | .013 | .030 | .002 |
| | T | .134 | .215 | . | .000 | .000 | .223 | .300 | .049 | .166 |
| | TMIN | .027 | .230 | .000 | . | .030 | .315 | .080 | .270 | .109 |
| | TMAX | .360 | .375 | .000 | .030 | . | .324 | .357 | .019 | .479 |
| | WINDS | .047 | .037 | .223 | .315 | .324 | . | .268 | .400 | .221 |
| | WINDD | .127 | .013 | .300 | .080 | .357 | .268 | . | .016 | .156 |
| | RAIN | .112 | .030 | .049 | .270 | .019 | .400 | .016 | . | .211 |
| | CLOUD | .041 | .002 | .166 | .109 | .479 | .221 | .156 | .211 | . |
| N | CASES | 20 | 20 | 20 | 20 | 20 | 20 | 20 | 20 | 20 |
| | UV | 20 | 20 | 20 | 20 | 20 | 20 | 20 | 20 | 20 |
| | T | 20 | 20 | 20 | 20 | 20 | 20 | 20 | 20 | 20 |
| | TMIN | 20 | 20 | 20 | 20 | 20 | 20 | 20 | 20 | 20 |
| | TMAX | 20 | 20 | 20 | 20 | 20 | 20 | 20 | 20 | 20 |
| | WINDS | 20 | 20 | 20 | 20 | 20 | 20 | 20 | 20 | 20 |
| | WINDD | 20 | 20 | 20 | 20 | 20 | 20 | 20 | 20 | 20 |
| | RAIN | 20 | 20 | 20 | 20 | 20 | 20 | 20 | 20 | 20 |
| | CLOUD | 20 | 20 | 20 | 20 | 20 | 20 | 20 | 20 | 20 |

**Variables Entered/Removed[a]**

| Model | Variables Entered | Variables Removed | Method |
|---|---|---|---|
| 1 | UV | . | Stepwise (Criteria: Probability-of-F-to-enter <= .050, Probability-of-F-to-remove >= .100). |

a. Dependent Variable: CASES

## Model Summary

| Model | R | R Square | Adjusted R Square | Std. Error of the Estimate | Change Statistics | | | | |
|---|---|---|---|---|---|---|---|---|---|
| | | | | | R Square Change | F Change | df1 | df2 | Sig. F Change |
| 1 | .648[a] | .420 | .388 | .30724 | .420 | 13.044 | 1 | 18 | .002 |

a. Predictors: (Constant), UV

## ANOVA[a]

| Model | | Sum of Squares | df | Mean Square | F | Sig. |
|---|---|---|---|---|---|---|
| 1 | Regression | 1.231 | 1 | 1.231 | 13.044 | .002[b] |
| | Residual | 1.699 | 18 | .094 | | |
| | Total | 2.930 | 19 | | | |

a. Dependent Variable: CASES

b. Predictors: (Constant), UV

## Coefficients[a]

| Model | | Unstandardized Coefficients B | Unstandardized Coefficients Std. Error | Standardized Coefficients Beta | t | Sig. | Correlations Zero-order | Correlations Partial | Correlations Part | Collinearity Statistics Tolerance | Collinearity Statistics VIF |
|---|---|---|---|---|---|---|---|---|---|---|---|
| 1 | (Constant) | 6.026 | 1.243 | | 4.847 | .000 | | | | | |
| | UV | -6.433 | 1.781 | -.648 | -3.612 | .002 | -.648 | -.648 | -.648 | 1.000 | 1.000 |

a. Dependent Variable: CASES

## Excluded Variables[a]

| Model | | Beta In | t | Sig. | Partial Correlation | Collinearity Statistics Tolerance | Collinearity Statistics VIF | Collinearity Statistics Minimum Tolerance |
|---|---|---|---|---|---|---|---|---|
| 1 | T | -.144[b] | -.781 | .446 | -.186 | .965 | 1.036 | .965 |
| | TMIN | -.334[b] | -1.976 | .065 | -.432 | .969 | 1.032 | .969 |
| | TMAX | -.037[b] | -.197 | .846 | -.048 | .994 | 1.006 | .994 |
| | WINDS | -.145[b] | -.730 | .475 | -.174 | .834 | 1.199 | .834 |
| | WINDD | .071[b] | .336 | .741 | .081 | .754 | 1.326 | .754 |
| | RAIN | -.010[b] | -.048 | .962 | -.012 | .818 | 1.223 | .818 |
| | CLOUD | .004[b] | .015 | .988 | .004 | .620 | 1.613 | .620 |

a. Dependent Variable: CASES

b. Predictors in the Model: (Constant), UV

## Collinearity Diagnostics[a]

| Model | Dimension | Eigenvalue | Condition Index | Variance Proportions |
|---|---|---|---|---|

| | | | | (Constant) | UV |
|---|---|---|---|---|---|
| 1 | 1 | 1.998 | 1.000 | .00 | .00 |
| | 2 | .002 | 36.165 | 1.00 | 1.00 |

a. Dependent Variable: CASES

*b) Results of stepwise multiple regression of cases for Spain by climatic variables with zero, three-day, seven-day and fourteen-day time lags.*

## Spain Zero Day Lag

**Descriptive Statistics**

| | Mean | Std. Deviation | N |
|---|---|---|---|
| CASES | 1.9263 | .34526 | 19 |
| UV | .7743 | .05073 | 19 |
| T | 2.4564 | .00351 | 19 |
| TMIN | 2.4518 | .00448 | 19 |
| TMAX | 2.4598 | .00329 | 19 |
| HUMIDITY | 1.8771 | .02233 | 19 |
| WINDS | .4526 | .10106 | 19 |
| WINDD | 2.1274 | .08436 | 19 |
| REAIN | -1.2485 | .59064 | 19 |
| CLOUDS | 1.5674 | .08462 | 19 |

## Correlations

| | | CASES | UV | T | TMIN | TMAX | HUMIDITY | WINDS | WINDD | REAIN | CLOUDS |
|---|---|---|---|---|---|---|---|---|---|---|---|
| Pearson Correlation | CASES | 1.000 | -.610 | -.782 | -.665 | -.804 | .363 | -.463 | -.136 | .080 | .214 |
| | UV | -.610 | 1.000 | .721 | .696 | .685 | -.654 | .514 | -.204 | -.179 | -.200 |
| | T | -.782 | .721 | 1.000 | .931 | .953 | -.521 | .366 | .152 | -.260 | -.310 |
| | TMIN | -.665 | .696 | .931 | 1.000 | .822 | -.453 | .447 | .042 | -.253 | -.477 |
| | TMAX | -.804 | .685 | .953 | .822 | 1.000 | -.592 | .269 | .192 | -.311 | -.179 |
| | HUMIDITY | .363 | -.654 | -.521 | -.453 | -.592 | 1.000 | -.044 | -.279 | .347 | .147 |
| | WINDS | -.463 | .514 | .366 | .447 | .269 | -.044 | 1.000 | -.281 | -.141 | .116 |
| | WINDD | -.136 | -.204 | .152 | .042 | .192 | -.279 | -.281 | 1.000 | .272 | -.214 |
| | REAIN | .080 | -.179 | -.260 | -.253 | -.311 | .347 | -.141 | .272 | 1.000 | -.303 |
| | CLOUDS | .214 | -.200 | -.310 | -.477 | -.179 | .147 | .116 | -.214 | -.303 | 1.000 |
| Sig. (1-tailed) | CASES | . | .003 | .000 | .001 | .000 | .063 | .023 | .290 | .372 | .189 |
| | UV | .003 | . | .000 | .000 | .001 | .001 | .012 | .201 | .232 | .206 |
| | T | .000 | .000 | . | .000 | .000 | .011 | .062 | .267 | .141 | .098 |
| | TMIN | .001 | .000 | .000 | . | .000 | .026 | .027 | .432 | .148 | .019 |
| | TMAX | .000 | .001 | .000 | .000 | . | .004 | .133 | .216 | .098 | .232 |
| | HUMIDITY | .063 | .001 | .011 | .026 | .004 | . | .430 | .124 | .073 | .274 |
| | WINDS | .023 | .012 | .062 | .027 | .133 | .430 | . | .122 | .283 | .318 |
| | WINDD | .290 | .201 | .267 | .432 | .216 | .124 | .122 | . | .130 | .189 |
| | REAIN | .372 | .232 | .141 | .148 | .098 | .073 | .283 | .130 | . | .103 |
| | CLOUDS | .189 | .206 | .098 | .019 | .232 | .274 | .318 | .189 | .103 | . |
| N | CASES | 19 | 19 | 19 | 19 | 19 | 19 | 19 | 19 | 19 | 19 |
| | UV | 19 | 19 | 19 | 19 | 19 | 19 | 19 | 19 | 19 | 19 |
| | T | 19 | 19 | 19 | 19 | 19 | 19 | 19 | 19 | 19 | 19 |
| | TMIN | 19 | 19 | 19 | 19 | 19 | 19 | 19 | 19 | 19 | 19 |
| | TMAX | 19 | 19 | 19 | 19 | 19 | 19 | 19 | 19 | 19 | 19 |
| | HUMIDITY | 19 | 19 | 19 | 19 | 19 | 19 | 19 | 19 | 19 | 19 |
| | WINDS | 19 | 19 | 19 | 19 | 19 | 19 | 19 | 19 | 19 | 19 |
| | WINDD | 19 | 19 | 19 | 19 | 19 | 19 | 19 | 19 | 19 | 19 |
| | REAIN | 19 | 19 | 19 | 19 | 19 | 19 | 19 | 19 | 19 | 19 |
| | CLOUDS | 19 | 19 | 19 | 19 | 19 | 19 | 19 | 19 | 19 | 19 |

### Variables Entered/Removed[a]

| Model | Variables Entered | Variables Removed | Method |
|---|---|---|---|
| 1 | TMAX | . | Stepwise (Criteria: Probability-of-F-to-enter <= .050, Probability-of-F-to-remove >= .100). |

a. Dependent Variable: CASES

**Model Summary**

| Model | R | R Square | Adjusted R Square | Std. Error of the Estimate | Change Statistics | | | | |
|---|---|---|---|---|---|---|---|---|---|
| | | | | | R Square Change | F Change | df1 | df2 | Sig. F Change |
| 1 | .804[a] | .647 | .626 | .21116 | .647 | 31.124 | 1 | 17 | .000 |

a. Predictors: (Constant), TMAX

**ANOVA[a]**

| Model | | Sum of Squares | df | Mean Square | F | Sig. |
|---|---|---|---|---|---|---|
| 1 | Regression | 1.388 | 1 | 1.388 | 31.124 | .000[b] |
| | Residual | .758 | 17 | .045 | | |
| | Total | 2.146 | 18 | | | |

a. Dependent Variable: CASES

b. Predictors: (Constant), TMAX

**Coefficients[a]**

| Model | | Unstandardized Coefficients | | Standardized Coefficients | t | Sig. | Correlations | | | Collinearity Statistics | |
|---|---|---|---|---|---|---|---|---|---|---|---|
| | | B | Std. Error | Beta | | | Zero-order | Partial | Part | Tolerance | VIF |
| 1 | (Constant) | 209.616 | 37.228 | | 5.631 | .000 | | | | | |
| | TMAX | -84.433 | 15.134 | -.804 | -5.579 | .000 | -.804 | -.804 | -.804 | 1.000 | 1.000 |

a. Dependent Variable: CASES

**Excluded Variables[a]**

| Model | | Beta In | t | Sig. | Partial Correlation | Collinearity Statistics Tolerance | VIF | Minimum Tolerance |
|---|---|---|---|---|---|---|---|---|
| 1 | UV | -.111[b] | -.548 | .592 | -.136 | .531 | 1.884 | .531 |
| | T | -.173[b] | -.354 | .728 | -.088 | .092 | 10.895 | .092 |
| | TMIN | -.012[b] | -.047 | .963 | -.012 | .325 | 3.077 | .325 |
| | HUMIDITY | -.174[b] | -.972 | .345 | -.236 | .650 | 1.539 | .650 |
| | WINDS | -.266[b] | -1.914 | .074 | -.432 | .928 | 1.078 | .928 |
| | WINDD | .019[b] | .128 | .900 | .032 | .963 | 1.038 | .963 |
| | REAIN | -.188[b] | -1.259 | .226 | -.300 | .903 | 1.107 | .903 |
| | CLOUDS | .073[b] | .485 | .634 | .120 | .968 | 1.033 | .968 |

a. Dependent Variable: CASES

b. Predictors in the Model: (Constant), TMAX

**Collinearity Diagnostics[a]**

| Model | Dimension | Eigenvalue | Condition Index | Variance Proportions (Constant) | TMAX |
|---|---|---|---|---|---|
| 1 | 1 | 2.000 | 1.000 | .00 | .00 |
| | 2 | 8.466E-7 | 1536.989 | 1.00 | 1.00 |

a. Dependent Variable: CASES

## Spain Zero Day Lag without Temperature

### Variables Entered/Removed[a]

| Model | Variables Entered | Variables Removed | Method |
|---|---|---|---|
| 1 | UV | . | Stepwise (Criteria: Probability-of-F-to-enter <= .050, Probability-of-F-to-remove >= .100). |

a. Dependent Variable: CASES

### Model Summary

| Model | R | R Square | Adjusted R Square | Std. Error of the Estimate | Change Statistics | | | | |
|---|---|---|---|---|---|---|---|---|---|
| | | | | | R Square Change | F Change | df1 | df2 | Sig. F Change |
| 1 | .610[a] | .372 | .335 | .28161 | .372 | 10.057 | 1 | 17 | .006 |

a. Predictors: (Constant), UV

### ANOVA[a]

| Model | | Sum of Squares | df | Mean Square | F | Sig. |
|---|---|---|---|---|---|---|
| 1 | Regression | .798 | 1 | .798 | 10.057 | .006[b] |
| | Residual | 1.348 | 17 | .079 | | |
| | Total | 2.146 | 18 | | | |

a. Dependent Variable: CASES

b. Predictors: (Constant), UV

## Excluded Variables[a]

| Model | | Beta In | t | Sig. | Partial Correlation | Collinearity Statistics Tolerance | VIF | Minimum Tolerance |
|---|---|---|---|---|---|---|---|---|
| 1 | HUMIDITY | -.063[b] | -.240 | .813 | -.060 | .572 | 1.748 | .572 |
| | WINDS | -.204[b] | -.905 | .379 | -.221 | .736 | 1.358 | .736 |
| | WINDD | -.272[b] | -1.425 | .173 | -.336 | .958 | 1.044 | .958 |
| | REAIN | -.030[b] | -.148 | .884 | -.037 | .968 | 1.033 | .968 |
| | CLOUDS | .096[b] | .480 | .637 | .119 | .960 | 1.042 | .960 |

a. Dependent Variable: CASES

b. Predictors in the Model: (Constant), UV

## Collinearity Diagnostics[a]

| Model | Dimension | Eigenvalue | Condition Index | Variance Proportions (Constant) | UV |
|---|---|---|---|---|---|
| 1 | 1 | 1.998 | 1.000 | .00 | .00 |
| | 2 | .002 | 31.392 | 1.00 | 1.00 |

a. Dependent Variable: CASES

**Coefficients[a]**

| Model | | Unstandardized Coefficients | | Standardized Coefficients | t | Sig. | Correlations | | | Collinearity Statistics | |
|---|---|---|---|---|---|---|---|---|---|---|---|
| | | B | Std. Error | Beta | | | Zero-order | Partial | Part | Tolerance | VIF |
| 1 | (Constant) | 5.139 | 1.015 | | 5.062 | .000 | | | | | |
| | UV | -4.149 | 1.308 | -.610 | -3.171 | .006 | -.610 | -.610 | -.610 | 1.000 | 1.000 |

a. Dependent Variable: CASES

# Spain 3 Day Lag

**Descriptive Statistics**

| | Mean | Std. Deviation | N |
|---|---|---|---|
| CASES | 1.9257 | .34726 | 19 |
| UV | .7743 | .05073 | 19 |
| T | 2.4564 | .00351 | 19 |
| TMIN | 2.4518 | .00448 | 19 |
| TMAX | 2.4598 | .00329 | 19 |
| HUMIDITY | 1.8771 | .02233 | 19 |
| WINDS | .4526 | .10106 | 19 |
| WINDD | 2.1274 | .08436 | 19 |
| RAIN | -1.2485 | .59064 | 19 |
| CLOUDS | 1.5674 | .08462 | 19 |

**Correlations**

| | CASES | UV | T | TMIN | TMAX | HUMIDITY | WINDS | WINDD | RAIN | CLOUDS |
|---|---|---|---|---|---|---|---|---|---|---|

| | | | | | | | | | | | |
|---|---|---|---|---|---|---|---|---|---|---|---|
| Pearson Correlation | CASES | 1.000 | -.607 | -.779 | -.663 | -.801 | .364 | -.464 | -.135 | .085 | .218 |
| | UV | -.607 | 1.000 | .721 | .696 | .685 | -.654 | .514 | -.204 | -.179 | -.200 |
| | T | -.779 | .721 | 1.000 | .931 | .953 | -.521 | .366 | .152 | -.260 | -.310 |
| | TMIN | -.663 | .696 | .931 | 1.000 | .822 | -.453 | .447 | .042 | -.253 | -.477 |
| | TMAX | -.801 | .685 | .953 | .822 | 1.000 | -.592 | .269 | .192 | -.311 | -.179 |
| | HUMIDITY | .364 | -.654 | -.521 | -.453 | -.592 | 1.000 | -.044 | -.279 | .347 | .147 |
| | WINDS | -.464 | .514 | .366 | .447 | .269 | -.044 | 1.000 | -.281 | -.141 | .116 |
| | WINDD | -.135 | -.204 | .152 | .042 | .192 | -.279 | -.281 | 1.000 | .272 | -.214 |
| | RAIN | .085 | -.179 | -.260 | -.253 | -.311 | .347 | -.141 | .272 | 1.000 | -.303 |
| | CLOUDS | .218 | -.200 | -.310 | -.477 | -.179 | .147 | .116 | -.214 | -.303 | 1.000 |
| Sig. (1-tailed) | CASES | . | .003 | .000 | .001 | .000 | .063 | .023 | .291 | .365 | .186 |
| | UV | .003 | . | .000 | .000 | .001 | .001 | .012 | .201 | .232 | .206 |
| | T | .000 | .000 | . | .000 | .000 | .011 | .062 | .267 | .141 | .098 |
| | TMIN | .001 | .000 | .000 | . | .000 | .026 | .027 | .432 | .148 | .019 |
| | TMAX | .000 | .001 | .000 | .000 | . | .004 | .133 | .216 | .098 | .232 |
| | HUMIDITY | .063 | .001 | .011 | .026 | .004 | . | .430 | .124 | .073 | .274 |
| | WINDS | .023 | .012 | .062 | .027 | .133 | .430 | . | .122 | .283 | .318 |
| | WINDD | .291 | .201 | .267 | .432 | .216 | .124 | .122 | . | .130 | .189 |
| | RAIN | .365 | .232 | .141 | .148 | .098 | .073 | .283 | .130 | . | .103 |
| | CLOUDS | .186 | .206 | .098 | .019 | .232 | .274 | .318 | .189 | .103 | . |
| N | CASES | 19 | 19 | 19 | 19 | 19 | 19 | 19 | 19 | 19 | 19 |
| | UV | 19 | 19 | 19 | 19 | 19 | 19 | 19 | 19 | 19 | 19 |
| | T | 19 | 19 | 19 | 19 | 19 | 19 | 19 | 19 | 19 | 19 |
| | TMIN | 19 | 19 | 19 | 19 | 19 | 19 | 19 | 19 | 19 | 19 |
| | TMAX | 19 | 19 | 19 | 19 | 19 | 19 | 19 | 19 | 19 | 19 |
| | HUMIDITY | 19 | 19 | 19 | 19 | 19 | 19 | 19 | 19 | 19 | 19 |
| | WINDS | 19 | 19 | 19 | 19 | 19 | 19 | 19 | 19 | 19 | 19 |
| | WINDD | 19 | 19 | 19 | 19 | 19 | 19 | 19 | 19 | 19 | 19 |
| | RAIN | 19 | 19 | 19 | 19 | 19 | 19 | 19 | 19 | 19 | 19 |
| | CLOUDS | 19 | 19 | 19 | 19 | 19 | 19 | 19 | 19 | 19 | 19 |

**Variables Entered/Removed[a]**

| Model | Variables Entered | Variables Removed | Method |
|---|---|---|---|
| 1 | TMAX | . | Stepwise (Criteria: Probability-of-F-to-enter <= .050, Probability-of-F-to-remove >= .100). |

a. Dependent Variable: CASES

**Model Summary**

| Model | R | R Square | Adjusted R Square | Std. Error of the Estimate | Change Statistics | | | | |
|---|---|---|---|---|---|---|---|---|---|
| | | | | | R Square Change | F Change | df1 | df2 | Sig. F Change |
| 1 | .801[a] | .642 | .621 | .21382 | .642 | 30.478 | 1 | 17 | .000 |

a. Predictors: (Constant), TMAX

**ANOVA[a]**

| Model | | Sum of Squares | df | Mean Square | F | Sig. |
|---|---|---|---|---|---|---|
| 1 | Regression | 1.393 | 1 | 1.393 | 30.478 | .000[b] |
| | Residual | .777 | 17 | .046 | | |
| | Total | 2.171 | 18 | | | |

a. Dependent Variable: CASES

b. Predictors: (Constant), TMAX

## Coefficients[a]

| Model | | Unstandardized Coefficients | | Standardized Coefficients | t | Sig. | Correlations | | | Collinearity Statistics | |
|---|---|---|---|---|---|---|---|---|---|---|---|
| | | B | Std. Error | Beta | | | Zero-order | Partial | Part | Tolerance | VIF |
| 1 | (Constant) | 210.040 | 37.697 | | 5.572 | .000 | | | | | |
| | TMAX | -84.605 | 15.325 | -.801 | -5.521 | .000 | -.801 | -.801 | -.801 | 1.000 | 1.000 |

a. Dependent Variable: CASES

## Excluded Variables[a]

| Model | | Beta In | t | Sig. | Partial Correlation | Collinearity Statistics | | |
|---|---|---|---|---|---|---|---|---|
| | | | | | | Tolerance | VIF | Minimum Tolerance |
| 1 | UV | -.110[b] | -.541 | .596 | -.134 | .531 | 1.884 | .531 |
| | T | -.171[b] | -.347 | .733 | -.086 | .092 | 10.895 | .092 |
| | TMIN | -.014[b] | -.052 | .959 | -.013 | .325 | 3.077 | .325 |
| | HUMIDITY | -.170[b] | -.938 | .362 | -.228 | .650 | 1.539 | .650 |
| | WINDS | -.268[b] | -1.909 | .074 | -.431 | .928 | 1.078 | .928 |
| | WINDD | .019[b] | .128 | .900 | .032 | .963 | 1.038 | .963 |
| | RAIN | -.182[b] | -1.205 | .246 | -.288 | .903 | 1.107 | .903 |
| | CLOUDS | .077[b] | .508 | .619 | .126 | .968 | 1.033 | .968 |

a. Dependent Variable: CASES

b. Predictors in the Model: (Constant), TMAX

### Collinearity Diagnostics[a]

| Model | Dimension | Eigenvalue | Condition Index | Variance Proportions (Constant) | TMAX |
|---|---|---|---|---|---|
| 1 | 1 | 2.000 | 1.000 | .00 | .00 |
| | 2 | 8.466E-7 | 1536.989 | 1.00 | 1.00 |

a. Dependent Variable: CASES

## Spain 3 Day Lag without Temperature

### Descriptive Statistics

| | Mean | Std. Deviation | N |
|---|---|---|---|
| CASES | 1.9257 | .34726 | 19 |
| UV | .7743 | .05073 | 19 |
| HUMIDITY | 1.8771 | .02233 | 19 |
| WINDS | .4526 | .10106 | 19 |
| WINDD | 2.1274 | .08436 | 19 |
| RAIN | -1.2485 | .59064 | 19 |
| CLOUDS | 1.5674 | .08462 | 19 |

## Correlations

| | | CASES | UV | HUMIDITY | WINDS | WINDD | RAIN | CLOUDS |
|---|---|---|---|---|---|---|---|---|
| Pearson Correlation | CASES | 1.000 | -.607 | .364 | -.464 | -.135 | .085 | .218 |
| | UV | -.607 | 1.000 | -.654 | .514 | -.204 | -.179 | -.200 |
| | HUMIDITY | .364 | -.654 | 1.000 | -.044 | -.279 | .347 | .147 |
| | WINDS | -.464 | .514 | -.044 | 1.000 | -.281 | -.141 | .116 |
| | WINDD | -.135 | -.204 | -.279 | -.281 | 1.000 | .272 | -.214 |
| | RAIN | .085 | -.179 | .347 | -.141 | .272 | 1.000 | -.303 |
| | CLOUDS | .218 | -.200 | .147 | .116 | -.214 | -.303 | 1.000 |
| Sig. (1-tailed) | CASES | . | .003 | .063 | .023 | .291 | .365 | .186 |
| | UV | .003 | . | .001 | .012 | .201 | .232 | .206 |
| | HUMIDITY | .063 | .001 | . | .430 | .124 | .073 | .274 |
| | WINDS | .023 | .012 | .430 | . | .122 | .283 | .318 |
| | WINDD | .291 | .201 | .124 | .122 | . | .130 | .189 |
| | RAIN | .365 | .232 | .073 | .283 | .130 | . | .103 |
| | CLOUDS | .186 | .206 | .274 | .318 | .189 | .103 | . |
| N | CASES | 19 | 19 | 19 | 19 | 19 | 19 | 19 |
| | UV | 19 | 19 | 19 | 19 | 19 | 19 | 19 |
| | HUMIDITY | 19 | 19 | 19 | 19 | 19 | 19 | 19 |
| | WINDS | 19 | 19 | 19 | 19 | 19 | 19 | 19 |
| | WINDD | 19 | 19 | 19 | 19 | 19 | 19 | 19 |
| | RAIN | 19 | 19 | 19 | 19 | 19 | 19 | 19 |
| | CLOUDS | 19 | 19 | 19 | 19 | 19 | 19 | 19 |

## Variables Entered/Removed[a]

| Model | Variables Entered | Variables Removed | Method |
|---|---|---|---|
| 1 | UV | . | Stepwise (Criteria: Probability-of-F-to-enter <= .050, Probability-of-F-to-remove >= .100). |

a. Dependent Variable: CASES

## Model Summary

| Model | R | R Square | Adjusted R Square | Std. Error of the Estimate | Change Statistics | | | | |
|---|---|---|---|---|---|---|---|---|---|
| | | | | | R Square Change | F Change | df1 | df2 | Sig. F Change |
| 1 | .607[a] | .369 | .332 | .28387 | .369 | 9.936 | 1 | 17 | .006 |

a. Predictors: (Constant), UV

## ANOVA[a]

| Model | | Sum of Squares | df | Mean Square | F | Sig. |
|---|---|---|---|---|---|---|
| 1 | Regression | .801 | 1 | .801 | 9.936 | .006[b] |
| | Residual | 1.370 | 17 | .081 | | |
| | Total | 2.171 | 18 | | | |

a. Dependent Variable: CASES

b. Predictors: (Constant), UV

**Coefficients[a]**

| Model | | Unstandardized Coefficients | | Standardized Coefficients | t | Sig. | Correlations | | | Collinearity Statistics | |
|---|---|---|---|---|---|---|---|---|---|---|---|
| | | B | Std. Error | Beta | | | Zero-order | Partial | Part | Tolerance | VIF |
| 1 | (Constant) | 5.145 | 1.023 | | 5.028 | .000 | | | | | |
| | UV | -4.157 | 1.319 | -.607 | -3.152 | .006 | -.607 | -.607 | -.607 | 1.000 | 1.000 |

a. Dependent Variable: CASES

**Excluded Variables[a]**

| Model | | Beta In | t | Sig. | Partial Correlation | Collinearity Statistics | | |
|---|---|---|---|---|---|---|---|---|
| | | | | | | Tolerance | VIF | Minimum Tolerance |
| 1 | HUMIDITY | -.058[b] | -.221 | .828 | -.055 | .572 | 1.748 | .572 |
| | WINDS | -.206[b] | -.913 | .375 | -.222 | .736 | 1.358 | .736 |
| | WINDD | -.270[b] | -1.414 | .177 | -.333 | .958 | 1.044 | .958 |
| | RAIN | -.025[b] | -.122 | .905 | -.030 | .968 | 1.033 | .968 |
| | CLOUDS | .100[b] | .498 | .625 | .124 | .960 | 1.042 | .960 |

a. Dependent Variable: CASES

b. Predictors in the Model: (Constant), UV

**Collinearity Diagnostics[a]**

| Model | Dimension | Eigenvalue | Condition Index | Variance Proportions (Constant) | UV |
|---|---|---|---|---|---|
| 1 | 1 | 1.998 | 1.000 | .00 | .00 |
| | 2 | .002 | 31.392 | 1.00 | 1.00 |

a. Dependent Variable: CASES

# Spain 7 Day Lag

**Descriptive Statistics**

| | Mean | Std. Deviation | N |
|---|---|---|---|
| CASES | 1.9270 | .35866 | 19 |
| UV | .7743 | .05073 | 19 |
| T | 2.4564 | .00351 | 19 |
| TMIN | 2.4518 | .00448 | 19 |
| TMAX | 2.4598 | .00329 | 19 |
| HUMIDITY | 1.8771 | .02233 | 19 |
| WINDS | .4526 | .10106 | 19 |
| WINDD | 2.1274 | .08436 | 19 |
| RAIN | -1.2485 | .59064 | 19 |
| CLOUDS | 1.5674 | .08462 | 19 |

**Correlations**

| | | CASES | UV | T | TMIN | TMAX | HUMIDITY | WINDS | WINDD | RAIN | CLOUDS |
|---|---|---|---|---|---|---|---|---|---|---|---|
| Pearson Correlation | CASES | 1.000 | -.618 | -.774 | -.662 | -.796 | .372 | -.458 | -.106 | .103 | .226 |
| | UV | -.618 | 1.000 | .721 | .696 | .685 | -.654 | .514 | -.204 | -.179 | -.200 |
| | T | -.774 | .721 | 1.000 | .931 | .953 | -.521 | .366 | .152 | -.260 | -.310 |
| | TMIN | -.662 | .696 | .931 | 1.000 | .822 | -.453 | .447 | .042 | -.253 | -.477 |
| | TMAX | -.796 | .685 | .953 | .822 | 1.000 | -.592 | .269 | .192 | -.311 | -.179 |
| | HUMIDITY | .372 | -.654 | -.521 | -.453 | -.592 | 1.000 | -.044 | -.279 | .347 | .147 |
| | WINDS | -.458 | .514 | .366 | .447 | .269 | -.044 | 1.000 | -.281 | -.141 | .116 |
| | WINDD | -.106 | -.204 | .152 | .042 | .192 | -.279 | -.281 | 1.000 | .272 | -.214 |
| | RAIN | .103 | -.179 | -.260 | -.253 | -.311 | .347 | -.141 | .272 | 1.000 | -.303 |
| | CLOUDS | .226 | -.200 | -.310 | -.477 | -.179 | .147 | .116 | -.214 | -.303 | 1.000 |
| Sig. (1-tailed) | CASES | . | .002 | .000 | .001 | .000 | .059 | .024 | .332 | .337 | .176 |
| | UV | .002 | . | .000 | .000 | .001 | .001 | .012 | .201 | .232 | .206 |
| | T | .000 | .000 | . | .000 | .000 | .011 | .062 | .267 | .141 | .098 |
| | TMIN | .001 | .000 | .000 | . | .000 | .026 | .027 | .432 | .148 | .019 |
| | TMAX | .000 | .001 | .000 | .000 | . | .004 | .133 | .216 | .098 | .232 |
| | HUMIDITY | .059 | .001 | .011 | .026 | .004 | . | .430 | .124 | .073 | .274 |
| | WINDS | .024 | .012 | .062 | .027 | .133 | .430 | . | .122 | .283 | .318 |
| | WINDD | .332 | .201 | .267 | .432 | .216 | .124 | .122 | . | .130 | .189 |
| | RAIN | .337 | .232 | .141 | .148 | .098 | .073 | .283 | .130 | . | .103 |
| | CLOUDS | .176 | .206 | .098 | .019 | .232 | .274 | .318 | .189 | .103 | . |
| N | CASES | 19 | 19 | 19 | 19 | 19 | 19 | 19 | 19 | 19 | 19 |
| | UV | 19 | 19 | 19 | 19 | 19 | 19 | 19 | 19 | 19 | 19 |
| | T | 19 | 19 | 19 | 19 | 19 | 19 | 19 | 19 | 19 | 19 |
| | TMIN | 19 | 19 | 19 | 19 | 19 | 19 | 19 | 19 | 19 | 19 |
| | TMAX | 19 | 19 | 19 | 19 | 19 | 19 | 19 | 19 | 19 | 19 |
| | HUMIDITY | 19 | 19 | 19 | 19 | 19 | 19 | 19 | 19 | 19 | 19 |
| | WINDS | 19 | 19 | 19 | 19 | 19 | 19 | 19 | 19 | 19 | 19 |
| | WINDD | 19 | 19 | 19 | 19 | 19 | 19 | 19 | 19 | 19 | 19 |
| | RAIN | 19 | 19 | 19 | 19 | 19 | 19 | 19 | 19 | 19 | 19 |
| | CLOUDS | 19 | 19 | 19 | 19 | 19 | 19 | 19 | 19 | 19 | 19 |

### Variables Entered/Removed[a]

| Model | Variables Entered | Variables Removed | Method |
|---|---|---|---|
| 1 | TMAX | . | Stepwise (Criteria: Probability-of-F-to-enter <= .050, Probability-of-F-to-remove >= .100). |

a. Dependent Variable: CASES

### Model Summary

| Model | R | R Square | Adjusted R Square | Std. Error of the Estimate | Change Statistics | | | | |
|---|---|---|---|---|---|---|---|---|---|
| | | | | | R Square Change | F Change | df1 | df2 | Sig. F Change |
| 1 | .796[a] | .634 | .613 | .22324 | .634 | 29.460 | 1 | 17 | .000 |

a. Predictors: (Constant), TMAX

### ANOVA[a]

| Model | | Sum of Squares | df | Mean Square | F | Sig. |
|---|---|---|---|---|---|---|
| 1 | Regression | 1.468 | 1 | 1.468 | 29.460 | .000[b] |
| | Residual | .847 | 17 | .050 | | |
| | Total | 2.315 | 18 | | | |

a. Dependent Variable: CASES

b. Predictors: (Constant), TMAX

## Coefficients[a]

| Model | | Unstandardized Coefficients | | Standardized Coefficients | | | Correlations | | | Collinearity Statistics | |
|---|---|---|---|---|---|---|---|---|---|---|---|
| | | B | Std. Error | Beta | t | Sig. | Zero-order | Partial | Part | Tolerance | VIF |
| 1 | (Constant) | 215.555 | 39.358 | | 5.477 | .000 | | | | | |
| | TMAX | -86.847 | 16.000 | -.796 | -5.428 | .000 | -.796 | -.796 | -.796 | 1.000 | 1.000 |

a. Dependent Variable: CASES

## Excluded Variables[a]

| Model | | Beta In | t | Sig. | Partial Correlation | Collinearity Statistics | | |
|---|---|---|---|---|---|---|---|---|
| | | | | | | Tolerance | VIF | Minimum Tolerance |
| 1 | UV | -.136[b] | -.664 | .516 | -.164 | .531 | 1.884 | .531 |
| | T | -.168[b] | -.338 | .740 | -.084 | .092 | 10.895 | .092 |
| | TMIN | -.025[b] | -.095 | .926 | -.024 | .325 | 3.077 | .325 |
| | HUMIDITY | -.153[b] | -.835 | .416 | -.204 | .650 | 1.539 | .650 |
| | WINDS | -.263[b] | -1.847 | .083 | -.419 | .928 | 1.078 | .928 |
| | WINDD | .048[b] | .314 | .758 | .078 | .963 | 1.038 | .963 |
| | RAIN | -.160[b] | -1.036 | .315 | -.251 | .903 | 1.107 | .903 |

| | | | | | | | |
|---|---|---|---|---|---|---|---|
| CLOUDS | .086[b] | .567 | .579 | .140 | .968 | 1.033 | .968 |

a. Dependent Variable: CASES

b. Predictors in the Model: (Constant), TMAX

**Collinearity Diagnostics[a]**

| Model | Dimension | Eigenvalue | Condition Index | Variance Proportions (Constant) | TMAX |
|---|---|---|---|---|---|
| 1 | 1 | 2.000 | 1.000 | .00 | .00 |
| | 2 | 8.466E-7 | 1536.989 | 1.00 | 1.00 |

a. Dependent Variable: CASES

# Spain 7 Day Lag without Temperature

**Descriptive Statistics**

| | Mean | Std. Deviation | N |
|---|---|---|---|
| CASES | 1.9270 | .35866 | 19 |
| UV | .7743 | .05073 | 19 |
| HUMIDITY | 1.8771 | .02233 | 19 |
| WINDS | .4526 | .10106 | 19 |
| WINDD | 2.1274 | .08436 | 19 |
| RAIN | -1.2485 | .59064 | 19 |
| CLOUDS | 1.5674 | .08462 | 19 |

**Correlations**

| | | CASES | UV | HUMIDITY | WINDS | WINDD | RAIN | CLOUDS |
|---|---|---|---|---|---|---|---|---|
| Pearson Correlation | CASES | 1.000 | -.618 | .372 | -.458 | -.106 | .103 | .226 |
| | UV | -.618 | 1.000 | -.654 | .514 | -.204 | -.179 | -.200 |
| | HUMIDITY | .372 | -.654 | 1.000 | -.044 | -.279 | .347 | .147 |
| | WINDS | -.458 | .514 | -.044 | 1.000 | -.281 | -.141 | .116 |
| | WINDD | -.106 | -.204 | -.279 | -.281 | 1.000 | .272 | -.214 |
| | RAIN | .103 | -.179 | .347 | -.141 | .272 | 1.000 | -.303 |
| | CLOUDS | .226 | -.200 | .147 | .116 | -.214 | -.303 | 1.000 |
| Sig. (1-tailed) | CASES | . | .002 | .059 | .024 | .332 | .337 | .176 |
| | UV | .002 | . | .001 | .012 | .201 | .232 | .206 |
| | HUMIDITY | .059 | .001 | . | .430 | .124 | .073 | .274 |
| | WINDS | .024 | .012 | .430 | . | .122 | .283 | .318 |
| | WINDD | .332 | .201 | .124 | .122 | . | .130 | .189 |
| | RAIN | .337 | .232 | .073 | .283 | .130 | . | .103 |
| | CLOUDS | .176 | .206 | .274 | .318 | .189 | .103 | . |
| N | CASES | 19 | 19 | 19 | 19 | 19 | 19 | 19 |
| | UV | 19 | 19 | 19 | 19 | 19 | 19 | 19 |
| | HUMIDITY | 19 | 19 | 19 | 19 | 19 | 19 | 19 |
| | WINDS | 19 | 19 | 19 | 19 | 19 | 19 | 19 |
| | WINDD | 19 | 19 | 19 | 19 | 19 | 19 | 19 |
| | RAIN | 19 | 19 | 19 | 19 | 19 | 19 | 19 |
| | CLOUDS | 19 | 19 | 19 | 19 | 19 | 19 | 19 |

**Variables Entered/Removed[a]**

| Model | Variables Entered | Variables Removed | Method |
|---|---|---|---|
| 1 | UV | . | Stepwise (Criteria: Probability-of-F-to-enter <= .050, Probability-of-F-to-remove >= .100). |

a. Dependent Variable: CASES

## Model Summary

| Model | R | R Square | Adjusted R Square | Std. Error of the Estimate | Change Statistics | | | | |
|---|---|---|---|---|---|---|---|---|---|
| | | | | | R Square Change | F Change | df1 | df2 | Sig. F Change |
| 1 | .618[a] | .382 | .345 | .29024 | .382 | 10.486 | 1 | 17 | .005 |

a. Predictors: (Constant), UV

## ANOVA[a]

| Model | | Sum of Squares | df | Mean Square | F | Sig. |
|---|---|---|---|---|---|---|
| 1 | Regression | .883 | 1 | .883 | 10.486 | .005[b] |
| | Residual | 1.432 | 17 | .084 | | |
| | Total | 2.315 | 18 | | | |

a. Dependent Variable: CASES

b. Predictors: (Constant), UV

**Coefficients[a]**

| Model | | Unstandardized Coefficients | | Standardized Coefficients | t | Sig. | Correlations | | | Collinearity Statistics | |
|---|---|---|---|---|---|---|---|---|---|---|---|
| | | B | Std. Error | Beta | | | Zero-order | Partial | Part | Tolerance | VIF |
| 1 | (Constant) | 5.308 | 1.046 | | 5.074 | .000 | | | | | |
| | UV | -4.367 | 1.348 | -.618 | -3.238 | .005 | -.618 | -.618 | -.618 | 1.000 | 1.000 |

a. Dependent Variable: CASES

**Excluded Variables[a]**

| Model | | Beta In | t | Sig. | Partial Correlation | Collinearity Statistics | | |
|---|---|---|---|---|---|---|---|---|
| | | | | | | Tolerance | VIF | Minimum Tolerance |
| 1 | HUMIDITY | -.056[b] | -.217 | .831 | -.054 | .572 | 1.748 | .572 |
| | WINDS | -.192[b] | -.855 | .405 | -.209 | .736 | 1.358 | .736 |
| | WINDD | -.243[b] | -1.269 | .223 | -.302 | .958 | 1.044 | .958 |
| | RAIN | -.007[b] | -.037 | .971 | -.009 | .968 | 1.033 | .968 |
| | CLOUDS | .107[b] | .538 | .598 | .133 | .960 | 1.042 | .960 |

a. Dependent Variable: CASES

b. Predictors in the Model: (Constant), UV

**Collinearity Diagnostics[a]**

| Model | Dimension | Eigenvalue | Condition Index | Variance Proportions |
|---|---|---|---|---|

| | | | | (Constant) | UV |
|---|---|---|---|---|---|
| 1 | 1 | 1.998 | 1.000 | .00 | .00 |
| | 2 | .002 | 31.392 | 1.00 | 1.00 |

a. Dependent Variable: CASES

# Spain 14 Day Lag

**Descriptive Statistics**

| | Mean | Std. Deviation | N |
|---|---|---|---|
| CASES | 1.8586 | .37721 | 19 |
| UV | .7743 | .05073 | 19 |
| T | 2.4564 | .00351 | 19 |
| TMIN | 2.4518 | .00448 | 19 |
| TMAX | 2.4598 | .00329 | 19 |
| HUMIDITY | 1.8771 | .02233 | 19 |
| WINDS | .4526 | .10106 | 19 |
| WINDD | 2.1274 | .08436 | 19 |
| RAIN | -1.2485 | .59064 | 19 |
| CLOUD | 1.5674 | .08462 | 19 |

**Correlations**

| | | CASES | UV | T | TMIN | TMAX | HUMIDITY | WINDS | WINDD | RAIN | CLOUD |
|---|---|---|---|---|---|---|---|---|---|---|---|
| Pearson Correlation | CASES | 1.000 | -.620 | -.772 | -.660 | -.790 | .377 | -.438 | -.083 | .118 | .250 |
| | UV | -.620 | 1.000 | .721 | .696 | .685 | -.654 | .514 | -.204 | -.179 | -.200 |
| | T | -.772 | .721 | 1.000 | .931 | .953 | -.521 | .366 | .152 | -.260 | -.310 |
| | TMIN | -.660 | .696 | .931 | 1.000 | .822 | -.453 | .447 | .042 | -.253 | -.477 |
| | TMAX | -.790 | .685 | .953 | .822 | 1.000 | -.592 | .269 | .192 | -.311 | -.179 |
| | HUMIDITY | .377 | -.654 | -.521 | -.453 | -.592 | 1.000 | -.044 | -.279 | .347 | .147 |
| | WINDS | -.438 | .514 | .366 | .447 | .269 | -.044 | 1.000 | -.281 | -.141 | .116 |
| | WINDD | -.083 | -.204 | .152 | .042 | .192 | -.279 | -.281 | 1.000 | .272 | -.214 |
| | RAIN | .118 | -.179 | -.260 | -.253 | -.311 | .347 | -.141 | .272 | 1.000 | -.303 |
| | CLOUD | .250 | -.200 | -.310 | -.477 | -.179 | .147 | .116 | -.214 | -.303 | 1.000 |
| Sig. (1-tailed) | CASES | . | .002 | .000 | .001 | .000 | .056 | .030 | .367 | .315 | .151 |
| | UV | .002 | . | .000 | .000 | .001 | .001 | .012 | .201 | .232 | .206 |
| | T | .000 | .000 | . | .000 | .000 | .011 | .062 | .267 | .141 | .098 |
| | TMIN | .001 | .000 | .000 | . | .000 | .026 | .027 | .432 | .148 | .019 |
| | TMAX | .000 | .001 | .000 | .000 | . | .004 | .133 | .216 | .098 | .232 |
| | HUMIDITY | .056 | .001 | .011 | .026 | .004 | . | .430 | .124 | .073 | .274 |
| | WINDS | .030 | .012 | .062 | .027 | .133 | .430 | . | .122 | .283 | .318 |
| | WINDD | .367 | .201 | .267 | .432 | .216 | .124 | .122 | . | .130 | .189 |
| | RAIN | .315 | .232 | .141 | .148 | .098 | .073 | .283 | .130 | . | .103 |
| | CLOUD | .151 | .206 | .098 | .019 | .232 | .274 | .318 | .189 | .103 | . |
| N | CASES | 19 | 19 | 19 | 19 | 19 | 19 | 19 | 19 | 19 | 19 |
| | UV | 19 | 19 | 19 | 19 | 19 | 19 | 19 | 19 | 19 | 19 |
| | T | 19 | 19 | 19 | 19 | 19 | 19 | 19 | 19 | 19 | 19 |
| | TMIN | 19 | 19 | 19 | 19 | 19 | 19 | 19 | 19 | 19 | 19 |
| | TMAX | 19 | 19 | 19 | 19 | 19 | 19 | 19 | 19 | 19 | 19 |
| | HUMIDITY | 19 | 19 | 19 | 19 | 19 | 19 | 19 | 19 | 19 | 19 |
| | WINDS | 19 | 19 | 19 | 19 | 19 | 19 | 19 | 19 | 19 | 19 |
| | WINDD | 19 | 19 | 19 | 19 | 19 | 19 | 19 | 19 | 19 | 19 |
| | RAIN | 19 | 19 | 19 | 19 | 19 | 19 | 19 | 19 | 19 | 19 |
| | CLOUD | 19 | 19 | 19 | 19 | 19 | 19 | 19 | 19 | 19 | 19 |

**Variables Entered/Removed[a]**

| Model | Variables Entered | Variables Removed | Method |
|---|---|---|---|
| 1 | TMAX | . | Stepwise (Criteria: Probability-of-F-to-enter <= .050, Probability-of-F-to-remove >= .100). |

a. Dependent Variable: CASES

**Model Summary**

| Model | R | R Square | Adjusted R Square | Std. Error of the Estimate | Change Statistics | | | | |
|---|---|---|---|---|---|---|---|---|---|
| | | | | | R Square Change | F Change | df1 | df2 | Sig. F Change |
| 1 | .790[a] | .625 | .602 | .23784 | .625 | 28.277 | 1 | 17 | .000 |

a. Predictors: (Constant), TMAX

**ANOVA[a]**

| Model | | Sum of Squares | df | Mean Square | F | Sig. |
|---|---|---|---|---|---|---|
| 1 | Regression | 1.600 | 1 | 1.600 | 28.277 | .000[b] |
| | Residual | .962 | 17 | .057 | | |
| | Total | 2.561 | 18 | | | |

a. Dependent Variable: CASES

b. Predictors: (Constant), TMAX

## Coefficients[a]

| Model | | Unstandardized Coefficients | | Standardized Coefficients | t | Sig. | Correlations | | | Collinearity Statistics | |
|---|---|---|---|---|---|---|---|---|---|---|---|
| | | B | Std. Error | Beta | | | Zero-order | Partial | Part | Tolerance | VIF |
| 1 | (Constant) | 224.839 | 41.932 | | 5.362 | .000 | | | | | |
| | TMAX | -90.649 | 17.047 | -.790 | -5.318 | .000 | -.790 | -.790 | -.790 | 1.000 | 1.000 |

a. Dependent Variable: CASES

## Excluded Variables[a]

| Model | | Beta In | t | Sig. | Partial Correlation | Collinearity Statistics | | |
|---|---|---|---|---|---|---|---|---|
| | | | | | | Tolerance | VIF | Minimum Tolerance |
| 1 | UV | -.147[b] | -.711 | .487 | -.175 | .531 | 1.884 | .531 |
| | T | -.203[b] | -.404 | .692 | -.100 | .092 | 10.895 | .092 |
| | TMIN | -.033[b] | -.122 | .905 | -.030 | .325 | 3.077 | .325 |
| | HUMIDITY | -.140[b] | -.752 | .463 | -.185 | .650 | 1.539 | .650 |
| | WINDS | -.244[b] | -1.657 | .117 | -.383 | .928 | 1.078 | .928 |
| | WINDD | .071[b] | .458 | .653 | .114 | .963 | 1.038 | .963 |
| | RAIN | -.141[b] | -.897 | .383 | -.219 | .903 | 1.107 | .903 |

| | | | | | | | |
|---|---|---|---|---|---|---|---|
| CLOUD | .112[b] | .730 | .476 | .180 | .968 | 1.033 | .968 |

a. Dependent Variable: CASES

b. Predictors in the Model: (Constant), TMAX

**Collinearity Diagnostics[a]**

| Model | Dimension | Eigenvalue | Condition Index | Variance Proportions (Constant) | TMAX |
|---|---|---|---|---|---|
| 1 | 1 | 2.000 | 1.000 | .00 | .00 |
| | 2 | 8.466E-7 | 1536.989 | 1.00 | 1.00 |

a. Dependent Variable: CASES

# Spain 14 Day Lag without Temperature

**Descriptive Statistics**

| | Mean | Std. Deviation | N |
|---|---|---|---|
| CASES | 1.8586 | .37721 | 19 |
| UV | .7743 | .05073 | 19 |
| HUMIDITY | 1.8771 | .02233 | 19 |
| WINDS | .4526 | .10106 | 19 |
| WINDD | 2.1274 | .08436 | 19 |
| RAIN | -1.2485 | .59064 | 19 |
| CLOUD | 1.5674 | .08462 | 19 |

**Correlations**

| | | CASES | UV | HUMIDITY | WINDS | WINDD | RAIN | CLOUD |
|---|---|---|---|---|---|---|---|---|
| Pearson Correlation | CASES | 1.000 | -.620 | .377 | -.438 | -.083 | .118 | .250 |
| | UV | -.620 | 1.000 | -.654 | .514 | -.204 | -.179 | -.200 |
| | HUMIDITY | .377 | -.654 | 1.000 | -.044 | -.279 | .347 | .147 |
| | WINDS | -.438 | .514 | -.044 | 1.000 | -.281 | -.141 | .116 |
| | WINDD | -.083 | -.204 | -.279 | -.281 | 1.000 | .272 | -.214 |
| | RAIN | .118 | -.179 | .347 | -.141 | .272 | 1.000 | -.303 |
| | CLOUD | .250 | -.200 | .147 | .116 | -.214 | -.303 | 1.000 |
| Sig. (1-tailed) | CASES | . | .002 | .056 | .030 | .367 | .315 | .151 |
| | UV | .002 | . | .001 | .012 | .201 | .232 | .206 |
| | HUMIDITY | .056 | .001 | . | .430 | .124 | .073 | .274 |
| | WINDS | .030 | .012 | .430 | . | .122 | .283 | .318 |
| | WINDD | .367 | .201 | .124 | .122 | . | .130 | .189 |
| | RAIN | .315 | .232 | .073 | .283 | .130 | . | .103 |
| | CLOUD | .151 | .206 | .274 | .318 | .189 | .103 | . |
| N | CASES | 19 | 19 | 19 | 19 | 19 | 19 | 19 |
| | UV | 19 | 19 | 19 | 19 | 19 | 19 | 19 |
| | HUMIDITY | 19 | 19 | 19 | 19 | 19 | 19 | 19 |
| | WINDS | 19 | 19 | 19 | 19 | 19 | 19 | 19 |
| | WINDD | 19 | 19 | 19 | 19 | 19 | 19 | 19 |
| | RAIN | 19 | 19 | 19 | 19 | 19 | 19 | 19 |
| | CLOUD | 19 | 19 | 19 | 19 | 19 | 19 | 19 |

**Variables Entered/Removed[a]**

| Model | Variables Entered | Variables Removed | Method |
|---|---|---|---|
| 1 | UV | . | Stepwise (Criteria: Probability-of-F-to-enter <= .050, Probability-of-F-to-remove >= .100). |

a. Dependent Variable: CASES

## Model Summary

| Model | R | R Square | Adjusted R Square | Std. Error of the Estimate | Change Statistics | | | | |
|---|---|---|---|---|---|---|---|---|---|
| | | | | | R Square Change | F Change | df1 | df2 | Sig. F Change |
| 1 | .620[a] | .384 | .348 | .30469 | .384 | 10.588 | 1 | 17 | .005 |

a. Predictors: (Constant), UV

## ANOVA[a]

| Model | | Sum of Squares | df | Mean Square | F | Sig. |
|---|---|---|---|---|---|---|
| 1 | Regression | .983 | 1 | .983 | 10.588 | .005[b] |
| | Residual | 1.578 | 17 | .093 | | |
| | Total | 2.561 | 18 | | | |

a. Dependent Variable: CASES

b. Predictors: (Constant), UV

**Coefficients[a]**

| Model | | Unstandardized Coefficients B | Unstandardized Coefficients Std. Error | Standardized Coefficients Beta | t | Sig. | Correlations Zero-order | Correlations Partial | Correlations Part | Collinearity Statistics Tolerance | Collinearity Statistics VIF |
|---|---|---|---|---|---|---|---|---|---|---|---|
| 1 | (Constant) | 5.425 | 1.098 | | 4.940 | .000 | | | | | |
| | UV | -4.606 | 1.416 | -.620 | -3.254 | .005 | -.620 | -.620 | -.620 | 1.000 | 1.000 |

a. Dependent Variable: CASES

**Excluded Variables[a]**

| Model | | Beta In | t | Sig. | Partial Correlation | Collinearity Statistics Tolerance | Collinearity Statistics VIF | Collinearity Statistics Minimum Tolerance |
|---|---|---|---|---|---|---|---|---|
| 1 | HUMIDITY | -.050[b] | -.194 | .849 | -.048 | .572 | 1.748 | .572 |
| | WINDS | -.163[b] | -.725 | .479 | -.178 | .736 | 1.358 | .736 |
| | WINDD | -.219[b] | -1.136 | .273 | -.273 | .958 | 1.044 | .958 |
| | RAIN | .008[b] | .038 | .970 | .009 | .968 | 1.033 | .968 |
| | CLOUD | .131[b] | .664 | .516 | .164 | .960 | 1.042 | .960 |

a. Dependent Variable: CASES

b. Predictors in the Model: (Constant), UV

**Collinearity Diagnostics[a]**

| Model | Dimension | Eigenvalue | Condition Index | Variance Proportions |
|---|---|---|---|---|

| | | | | (Constant) | UV |
|---|---|---|---|---|---|
| 1 | 1 | 1.998 | 1.000 | .00 | .00 |
| | 2 | .002 | 31.392 | 1.00 | 1.00 |

a. Dependent Variable: CASES

---------

---------

*c) Results of stepwise multiple regression of cases for Australia by climatic variables with zero, three-day, seven-day and fourteen-day time lags.*

## Australia Zero Day Lag

**Descriptive Statistics**

| | Mean | Std. Deviation | N |
|---|---|---|---|
| CASES | .5957 | .18038 | 8 |
| UV | .7437 | .17275 | 8 |
| T | 2.4638 | .00797 | 8 |
| TMIN | 2.4604 | .00827 | 8 |
| TMAX | 2.4665 | .00765 | 8 |
| HUMIDITY | 1.8262 | .03471 | 8 |
| WINDS | .6081 | .06747 | 8 |
| WINDD | 2.2773 | .09537 | 8 |
| RAIN | -1.0103 | .37387 | 8 |
| CLOUDS | 1.5015 | .18994 | 8 |

**Correlations**

| | | CASES | UV | T | TMIN | TMAX | HUMIDITY | WINDS | WINDD | RAIN | CLOUDS |
|---|---|---|---|---|---|---|---|---|---|---|---|
| Pearson Correlation | CASES | 1.000 | -.862 | -.778 | -.784 | -.755 | -.084 | .679 | .761 | .634 | .136 |
| | UV | -.862 | 1.000 | .942 | .941 | .928 | -.053 | -.792 | -.868 | -.665 | -.399 |
| | T | -.778 | .942 | 1.000 | .999 | .998 | -.245 | -.702 | -.838 | -.691 | -.380 |
| | TMIN | -.784 | .941 | .999 | 1.000 | .995 | -.226 | -.702 | -.819 | -.686 | -.374 |
| | TMAX | -.755 | .928 | .998 | .995 | 1.000 | -.263 | -.683 | -.834 | -.705 | -.396 |
| | HUMIDITY | -.084 | -.053 | -.245 | -.226 | -.263 | 1.000 | -.265 | .419 | .417 | .488 |
| | WINDS | .679 | -.792 | -.702 | -.702 | -.683 | -.265 | 1.000 | .572 | .454 | .056 |
| | WINDD | .761 | -.868 | -.838 | -.819 | -.834 | .419 | .572 | 1.000 | .747 | .493 |
| | RAIN | .634 | -.665 | -.691 | -.686 | -.705 | .417 | .454 | .747 | 1.000 | .692 |
| | CLOUDS | .136 | -.399 | -.380 | -.374 | -.396 | .488 | .056 | .493 | .692 | 1.000 |
| Sig. (1-tailed) | CASES | . | .003 | .011 | .011 | .015 | .422 | .032 | .014 | .046 | .374 |
| | UV | .003 | . | .000 | .000 | .000 | .451 | .010 | .003 | .036 | .164 |
| | T | .011 | .000 | . | .000 | .000 | .279 | .026 | .005 | .029 | .177 |
| | TMIN | .011 | .000 | .000 | . | .000 | .296 | .026 | .006 | .030 | .181 |
| | TMAX | .015 | .000 | .000 | .000 | . | .264 | .031 | .005 | .025 | .165 |
| | HUMIDITY | .422 | .451 | .279 | .296 | .264 | . | .263 | .151 | .152 | .110 |
| | WINDS | .032 | .010 | .026 | .026 | .031 | .263 | . | .069 | .130 | .448 |
| | WINDD | .014 | .003 | .005 | .006 | .005 | .151 | .069 | . | .017 | .107 |
| | RAIN | .046 | .036 | .029 | .030 | .025 | .152 | .130 | .017 | . | .028 |
| | CLOUDS | .374 | .164 | .177 | .181 | .165 | .110 | .448 | .107 | .028 | . |
| N | CASES | 8 | 8 | 8 | 8 | 8 | 8 | 8 | 8 | 8 | 8 |
| | UV | 8 | 8 | 8 | 8 | 8 | 8 | 8 | 8 | 8 | 8 |
| | T | 8 | 8 | 8 | 8 | 8 | 8 | 8 | 8 | 8 | 8 |
| | TMIN | 8 | 8 | 8 | 8 | 8 | 8 | 8 | 8 | 8 | 8 |
| | TMAX | 8 | 8 | 8 | 8 | 8 | 8 | 8 | 8 | 8 | 8 |
| | HUMIDITY | 8 | 8 | 8 | 8 | 8 | 8 | 8 | 8 | 8 | 8 |
| | WINDS | 8 | 8 | 8 | 8 | 8 | 8 | 8 | 8 | 8 | 8 |
| | WINDD | 8 | 8 | 8 | 8 | 8 | 8 | 8 | 8 | 8 | 8 |
| | RAIN | 8 | 8 | 8 | 8 | 8 | 8 | 8 | 8 | 8 | 8 |
| | CLOUDS | 8 | 8 | 8 | 8 | 8 | 8 | 8 | 8 | 8 | 8 |

### Variables Entered/Removed[a]

| Model | Variables Entered | Variables Removed | Method |
|---|---|---|---|
| 1 | UV | . | Stepwise (Criteria: Probability-of-F-to-enter <= .050, Probability-of-F-to-remove >= .100). |

a. Dependent Variable: CASES

### Model Summary

| Model | R | R Square | | | Change Statistics |
|---|---|---|---|---|---|

| | | | Adjusted R Square | Std. Error of the Estimate | R Square Change | F Change | df1 | df2 | Sig. F Change |
|---|---|---|---|---|---|---|---|---|---|
| 1 | .862[a] | .744 | .701 | .09860 | .744 | 17.425 | 1 | 6 | .006 |

a. Predictors: (Constant), UV

## ANOVA[a]

| Model | | Sum of Squares | df | Mean Square | F | Sig. |
|---|---|---|---|---|---|---|
| 1 | Regression | .169 | 1 | .169 | 17.425 | .006[b] |
| | Residual | .058 | 6 | .010 | | |
| | Total | .228 | 7 | | | |

a. Dependent Variable: CASES

b. Predictors: (Constant), UV

## Coefficients[a]

| Model | | Unstandardized Coefficients | | Standardized Coefficients | t | Sig. | Correlations | | | Collinearity Statistics | |
|---|---|---|---|---|---|---|---|---|---|---|---|
| | | B | Std. Error | Beta | | | Zero-order | Partial | Part | Tolerance | VIF |
| 1 | (Constant) | 1.265 | .164 | | 7.707 | .000 | | | | | |
| | UV | -.901 | .216 | -.862 | -4.174 | .006 | -.862 | -.862 | -.862 | 1.000 | 1.000 |

a. Dependent Variable: CASES

## Excluded Variables[a]

| Model | Beta In | t | Sig. | Partial Correlation | Collinearity Statistics |
|---|---|---|---|---|---|

| | | | | | | Tolerance | VIF | Minimum Tolerance |
|---|---|---|---|---|---|---|---|---|
| 1 | T | .307[b] | .464 | .662 | .203 | .112 | 8.900 | .112 |
| | TMIN | .243[b] | .366 | .729 | .162 | .114 | 8.801 | .114 |
| | TMAX | .324[b] | .552 | .605 | .240 | .140 | 7.165 | .140 |
| | HUMIDITY | -.130[b] | -.592 | .579 | -.256 | .997 | 1.003 | .997 |
| | WINDS | -.010[b] | -.028 | .979 | -.012 | .373 | 2.682 | .373 |
| | WINDD | .050[b] | .109 | .917 | .049 | .247 | 4.057 | .247 |
| | RAIN | .108[b] | .361 | .733 | .159 | .557 | 1.795 | .557 |
| | CLOUDS | -.248[b] | -1.124 | .312 | -.449 | .841 | 1.189 | .841 |

a. Dependent Variable: CASES

b. Predictors in the Model: (Constant), UV

**Collinearity Diagnostics[a]**

| Model | Dimension | Eigenvalue | Condition Index | Variance Proportions | |
|---|---|---|---|---|---|
| | | | | (Constant) | UV |
| 1 | 1 | 1.977 | 1.000 | .01 | .01 |
| | 2 | .023 | 9.312 | .99 | .99 |

a. Dependent Variable: CASES

# Australia 3 Day Lag

**Descriptive Statistics**

| | Mean | Std. Deviation | N |
|---|---|---|---|
| CASES | .5290 | .18180 | 8 |
| UV | .7437 | .17275 | 8 |
| T | 2.4638 | .00797 | 8 |
| TMIN | 2.4604 | .00827 | 8 |
| TMAX | 2.4665 | .00765 | 8 |
| HUMIDITY | 1.8262 | .03471 | 8 |
| WINDS | .6081 | .06747 | 8 |
| WINDD | 2.2773 | .09537 | 8 |
| RAIN | -1.0103 | .37387 | 8 |
| CLOUD | 1.5015 | .18994 | 8 |

**Correlations**

| | | CASES | UV | T | TMIN | TMAX | HUMIDITY | WINDS | WINDD | RAIN | CLOUD |
|---|---|---|---|---|---|---|---|---|---|---|---|
| Pearson Correlation | CASES | 1.000 | -.849 | -.724 | -.725 | -.705 | -.119 | .645 | .784 | .628 | .188 |
| | UV | -.849 | 1.000 | .942 | .941 | .928 | -.053 | -.792 | -.868 | -.665 | -.399 |
| | T | -.724 | .942 | 1.000 | .999 | .998 | -.245 | -.702 | -.838 | -.691 | -.380 |
| | TMIN | -.725 | .941 | .999 | 1.000 | .995 | -.226 | -.702 | -.819 | -.686 | -.374 |
| | TMAX | -.705 | .928 | .998 | .995 | 1.000 | -.263 | -.683 | -.834 | -.705 | -.396 |
| | HUMIDITY | -.119 | -.053 | -.245 | -.226 | -.263 | 1.000 | -.265 | .419 | .417 | .488 |
| | WINDS | .645 | -.792 | -.702 | -.702 | -.683 | -.265 | 1.000 | .572 | .454 | .056 |
| | WINDD | .784 | -.868 | -.838 | -.819 | -.834 | .419 | .572 | 1.000 | .747 | .493 |
| | RAIN | .628 | -.665 | -.691 | -.686 | -.705 | .417 | .454 | .747 | 1.000 | .692 |
| | CLOUD | .188 | -.399 | -.380 | -.374 | -.396 | .488 | .056 | .493 | .692 | 1.000 |
| Sig. (1-tailed) | CASES | . | .004 | .021 | .021 | .025 | .389 | .042 | .011 | .048 | .328 |
| | UV | .004 | . | .000 | .000 | .000 | .451 | .010 | .003 | .036 | .164 |
| | T | .021 | .000 | . | .000 | .000 | .279 | .026 | .005 | .029 | .177 |
| | TMIN | .021 | .000 | .000 | . | .000 | .296 | .026 | .006 | .030 | .181 |
| | TMAX | .025 | .000 | .000 | .000 | . | .264 | .031 | .005 | .025 | .165 |
| | HUMIDITY | .389 | .451 | .279 | .296 | .264 | . | .263 | .151 | .152 | .110 |
| | WINDS | .042 | .010 | .026 | .026 | .031 | .263 | . | .069 | .130 | .448 |
| | WINDD | .011 | .003 | .005 | .006 | .005 | .151 | .069 | . | .017 | .107 |
| | RAIN | .048 | .036 | .029 | .030 | .025 | .152 | .130 | .017 | . | .028 |
| | CLOUD | .328 | .164 | .177 | .181 | .165 | .110 | .448 | .107 | .028 | . |
| N | CASES | 8 | 8 | 8 | 8 | 8 | 8 | 8 | 8 | 8 | 8 |
| | UV | 8 | 8 | 8 | 8 | 8 | 8 | 8 | 8 | 8 | 8 |
| | T | 8 | 8 | 8 | 8 | 8 | 8 | 8 | 8 | 8 | 8 |
| | TMIN | 8 | 8 | 8 | 8 | 8 | 8 | 8 | 8 | 8 | 8 |
| | TMAX | 8 | 8 | 8 | 8 | 8 | 8 | 8 | 8 | 8 | 8 |
| | HUMIDITY | 8 | 8 | 8 | 8 | 8 | 8 | 8 | 8 | 8 | 8 |
| | WINDS | 8 | 8 | 8 | 8 | 8 | 8 | 8 | 8 | 8 | 8 |
| | WINDD | 8 | 8 | 8 | 8 | 8 | 8 | 8 | 8 | 8 | 8 |
| | RAIN | 8 | 8 | 8 | 8 | 8 | 8 | 8 | 8 | 8 | 8 |
| | CLOUD | 8 | 8 | 8 | 8 | 8 | 8 | 8 | 8 | 8 | 8 |

### Variables Entered/Removed[a]

| Model | Variables Entered | Variables Removed | Method |
|---|---|---|---|
| 1 | UV | . | Stepwise (Criteria: Probability-of-F-to-enter <= .050, Probability-of-F-to-remove >= .100). |

a. Dependent Variable: CASES

**Model Summary**

| Model | R | R Square | Adjusted R Square | Std. Error of the Estimate | Change Statistics | | | | |
|---|---|---|---|---|---|---|---|---|---|
| | | | | | R Square Change | F Change | df1 | df2 | Sig. F Change |
| 1 | .849[a] | .721 | .675 | .10367 | .721 | 15.526 | 1 | 6 | .008 |

a. Predictors: (Constant), UV

**ANOVA[a]**

| Model | | Sum of Squares | df | Mean Square | F | Sig. |
|---|---|---|---|---|---|---|
| 1 | Regression | .167 | 1 | .167 | 15.526 | .008[b] |
| | Residual | .064 | 6 | .011 | | |
| | Total | .231 | 7 | | | |

a. Dependent Variable: CASES

b. Predictors: (Constant), UV

**Coefficients[a]**

| Model | | Unstandardized Coefficients | | Standardized Coefficients | t | Sig. | Correlations | | | Collinearity Statistics | |
|---|---|---|---|---|---|---|---|---|---|---|---|
| | | B | Std. Error | Beta | | | Zero-order | Partial | Part | Tolerance | VIF |
| 1 | (Constant) | 1.194 | .173 | | 6.915 | .000 | | | | | |
| | UV | -.894 | .227 | -.849 | -3.940 | .008 | -.849 | -.849 | -.849 | 1.000 | 1.000 |

a. Dependent Variable: CASES

## Excluded Variables[a]

| Model | | Beta In | t | Sig. | Partial Correlation | Collinearity Statistics | | |
|---|---|---|---|---|---|---|---|---|
| | | | | | | Tolerance | VIF | Minimum Tolerance |
| 1 | T | .677[b] | 1.065 | .336 | .430 | .112 | 8.900 | .112 |
| | TMIN | .657[b] | 1.034 | .349 | .420 | .114 | 8.801 | .114 |
| | TMAX | .594[b] | 1.037 | .347 | .421 | .140 | 7.165 | .140 |
| | HUMIDITY | -.164[b] | -.732 | .497 | -.311 | .997 | 1.003 | .997 |
| | WINDS | -.073[b] | -.191 | .856 | -.085 | .373 | 2.682 | .373 |
| | WINDD | .189[b] | .403 | .703 | .177 | .247 | 4.057 | .247 |
| | RAIN | .113[b] | .361 | .733 | .159 | .557 | 1.795 | .557 |
| | CLOUD | -.180[b] | -.734 | .496 | -.312 | .841 | 1.189 | .841 |

a. Dependent Variable: CASES

b. Predictors in the Model: (Constant), UV

## Collinearity Diagnostics[a]

| Model | Dimension | Eigenvalue | Condition Index | Variance Proportions |
|---|---|---|---|---|

| | | | | (Constant) | UV |
|---|---|---|---|---|---|
| 1 | 1 | 1.977 | 1.000 | .01 | .01 |
| | 2 | .023 | 9.312 | .99 | .99 |

a. Dependent Variable: CASES

# Australia 7 Day Lag

**Descriptive Statistics**

| | Mean | Std. Deviation | N |
|---|---|---|---|
| CASES | .3271 | .23194 | 8 |
| UV | .7437 | .17275 | 8 |
| T | 2.4638 | .00797 | 8 |
| TMIN | 2.4604 | .00827 | 8 |
| TMAX | 2.4665 | .00765 | 8 |
| HUMIDITY | 1.8262 | .03471 | 8 |
| WINDS | .6081 | .06747 | 8 |
| WINDD | 2.2773 | .09537 | 8 |
| RAIN | -1.0103 | .37387 | 8 |
| CLOUDS | 1.5015 | .18994 | 8 |

**Correlations**

| | | CASES | UV | T | TMIN | TMAX | HUMIDITY | WINDS | WINDD | RAIN | CLOUDS |
|---|---|---|---|---|---|---|---|---|---|---|---|
| Pearson Correlation | CASES | 1.000 | -.792 | -.582 | -.579 | -.554 | -.173 | .596 | .758 | .492 | .227 |
| | UV | -.792 | 1.000 | .942 | .941 | .928 | -.053 | -.792 | -.868 | -.665 | -.399 |
| | T | -.582 | .942 | 1.000 | .999 | .998 | -.245 | -.702 | -.838 | -.691 | -.380 |
| | TMIN | -.579 | .941 | .999 | 1.000 | .995 | -.226 | -.702 | -.819 | -.686 | -.374 |
| | TMAX | -.554 | .928 | .998 | .995 | 1.000 | -.263 | -.683 | -.834 | -.705 | -.396 |
| | HUMIDITY | -.173 | -.053 | -.245 | -.226 | -.263 | 1.000 | -.265 | .419 | .417 | .488 |
| | WINDS | .596 | -.792 | -.702 | -.702 | -.683 | -.265 | 1.000 | .572 | .454 | .056 |
| | WINDD | .758 | -.868 | -.838 | -.819 | -.834 | .419 | .572 | 1.000 | .747 | .493 |
| | RAIN | .492 | -.665 | -.691 | -.686 | -.705 | .417 | .454 | .747 | 1.000 | .692 |
| | CLOUDS | .227 | -.399 | -.380 | -.374 | -.396 | .488 | .056 | .493 | .692 | 1.000 |
| Sig. (1-tailed) | CASES | . | .010 | .065 | .066 | .077 | .341 | .060 | .015 | .108 | .295 |
| | UV | .010 | . | .000 | .000 | .000 | .451 | .010 | .003 | .036 | .164 |
| | T | .065 | .000 | . | .000 | .000 | .279 | .026 | .005 | .029 | .177 |
| | TMIN | .066 | .000 | .000 | . | .000 | .296 | .026 | .006 | .030 | .181 |
| | TMAX | .077 | .000 | .000 | .000 | . | .264 | .031 | .005 | .025 | .165 |
| | HUMIDITY | .341 | .451 | .279 | .296 | .264 | . | .263 | .151 | .152 | .110 |
| | WINDS | .060 | .010 | .026 | .026 | .031 | .263 | . | .069 | .130 | .448 |
| | WINDD | .015 | .003 | .005 | .006 | .005 | .151 | .069 | . | .017 | .107 |
| | RAIN | .108 | .036 | .029 | .030 | .025 | .152 | .130 | .017 | . | .028 |
| | CLOUDS | .295 | .164 | .177 | .181 | .165 | .110 | .448 | .107 | .028 | . |
| N | CASES | 8 | 8 | 8 | 8 | 8 | 8 | 8 | 8 | 8 | 8 |
| | UV | 8 | 8 | 8 | 8 | 8 | 8 | 8 | 8 | 8 | 8 |
| | T | 8 | 8 | 8 | 8 | 8 | 8 | 8 | 8 | 8 | 8 |
| | TMIN | 8 | 8 | 8 | 8 | 8 | 8 | 8 | 8 | 8 | 8 |
| | TMAX | 8 | 8 | 8 | 8 | 8 | 8 | 8 | 8 | 8 | 8 |
| | HUMIDITY | 8 | 8 | 8 | 8 | 8 | 8 | 8 | 8 | 8 | 8 |
| | WINDS | 8 | 8 | 8 | 8 | 8 | 8 | 8 | 8 | 8 | 8 |
| | WINDD | 8 | 8 | 8 | 8 | 8 | 8 | 8 | 8 | 8 | 8 |
| | RAIN | 8 | 8 | 8 | 8 | 8 | 8 | 8 | 8 | 8 | 8 |
| | CLOUDS | 8 | 8 | 8 | 8 | 8 | 8 | 8 | 8 | 8 | 8 |

**Variables Entered/Removed[a]**

| Model | Variables Entered | Variables Removed | Method |
|---|---|---|---|
| 1 | UV | . | Stepwise (Criteria: Probability-of-F-to-enter <= .050, Probability-of-F-to-remove >= .100). |
| 2 | TMIN | . | Stepwise (Criteria: Probability-of-F-to-enter <= .050, Probability-of-F-to-remove >= .100). |

a. Dependent Variable: CASES

**Model Summary**

| Model | R | R Square | Adjusted R Square | Std. Error of the Estimate | Change Statistics | | | | |
|---|---|---|---|---|---|---|---|---|---|
| | | | | | R Square Change | F Change | df1 | df2 | Sig. F Change |
| 1 | .792[a] | .627 | .565 | .15305 | .627 | 10.076 | 1 | 6 | .019 |
| 2 | .933[b] | .870 | .818 | .09902 | .243 | 9.336 | 1 | 5 | .028 |

a. Predictors: (Constant), UV

b. Predictors: (Constant), UV, TMIN

**ANOVA[a]**

| Model | | Sum of Squares | df | Mean Square | F | Sig. |
|---|---|---|---|---|---|---|
| 1 | Regression | .236 | 1 | .236 | 10.076 | .019[b] |
| | Residual | .141 | 6 | .023 | | |
| | Total | .377 | 7 | | | |
| 2 | Regression | .328 | 2 | .164 | 16.704 | .006[c] |
| | Residual | .049 | 5 | .010 | | |
| | Total | .377 | 7 | | | |

a. Dependent Variable: CASES

b. Predictors: (Constant), UV

c. Predictors: (Constant), UV, TMIN

**Coefficients[a]**

| Model | | Unstandardized Coefficients | | Standardized Coefficients | t | Sig. | Correlations | | | Collinearity Statistics | |
|---|---|---|---|---|---|---|---|---|---|---|---|
| | | B | Std. Error | Beta | | | Zero-order | Partial | Part | Tolerance | VIF |
| 1 | (Constant) | 1.118 | .255 | | 4.385 | .005 | | | | | |
| | UV | -1.063 | .335 | -.792 | -3.174 | .019 | -.792 | -.792 | -.792 | 1.000 | 1.000 |
| 2 | (Constant) | -98.395 | 32.570 | | -3.021 | .029 | | | | | |
| | UV | -2.912 | .643 | -2.169 | -4.530 | .006 | -.792 | -.897 | -.731 | .114 | 8.801 |

| | | | | | | | | | | |
|---|---|---|---|---|---|---|---|---|---|---|
| TMIN | 41.004 | 13.420 | 1.463 | 3.055 | .028 | -.579 | .807 | .493 | .114 | 8.801 |

a. Dependent Variable: CASES

## Excluded Variables[a]

| Model | | Beta In | t | Sig. | Partial Correlation | Collinearity Statistics | | |
|---|---|---|---|---|---|---|---|---|
| | | | | | | Tolerance | VIF | Minimum Tolerance |
| 1 | T | 1.454[b] | 2.961 | .031 | .798 | .112 | 8.900 | .112 |
| | TMIN | 1.463[b] | 3.055 | .028 | .807 | .114 | 8.801 | .114 |
| | TMAX | 1.291[b] | 2.875 | .035 | .789 | .140 | 7.165 | .140 |
| | HUMIDITY | -.215[b] | -.840 | .439 | -.352 | .997 | 1.003 | .997 |
| | WINDS | -.084[b] | -.189 | .858 | -.084 | .373 | 2.682 | .373 |
| | WINDD | .285[b] | .533 | .617 | .232 | .247 | 4.057 | .247 |
| | RAIN | -.063[b] | -.173 | .869 | -.077 | .557 | 1.795 | .557 |
| | CLOUDS | -.106[b] | -.360 | .734 | -.159 | .841 | 1.189 | .841 |
| 2 | T | -.060[c] | -.016 | .988 | -.008 | .002 | 436.185 | .002 |
| | TMAX | .266[c] | .147 | .890 | .073 | .010 | 100.729 | .008 |
| | HUMIDITY | .059[c] | .282 | .792 | .139 | .725 | 1.380 | .083 |
| | WINDS | -.265[c] | -.977 | .384 | -.439 | .356 | 2.805 | .080 |
| | WINDD | .295[c] | .890 | .424 | .407 | .246 | 4.057 | .085 |
| | RAIN | .098[c] | .403 | .707 | .198 | .526 | 1.901 | .107 |
| | CLOUDS | -.109[c] | -.575 | .596 | -.276 | .841 | 1.189 | .111 |

a. Dependent Variable: CASES

b. Predictors in the Model: (Constant), UV

c. Predictors in the Model: (Constant), UV, TMIN

## Collinearity Diagnostics[a]

| Model | Dimension | Eigenvalue | Condition Index | Variance Proportions (Constant) | UV | TMIN |
|---|---|---|---|---|---|---|
| 1 | 1 | 1.977 | 1.000 | .01 | .01 | |
| | 2 | .023 | 9.312 | .99 | .99 | |
| 2 | 1 | 2.970 | 1.000 | .00 | .00 | .00 |
| | 2 | .030 | 9.958 | .00 | .11 | .00 |
| | 3 | 5.697E-7 | 2283.272 | 1.00 | .88 | 1.00 |

a. Dependent Variable: CASES

# Australia 14 Day Lag

## Descriptive Statistics

| | Mean | Std. Deviation | N |
|---|---|---|---|
| CASES | -.0912 | .45140 | 8 |
| UV | .7437 | .17275 | 8 |
| T | 2.4638 | .00797 | 8 |
| TMIN | 2.4604 | .00827 | 8 |
| TMAX | 2.4665 | .00765 | 8 |
| HUMIDITY | 1.8262 | .03471 | 8 |
| WINDS | .6081 | .06747 | 8 |
| WINDD | 2.2773 | .09537 | 8 |
| RAIN | -1.0103 | .37387 | 8 |
| CLOUDS | 1.5015 | .18994 | 8 |

## Correlations

| | | CASES | UV | T | TMIN | TMAX | HUMIDITY | WINDS | WINDD | RAIN | CLOUDS |
|---|---|---|---|---|---|---|---|---|---|---|---|
| Pearson Correlation | CASES | 1.000 | -.861 | -.697 | -.691 | -.666 | -.055 | .632 | .827 | .414 | .197 |
| | UV | -.861 | 1.000 | .942 | .941 | .928 | -.053 | -.792 | -.868 | -.665 | -.399 |
| | T | -.697 | .942 | 1.000 | .999 | .998 | -.245 | -.702 | -.838 | -.691 | -.380 |
| | TMIN | -.691 | .941 | .999 | 1.000 | .995 | -.226 | -.702 | -.819 | -.686 | -.374 |
| | TMAX | -.666 | .928 | .998 | .995 | 1.000 | -.263 | -.683 | -.834 | -.705 | -.396 |
| | HUMIDITY | -.055 | -.053 | -.245 | -.226 | -.263 | 1.000 | -.265 | .419 | .417 | .488 |
| | WINDS | .632 | -.792 | -.702 | -.702 | -.683 | -.265 | 1.000 | .572 | .454 | .056 |
| | WINDD | .827 | -.868 | -.838 | -.819 | -.834 | .419 | .572 | 1.000 | .747 | .493 |
| | RAIN | .414 | -.665 | -.691 | -.686 | -.705 | .417 | .454 | .747 | 1.000 | .692 |
| | CLOUDS | .197 | -.399 | -.380 | -.374 | -.396 | .488 | .056 | .493 | .692 | 1.000 |
| Sig. (1-tailed) | CASES | . | .003 | .027 | .029 | .036 | .449 | .046 | .006 | .154 | .320 |
| | UV | .003 | . | .000 | .000 | .000 | .451 | .010 | .003 | .036 | .164 |
| | T | .027 | .000 | . | .000 | .000 | .279 | .026 | .005 | .029 | .177 |
| | TMIN | .029 | .000 | .000 | . | .000 | .296 | .026 | .006 | .030 | .181 |
| | TMAX | .036 | .000 | .000 | .000 | . | .264 | .031 | .005 | .025 | .165 |
| | HUMIDITY | .449 | .451 | .279 | .296 | .264 | . | .263 | .151 | .152 | .110 |
| | WINDS | .046 | .010 | .026 | .026 | .031 | .263 | . | .069 | .130 | .448 |
| | WINDD | .006 | .003 | .005 | .006 | .005 | .151 | .069 | . | .017 | .107 |
| | RAIN | .154 | .036 | .029 | .030 | .025 | .152 | .130 | .017 | . | .028 |
| | CLOUDS | .320 | .164 | .177 | .181 | .165 | .110 | .448 | .107 | .028 | . |
| N | CASES | 8 | 8 | 8 | 8 | 8 | 8 | 8 | 8 | 8 | 8 |
| | UV | 8 | 8 | 8 | 8 | 8 | 8 | 8 | 8 | 8 | 8 |
| | T | 8 | 8 | 8 | 8 | 8 | 8 | 8 | 8 | 8 | 8 |
| | TMIN | 8 | 8 | 8 | 8 | 8 | 8 | 8 | 8 | 8 | 8 |
| | TMAX | 8 | 8 | 8 | 8 | 8 | 8 | 8 | 8 | 8 | 8 |
| | HUMIDITY | 8 | 8 | 8 | 8 | 8 | 8 | 8 | 8 | 8 | 8 |
| | WINDS | 8 | 8 | 8 | 8 | 8 | 8 | 8 | 8 | 8 | 8 |
| | WINDD | 8 | 8 | 8 | 8 | 8 | 8 | 8 | 8 | 8 | 8 |
| | RAIN | 8 | 8 | 8 | 8 | 8 | 8 | 8 | 8 | 8 | 8 |
| | CLOUDS | 8 | 8 | 8 | 8 | 8 | 8 | 8 | 8 | 8 | 8 |

**Variables Entered/Removed[a]**

| Model | Variables Entered | Variables Removed | Method |
|---|---|---|---|
| 1 | UV | . | Stepwise (Criteria: Probability-of-F-to-enter <= .050, Probability-of-F-to-remove >= .100). |

a. Dependent Variable: CASES

**Model Summary**

| Model | R | R Square | Adjusted R Square | Std. Error of the Estimate | Change Statistics | | | | |
|---|---|---|---|---|---|---|---|---|---|
| | | | | | R Square Change | F Change | df1 | df2 | Sig. F Change |
| 1 | .861[a] | .742 | .699 | .24777 | .742 | 17.235 | 1 | 6 | .006 |

a. Predictors: (Constant), UV

**ANOVA[a]**

| Model | | Sum of Squares | df | Mean Square | F | Sig. |
|---|---|---|---|---|---|---|
| 1 | Regression | 1.058 | 1 | 1.058 | 17.235 | .006[b] |
| | Residual | .368 | 6 | .061 | | |
| | Total | 1.426 | 7 | | | |

a. Dependent Variable: CASES

b. Predictors: (Constant), UV

**Coefficients[a]**

| Model | | Unstandardized Coefficients | | Standardized Coefficients | t | Sig. | Correlations | | | Collinearity Statistics | |
|---|---|---|---|---|---|---|---|---|---|---|---|
| | | B | Std. Error | Beta | | | Zero-order | Partial | Part | Tolerance | VIF |
| 1 | (Constant) | 1.582 | .413 | | 3.836 | .009 | | | | | |
| | UV | -2.250 | .542 | -.861 | -4.152 | .006 | -.861 | -.861 | -.861 | 1.000 | 1.000 |

a. Dependent Variable: CASES

**Excluded Variables[a]**

| Model | | Beta In | t | Sig. | Partial Correlation | Collinearity Statistics | | |
|---|---|---|---|---|---|---|---|---|
| | | | | | | Tolerance | VIF | Minimum Tolerance |
| 1 | T | 1.014[b] | 2.013 | .100 | .669 | .112 | 8.900 | .112 |

| | | | | | | | |
|---|---|---|---|---|---|---|---|
| | TMIN | 1.053[b] | 2.183 | .081 | .699 | .114 | 8.801 | .114 |
| | TMAX | .949[b] | 2.176 | .082 | .697 | .140 | 7.165 | .140 |
| | HUMIDITY | -.100[b] | -.450 | .672 | -.197 | .997 | 1.003 | .997 |
| | WINDS | -.135[b] | -.368 | .728 | -.162 | .373 | 2.682 | .373 |
| | WINDD | .321[b] | .738 | .494 | .313 | .247 | 4.057 | .247 |
| | RAIN | -.286[b] | -1.035 | .348 | -.420 | .557 | 1.795 | .557 |
| | CLOUDS | -.174[b] | -.738 | .494 | -.313 | .841 | 1.189 | .841 |

a. Dependent Variable: CASES

b. Predictors in the Model: (Constant), UV

**Collinearity Diagnostics[a]**

| Model | Dimension | Eigenvalue | Condition Index | Variance Proportions (Constant) | Variance Proportions UV |
|---|---|---|---|---|---|
| 1 | 1 | 1.977 | 1.000 | .01 | .01 |
| | 2 | .023 | 9.312 | .99 | .99 |

a. Dependent Variable: CASES